# Spike Pattern Structure Influences Efficacy Variability under STDP and Synaptic Homeostasis

Zedong Bi[1,2]*, Changsong Zhou[2,3,4,5], Hai-Jun Zhou[1]

[1]State Key Laboratory of Theoretical Physics, Institute of Theoretical Physics, Chinese Academy of Sciences, Beijing 100190, China

[2]Department of Physics, Hong Kong Baptist University, Kowloon Tong, Hong Kong, China

[3]Centre for Nonlinear Studies, and Beijing-Hong Kong-Singapore Joint Centre for Nonlinear and Complex Systems (Hong Kong), Institute of Computational and Theoretical Studies, Hong Kong Baptist University, Kowloon Tong, Hong Kong, China

[4]Beijing Computational Science Research Center, Beijing, China

[5]Research Centre, HKBU Institute of Research and Continuing Education, Virtual University Park Building, South Area Hi-tech Industrial Park, Shenzhen, China

*Corresponding author

Email: zedong.bi@gmail.com.

## Abstract

Neuronal spike trains typically exhibit spatial heterogeneity and temporal stochasticity. This variability are typically due to inherent properties of neurons, synapses and networks, so that the exact spike patterns are hard to be externally manipulated in a detailed way. As synaptic plasticity is usually driven by spike trains, the uncontrollable variability of spike trains should result in uncontrollable variability of synapses during plasticity. However, how the variability of spike trains influences this *efficacy variability* of synapses remains unclear. Here, we systematically study this influence when spike patterns possess four aspects of statistical features, i.e. synchronous firing, auto-temporal structure, heterogeneity of rates and heterogeneity of cross-correlations, under spike-timing dependent plasticity (STDP) after dynamically bounding the mean strength of plastic synapses into or out of a neuron (synaptic homeostasis). Specifically, we first studied how different pattern structures influence the efficacy variability using simple network motifs, i.e. dendritic motif (one neuron receiving from many other neurons) and axonal motifs (many neurons receiving from one), including cases when dendritic and axonal motifs are coupled together. Spikes of neurons in motifs were generated using statistical models, so that pattern structures could be explicitly controlled. We then studied a biologically plausible leaky integrate-and-fire (LIF) neuronal networks, and understood the changes of the efficacy variability based on the motif studies after destroying different aspects of pattern structures using sophisticated spike shuffling methods. We then performed simulations to show that the capability of neuronal networks for faithfully encoding and long-termly maintaining connection patterns is inversely correlated with the efficacy variability, and the efficacy variability is important to understand the competition process driven by retinal waves during the early development of primary visual systems. From motifs to LIF networks and then to biological meanings, we provide a bottom-up framework to understand the efficacy variability.

## Author Summary

In neural systems, synaptic plasticity is usually driven by spike trains. Due to the inherent noises of neurons, synapses and networks, spike trains typically exhibit externally uncontrollable variability such as spatial heterogeneity and temporal stochasticity, resulting in variability of synapses, which we call *efficacy variability*. Spike patterns with the same population rate but inducing different efficacy variability may result in neuronal networks with sharply different structures and functions. However, how the variability of spike trains influences the efficacy variability remains unclear. Here, we systematically study this influence when spike patterns possess four aspects of statistical features, i.e. synchronous firing, auto-temporal structure, heterogeneity of rates and heterogeneity of cross-correlations, under spike-timing dependent plasticity (STDP) after dynamically bounding the mean strength of plastic synapses into or out of a neuron (synaptic homeostasis). We then show the functional importance of efficacy variability on the encoding and maintenance of connection patterns and on the early development of primary visual systems driven by retinal waves. We anticipate our work brings a fresh perspective to the understanding of the interaction between synaptic plasticity and dynamical spike patterns in functional processes of neural systems.

# Introduction

Neuronal spike trains usually exhibit spatial heterogeneity and temporal stochasticity. For example, firing rates are long-tailed distributed in many brain areas [1-3], spatio-temporal correlations within neuronal population often exhibit rich structures [4-7]; and two neurons will not emit the same spike train even if they are receiving exactly the same stimuli [8-10]. The spatial heterogeneity may emerge from neuronal response properties and connection details [11,12], and the temporal stochasticity may be due to the inner stochasticity of neurons and synapses [8-10], both of which are inherent properties of neurons, synapses or networks so that the exact spike patterns are hard to be externally manipulated in a detailed way. As synaptic plasticity is usually driven by spike trains, the variability of spike trains should result in variability of synapses, i.e. the synaptic efficacies can get uncontrollably dissimilar after plasticity even if they start from uniformity. We call this dissimilarity *efficacy variability* of synapses. Synaptic efficacy was observed to be widely distributed *in vivo* [3,13], but this may be induced by deterministic rules. For example, in Hopfield model [14], the connection strength between a pair of neurons participating in 100 memory patterns should be very different from a pair participating in a single pattern. Here, by *efficacy variability*, we emphasize the dissimilarity caused by the uncontrollable spatio-temporal noises during plasticity.

Efficacy variability may have important biological implications. For example, suppose a function of a neuronal network, say memory [15] or spike sequence generation [16], requires a connection pattern in which a few synapses (foreground synapses) have stronger efficacies than the others (background synapses). When the efficacy variability is small, both the foreground and background synapses tend to be uniform around their mean values respectively, thus the connection pattern is clear-cut. However, when the efficacy variability is large, some foreground synapses can be very weak and some background ones can be very strong, which destroys the connection pattern even if the mean strength of the foreground synapses is still larger than that of the background ones (**Fig. 1A**). As another example, synaptic competition and elimination is a classical scenario for the formation of neural network structure during development, when synapses compete with each other for strength and those that are too weak will disappear [17]. In this case, efficacy variability quantifies the degree of competition. If we suppose that the total synaptic strength before elimination is constrained by, say, synaptic homeostasis [18], then when the efficacy variability is small, only a few synapses are below the elimination threshold and get eliminated, and those left also have similar strength; when the efficacy variability is large, a larger portion of synapses get below the elimination threshold, while the remaining ones have a wider efficacy distribution with also a larger mean value than the case of small efficacy variability (**Fig. 1B**). This is consistent with the scenario found in the early development of auditory cortex [19]: if the spontaneous activity of medial nucleus of the trapezoid body (MNTB) is changed using genetic method, then its feedforward projection to lateral superior olive (LSO) becomes denser and weaker, which suggests that the normal pattern induces stronger efficacy variability. Due to its important biological implications, it is a surprise that efficacy variability has not become a key concept and attracted sufficient research attention in neuroscience.

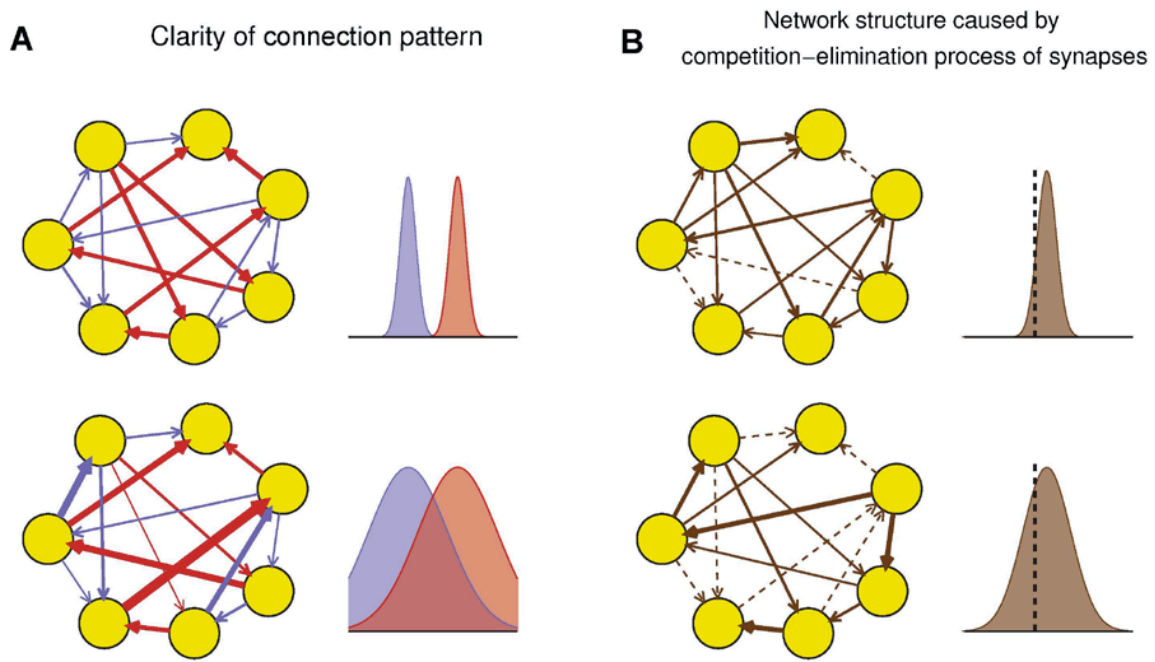

**Fig. 1. Biological implications of efficacy variability.** (**A**) A connection pattern used for, say, memory or spike sequence generation is defined as a few synapses (red) being stronger than the others (blue). When the efficacy variability is small (upper), the connection pattern is clear-cut; when the efficacy variability is large (lower), the connection pattern is destroyed even if the mean strength of the red synapses is still larger than the blue ones. Widths of arrows indicate synaptic strengths. (**B**) Efficacy variability causes different network structures by controlling the degree of synaptic competition. When the efficacy variability is small (upper), only a few synapses is weaker than the elimination threshold (black dashed vertical line), so most synapses are left and their strengths tend to be uniform; when the efficacy variability is large (lower), more synapses are eliminated, and the left ones are more heterogeneous and also stronger than the upper case on average. Dashed arrows represent eliminated synapses.

    Under temporal stochasticity and spatial heterogeneity, spike trains may exhibit a variety of statistical features, which form rich spike pattern structures. Groups of neurons may spurt firing activity (synchronous firing) [20-22], the spike train of a single neuron can be bursty or regular (auto-temporal structure) [23-25], firing rates of cortical neurons are typically long-tailed distributed *in vivo* (heterogeneity of rates) [1-3], and spike trains of different neurons also reveal rich interdependences (heterogeneity of cross-correlations) [4,7,12]. As synaptic plasticity is driven by spike trains, spike pattern structure must have strong influence on efficacy variability, inducing neuronal networks with sharply different structures even under the same population rate.

    To understand how different spike patterns influence efficacy variability, it is helpful to first consider a group of particles doing 1-dimensional Brownian motion driven by noises, starting from the zero point. If the noises imposed on different particles have different biases, for example the noises on particle 1 prefer the positive direction while those on particle 2 prefer the

negative direction, then the Brownian motions of different particles will have different drift velocities, causing displacement variability. If the noises on all the particles have zero bias, the displacements of these particles can also be different due to diffusion. The variability caused by diffusion not only depends on the strength of noises, but also on their cross-correlation and auto-correlation. Cross-correlated noises can push all the particles to simultaneously move positively or negatively, reducing the displacement variability. Auto-correlated noises can push a particle to jump toward the same direction in several adjacent steps within the time scale of the auto-correlation $\tau_{auto}$, increasing the increment of the displacement variance $\Delta\sigma^2$ during $\tau_{auto}$; as the noises separated apart farther than $\tau_{auto}$ are largely independent, the total variance after $t$ time of running is about $\Delta\sigma^2 t / \tau_{auto}$, which increases with $\Delta\sigma^2$. In general, the total variance (ToV) can be written as the summation of the variance caused by drift velocities (DrV) and the variance caused by diffusion (DiV) (**S1 Text Section S1**)

$$\text{ToV} = \text{DrV} + \text{DiV}, \qquad (1)$$

and during $t$ time of evolution $\text{DrV} \propto t^2$ while $\text{DiV} \propto t$. During plasticity, DrV is usually caused by the spatial heterogeneity of spike trains. For example, in classical Hebbian learning synapses sharing the same presynaptic neuron can have different learning rates depending on the firing rates of the post-synaptic neurons; if the plasticity is spike-timing dependent, the heterogeneity of cross-correlations can induce different learning rates even if the firing rates are the same. Because of the inner stochasticity of neurons and synapses, even two neurons receiving exactly the same stimuli emit different spike trains, causing DiV.

In this paper, we systematically study how four aspects of pattern structure, i.e. synchronous firing, auto-temporal structure, heterogeneity of rates and heterogeneity of cross-correlations as well as their interactions influence efficacy variability by taking spike-timing dependent plasticity (STDP) [26] as an example (**Fig. 2AB**). To only focus on the efficacy variability without worrying about the change of the mean, we also introduce synaptic homeostasis, so that the mean strength of the plastic synapses into or out of a neuron is dynamically bounded (dendritic or axonal homeostasis, **Fig. 2C**) (see **Methods** for model details). Physiologically, dendritic homeostasis can be due to activity-dependent protein synthesis in post-synaptic neurons [18,27,28], and axonal homeostasis may be induced by the constraint and reallocation of pre-synaptic resources [29,30]. In the following part of the paper, we will first do our study on dendritic and axonal motifs and their coupling (**Fig. 2DE**), then study biologically plausible conductance-based neuronal models and finally investigate the functional implications of efficacy variability.

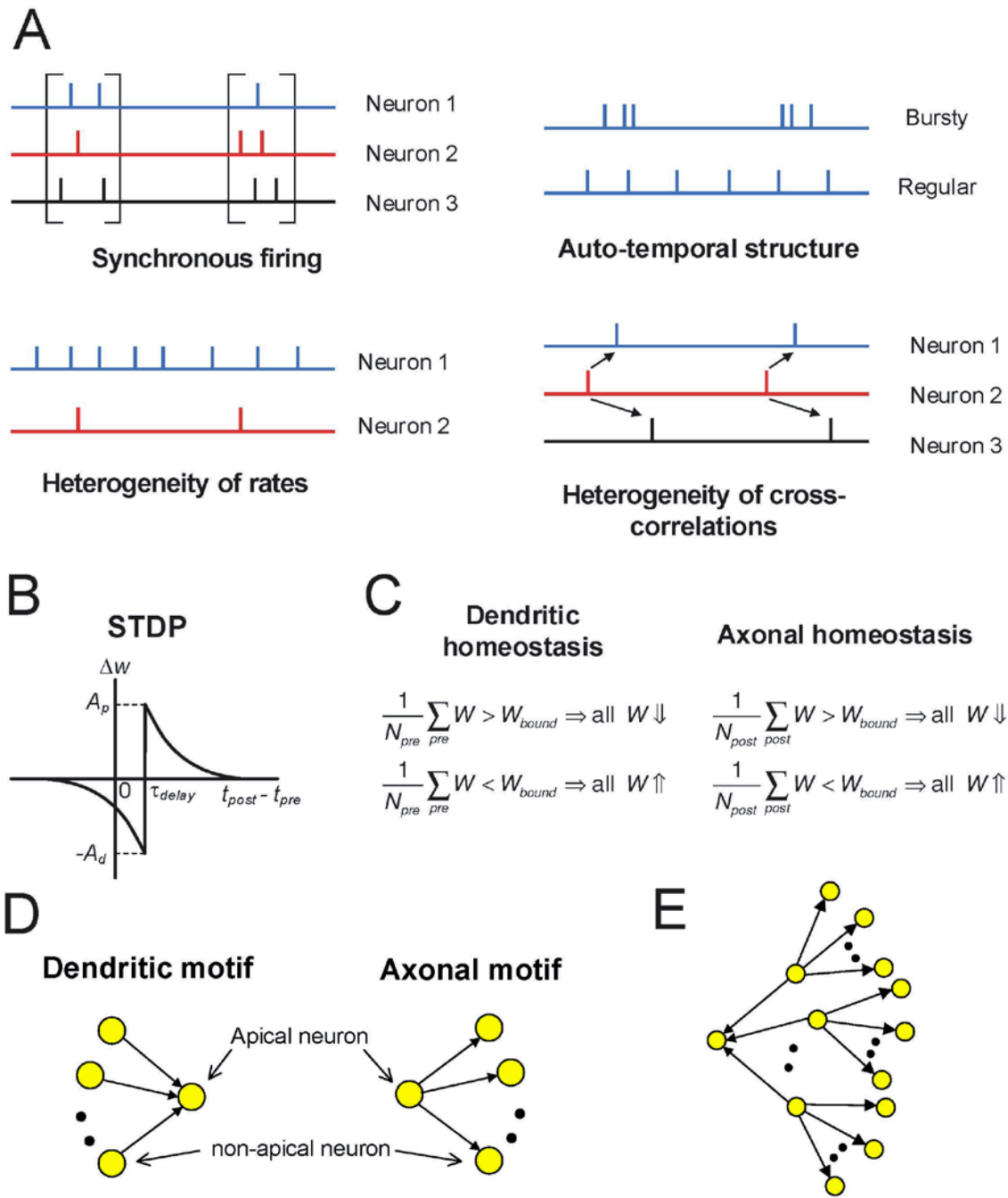

**Fig. 2. Schematic of key concepts in our modeling work.** (**A**) The four aspects of pattern structure studied in this paper. (**B**) The STDP time window used in our work. Note that the axons in our work have time delay $\tau_{delay}$, and the synapses are updated according to the spike time of the post-synaptic neuron and the time that the pre-synaptic spike arrives at the terminal. The STDP updatings of all spike pairs are summed together. (**C**) Dendritic homeostasis and axonal homeostasis. The synapses at a neuron are subject to a soft bound on their mean strength: when the mean strength of the synapses into (or out of) a neuron is different from this bound, all the incoming (or outgoing) synapses of that neuron undergo a slight adjustment. (**D**) Dendritic motif and axonal motif. Dendritic (axonal) homeostasis is imposed onto the apical neuron of a dendritic (axonal) motif. (**E**) A dendritic motif coupled with many axonal motifs. Modeling details are presented in **Methods**.

# Results

## Efficacy Variability in Dendritic Motifs

Dendritic or axonal motifs are simple networks in which one neuron receives from many other neurons or many neurons receive from one (**Fig. 2D**). In this work, activities of these neurons were generated using statistical models, so that we could explicitly control different aspects of pattern structure while keeping population rate constant, and study their influences on the efficacy variability without worrying about the feedback of synaptic changes onto spike patterns as usually happens in biologically more realistic models. We mainly focused on dendritic motifs, as results for axonal motifs are similar (**S1 Text Section S2.1**). In the main text, we focus to explain the mechanisms of how pattern structures influence the efficacy variability, validations using spike generating models are presented in details in **S1 Text Section S2**.

Let us first consider synchronous firing. We use $p$ to represent the number of spikes per neuron during a firing event and $\tau_{cross}$ to represent the duration of a firing event. In a dendritic motif, if all the neurons have the same firing rate, then synchronous firing influences DiV under STDP by three factors: *spike gathering*, *synapse splitting* and *synapse correlating*. Spike gathering means that if $p$ increases, the spikes of the apical and non-apical neurons are gathered closer by synchronous firing, which results in a stronger efficacy change in each STDP updating, thereby increasing DiV. To understand synapse splitting and synapses correlating, suppose a firing event happening during $[t_1, t_2]$ ($t_2 = t_1 + \tau_{cross}$), then the apical neuron will receive its afferents during $[t_1 + \tau_{delay}, t_2 + \tau_{delay}]$, with $\tau_{delay}$ being the axonal delay. If the apical neuron itself fires at $t_0$ with $t_1 + t_{delay} < t_0 < t_2 + t_{delay}$, then all the in-coming spikes during $[t_1 + t_{delay}, t_0)$ potentiate the corresponding synapses, and all the in-coming spikes during $(t_0, t_2 + t_{delay}]$ depress the corresponding synapses, which splits the synapses into different directions (synapse splitting), increasing the efficacy variability. However, if $t_0 < t_1 + t_{delay}$ or $t_0 > t_2 + t_{delay}$, the spikes of the non-apical neurons depress or potentiate their out-going synapses simultaneously. In this case, if the depression or potentiation on these synapses are similar, then the efficacy variability can be reduced (synapse correlating). This similarity of depression or potentiation can strongly depend on the homogeneity of the spike numbers of the non-apical neurons in a firing event. As an example, suppose each non-apical neuron fires one spike in a firing event, then when $t_0 < t_1 + t_{delay}$ or $t_0 > t_2 + t_{delay}$, the potentiation or depression on all the synapses are similar after the firing event; but if half of the non-apical neurons fire no spike, and the other half fire two spikes, then the potentiation or depression will be heterogeneous among the synapses, which may fail the mechanism of synapse correlating to reduce the efficacy variability. When $t_0 < t_1 + t_{delay}$ or $t_0 > t_2 + t_{delay}$, large $\tau_{cross}$ can also make the synaptic updatings heterogeneous, thereby discounting the reduction of the efficacy variability caused by synapse correlating. See **S1 Text Section S2.2** and **S1 Fig** for modeling details.

Now let us add rate heterogeneity into synchronous firing. During STDP, both the strengths of the potentiation and depression processes are proportional to the firing rates of the pre- and post-synaptic neurons. Therefore, the trial expectation of the change of the *a*th synapse in a dendritic motif $\langle \Delta w_a \rangle \propto (S_p - S_d) r_a r_0$, with $r_0$ and $r_a$ being the rates of the apical and *a*th non-apical neuron, and $S_p - S_d$ quantifying the imbalance of potentiation and depression (*P-D imbalance*). When $S_p \neq S_d$, $\langle \Delta w_a \rangle \propto r_a$ with a non-zero coefficient, so that the heterogeneity of $r_a$ will make $\Delta w_a$ drift in different velocities for different *a*, thereby inducing DrV. When the spike trains are homogeneous Poisson, $S_p = A_p$ and $S_d = A_d$, with $A_p$ and $A_d$ being the strengths of the exponentially decayed STDP windows for potentiation and depression (**Fig. 2B**). However, after adding synchronous firing, $S_p$ and $S_d$ can also be influenced by the relative timing of the spike of the apical neuron within a firing event. As an extreme example, when $t_0 < t_1 + t_{delay}$ or $t_0 > t_2 + t_{delay}$, all the synapses are simultaneously depressed or potentiated, strongly changing P-D imbalance; so DrV can be accordingly changed. See **S1 Text Section S2.3** and **S2 Fig** for modeling details.

Both burstiness and strong regularity in auto-temporal structure increase the efficacy variability. To understand the effect of burstiness, consider two adjacent spikes of the apical neuron $\{t_{0,1}, t_{0,2}\}$ and the spike sequences of the non-apical neurons $\mathcal{P}_a = \{t_{a,1}, t_{a,2}, \cdots, t_{a,l}\}$ in the neighborhood of $t_{0,1}$, with $a = 1, 2, \cdots$ being the indexes of the non-apical neurons. So $t_{0,1}$ contributes to the efficacy changes mainly from its interaction with $\mathcal{P}_a$. As our STDP is additive (**Methods**), the efficacy variance caused by $t_{0,1}$ is [31]

$$\text{Var}(t_{0,1}) = \sum_{i=1}^{l} \text{Var}_a \left( \Delta w_a \left( t_{0,1}, t_{a,i} \right) \right) + \sum_{i \neq j}^{l} c_{ij} \sqrt{\text{Var}_a \left( \Delta w_a \left( t_{0,1}, t_{a,i} \right) \right) \cdot \text{Var}_a \left( \Delta w_a \left( t_{0,1}, t_{a,j} \right) \right)} \quad (2)$$

with $\Delta w_a (t_{0,1}, t_{a,i})$ being the efficacy change of the *a*th synapse caused by the pairing of the two spikes $t_{0,1}$ and $t_{a,i}$ using STDP, and $c_{ij}$ is the correlation coefficient between $\Delta w_a (t_{0,1}, t_{a,i})$ and $\Delta w_a (t_{0,1}, t_{a,j})$. To understand the effect of burstiness, note that on the one hand, when $\mathcal{P}_a$ shows strong burstiness, it is clustered into bursting events, which can greatly increase $c_{ij}$ when *i* and *j* are nearby in time, thereby increasing $\text{Var}(t_{0,1})$. On the other hand, the burstiness of the apical neuron itself may gather $t_{0,1}$ and $t_{0,2}$ closer, thereby correlating the STDP updatings caused by these two spikes, i.e. $\sum_i \Delta w_a (t_{0,1}, t_{a,i})$ and $\sum_i \Delta w_a (t_{0,2}, t_{a,i})$, for each *a*. This correlation increases the increment of the efficacy variance $\Delta \sigma^2$ during the time scale $\tau_{auto}$ of the bursting events of the apical neuron. As spikes separated apart farther than $\tau_{auto}$ are largely independent, the total efficacy variance after *t* time of running is approximately $\Delta \sigma^2 t / \tau_{auto}$, which increases with $\Delta\sigma^2$. To understand the effect of strong regularity, consider two adjacent spikes of the apical neuron $\{t_{0,1}, t_{0,2}\}$ and two adjacent spikes of the $a$th non-apical neuron $\{t_{a,1}, t_{a,2}\}$. Suppose $t_{0,1} < t_{a,1}$, then under strong regularity and equal firing rate it is very likely that $t_{0,2} < t_{a,2}$, too. This *transient cross-correlation* correlates the efficacy changes caused by adjacent spikes of the apical neuron, increasing the efficacy variability. Our simulation suggested that the efficacy variability is smallest when *CV* is in the range 0.3~0.7, which is the range most neurons lie within [23]. The efficacy variability caused by auto-temporal structure is of DiV nature. See **S1 Text Section S2.4** and **S3 Fig** for modeling details.

Heterogeneity of cross-correlations mainly influences efficacy variability in DrV manner, and this influence depends on the structure of cross-correlations in spike patterns. In a dendritic motif, the synapses which tend to inject spikes before or after the firing of the apical neuron get weaker or stronger, and the strengths of potentiation and depression also depend on the size, position and duration of the time window of the cross-correlation. Strongest DrV happens when some cross-correlations concentrate onto the negative side of the sharp change point of the STDP time window ($t_{post} - t_{pre} = \tau_{delay}$, see **Fig. 2B**), strongly depressing the corresponding synapses, while the others concentrating onto the positive side, strongly potentiating the corresponding synapses. See **S1 Text Section S2.5** and **S4 Fig** for modeling details.

When these aspects of pattern structure coexist, the above mechanisms how they influence efficacy variability still remain valid, but these mechanisms may interact with each other, inducing more complicated coupling effects. We discuss these interactions in details in **S1 Text Section S2.6-S2.9** (also see **S5-S7 Figs**). An important point here is how to couple synchronous firing and auto-temporal structure in spike patterns together. In our model, the population rate $r(t)$ is determined by the occurrence of firing events, and a rescaled time is then defined as the cumulative function of $r(t)$ [32]

$$\Lambda(t) = \int_0^t r(s) \mathrm{d}s, \tag{S18}$$

which stretches the inter-spike intervals in proportion to the firing rate. Auto-temporal structure then comes into the picture in two ways: the auto-temporal structure of the spikes in the rescaled time, quantified by the *CV* value $CV_{rescale}$, and the temporal structure of the occurrence of firing events, quantified by $CV_{events}$ (**Fig. 3**).

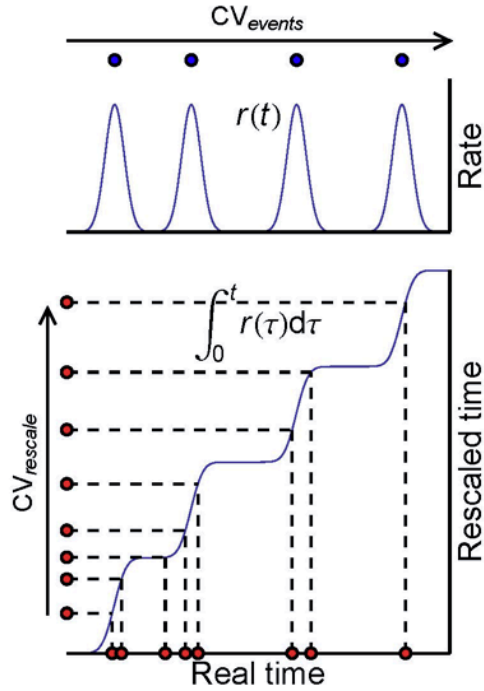

**Fig. 3. The scheme to couple synchronous firing and auto-temporal structure.**
Blue curves represent firing rate (upper) and accumulative function of firing rate (lower) in the real time. Blue dots (upper) represent times of firing events, whose temporal structure is quantified by their *CV* value $CV_{events}$. Red dots (lower) represent spikes in the real time and their correspondences in the rescaled time, whose auto-temporal structure is quantified by their *CV* value $CV_{rescale}$.

To understand how dendritic and axonal homeostasis interact with each other, we consider a dendritic motif coupled with many axonal motifs (**Fig. 2E**), so that the synapses of the dendritic motif are also subject to the axonal homeostasis imposed on the axonal motifs. We denote the strength of the link from the *a*th non-apical neuron in the coupled dendritic motif to the apical neuron as $w_{0a}$, and the mean synaptic strength within the *a*th axonal motif as $\bar{w}_a$. If $w_{0a}$ increases, and $\bar{w}_a$ positively correlates with $w_{0a}$, then $w_{0a}$ can be dragged back by the axonal homeostasis imposed on $\bar{w}_a$. Therefore, if the correlation $\text{Corr}(\Delta w_{0a}, \Delta \bar{w}_a)$ of the STDP updatings onto $w_{0a}$ and $\bar{w}_a$ is positively strong, then the efficacy variability in the coupled dendritic motif can get smaller than that in the free one. On the contrary, if this correlation is negative, then the efficacy variability in the coupled dendritic motif can get larger than that in the free one.

Synchronous firing increases $\text{Corr}(\Delta w_{0a}, \Delta \bar{w}_a)$, because both the changes of $w_{0a}$ and $\bar{w}_a$ after a firing event have a dependence on the relative spike timing of the *a*th non-apical neuron within the firing event: when the *a*th non-apical neuron fires at the beginning of the firing event, both $w_{0a}$ and $\bar{w}_a$ tend to be potentiated; when it fires at the end of the firing event, both $w_{0a}$ and $\bar{w}_a$ tend to be depressed. Heterogeneity of rates increases $\text{Corr}(\Delta w_{0a}, \Delta \bar{w}_a)$ at P-D imbalance, because both $\langle \Delta w_{0a} \rangle$ and $\langle \Delta \bar{w}_a \rangle$ (with $\langle \cdot \rangle$ denoting trial expectation) are proportional to the firing rate of the *a*th non-apical neuron, and the proportional coefficients have the same +/- sign and are non-zero at P-D imbalance. Heterogeneity of cross-correlations may also change $\text{Corr}(\Delta w_{0a}, \Delta \bar{w}_a)$ by introducing a correlation between $\langle \Delta w_{0a} \rangle$ and $\langle \Delta \bar{w}_a \rangle$, but this correlation can be positive or negative, depending on the details of the cross-correlation structure in the spike pattern. Auto-temporal structure, however, hardly has effect when the other aspects of pattern structure are absent. See **S1 Text Section S2.10** and **S8-S10 Figs** for more information.

Generally, when the motif size is large, the condition that the efficacy variability in the coupled dendritic motif is smaller than that in the free one is (**S1 Text eq. S22**)

$$\text{Corr}(\Delta w_{0a}, \Delta \bar{w}_a) > \frac{1}{2} \sqrt{\frac{\text{Var}_a(\Delta \bar{w}_a)}{\text{Var}_a(\Delta w_{0a})}} \tag{3}$$

We summarize the key points in our motifs studies in **Fig. 4**.

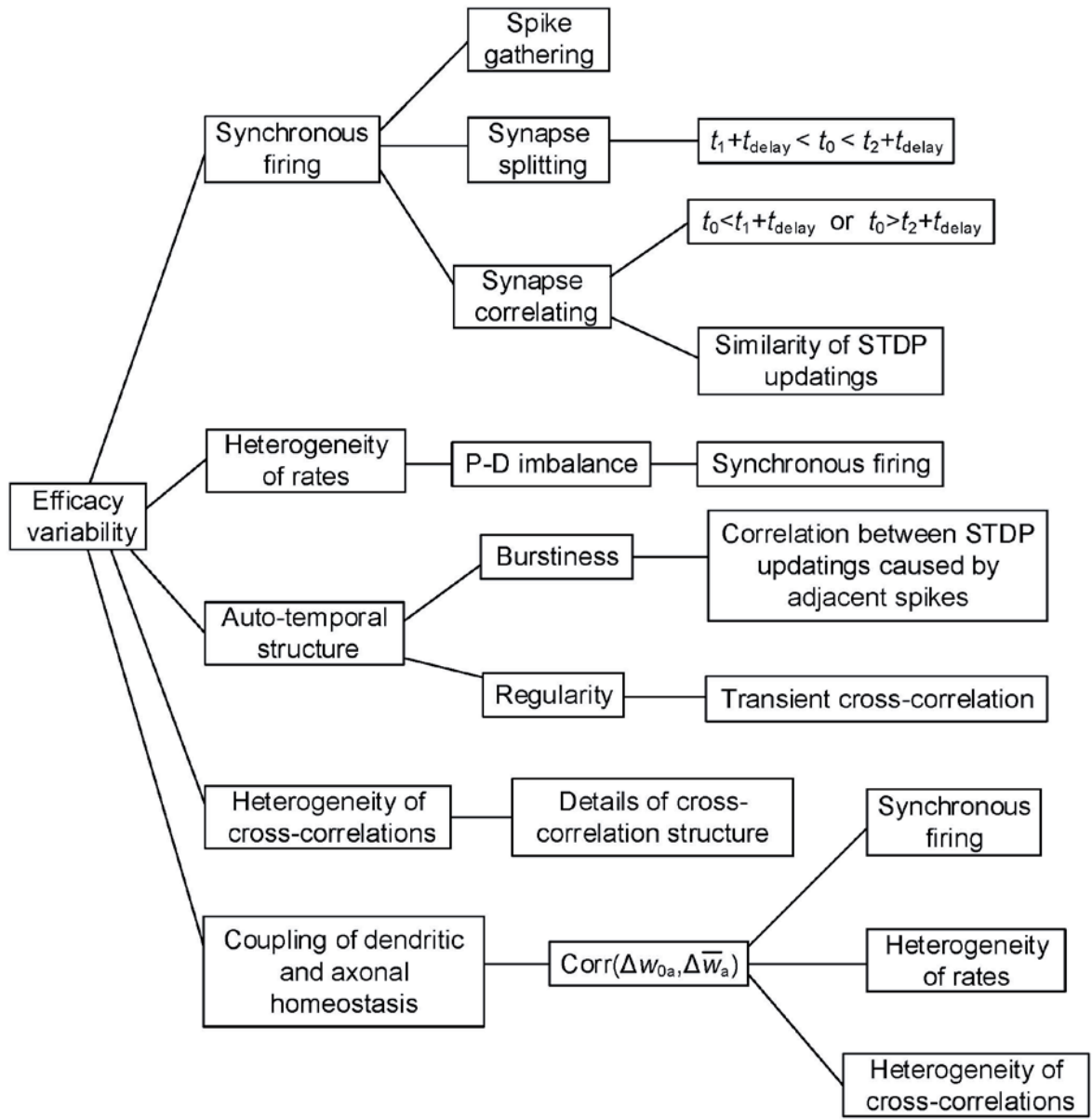

**Fig. 4. Key points to understand the mechanisms of how the four aspects of spike pattern structure influence the efficacy variability under STDP and synaptic homeostasis.** Explanations are presented in the main text. Models to validate these mechanisms and the cases when different aspects of spike pattern structure interact are discussed in details in **SI Text Section S2** and **S1-S10 Figs**.

## Efficacy Variability in LIF Networks

Our next goal is to examine whether our results obtained from studying motifs using spike generating models can still be valid in a more biologically plausible manner. To do this, we simulated a conductance-based leaky integrate-and-fire (LIF) neuronal random network which contained 2000 excitatory neurons and 500 inhibitory neurons with link probability 0.2. We kept the mean rate of the excitatory population at 20Hz and the time scale of the excitatory synaptic conductance at 4ms (see **Methods** for model details). When changing the time scale of the inhibitory synaptic conductance $\tau_{d,I}$ as the integer values from 3ms to 14ms, we found the network transited from asynchronous to weak synchronous and then to synchronously bursting state (**Fig. 5A**).

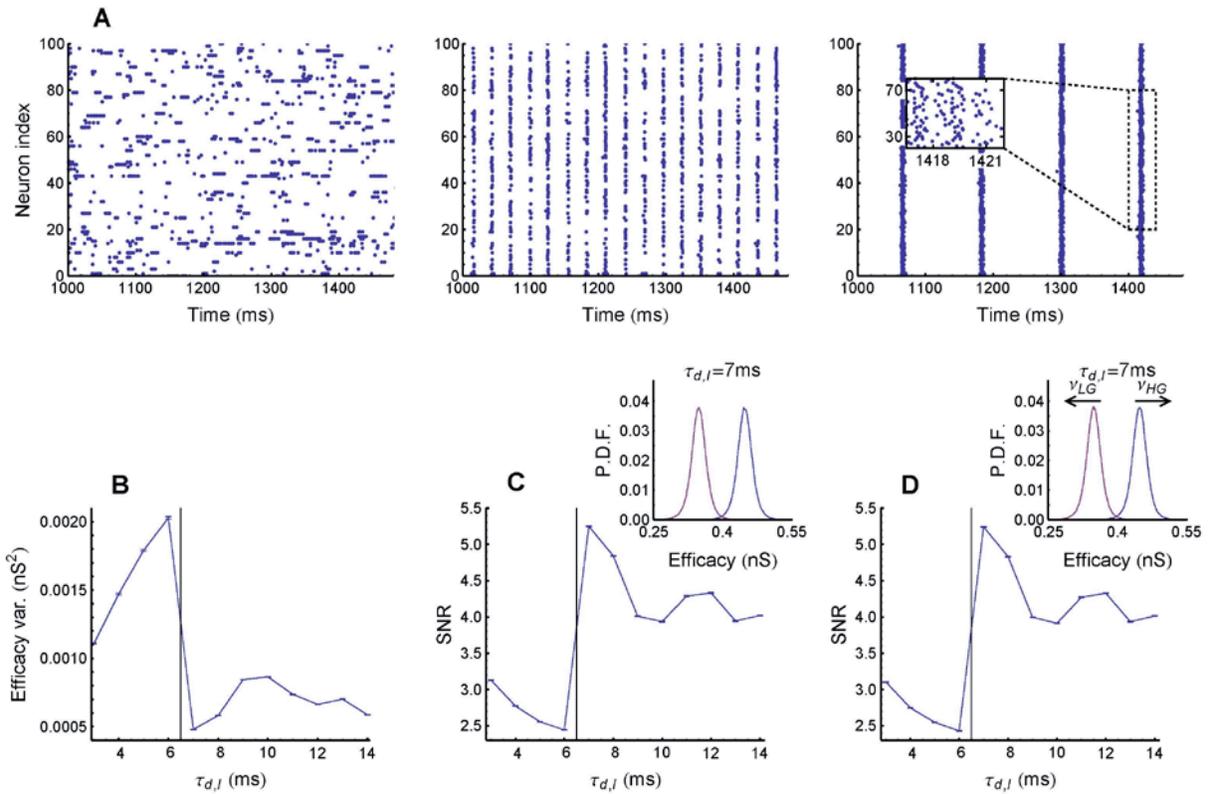

**Fig. 5. Spike patterns of the LIF network influence the efficacy variability, thereby influencing the performance of the network for encoding and maintaining connection patterns.** (**A**) Spike patterns at asynchronous (left, $\tau_{d,I} = 3\text{ms}$), weak synchronous (middle, $\tau_{d,I} = 7\text{ms}$) and synchronously bursting (right, $\tau_{d,I} = 14\text{ms}$) states. (**B**) Efficacy variance as a function of $\tau_{d,I}$ when the excitatory-to-excitatory (E-E) synapses are evolved according to the original recorded spike patterns under STDP and both dendritic and axonal homeostasis. (**C**) The capability of the network to long-termly maintaining connection patterns (quantified by *SNR*) changes with $\tau_{d,I}$ in an inverse way against the efficacy variance (compared with **B**). Inset: the efficacy distribution of the high group (blue) and low group (red) links for $\tau_{d,I} = 7\text{ms}$ at the end of the simulation. (**D**) The same as **C**, but for the capability of faithfully encoding connection-patterns. High

group and low group synapses were subject to different artificial velocities ($v_{HG}$ and $v_{LG}$) during plasticity. In **B-D**, error bars indicate s.e.m. over 24 trials, vertical black lines indicate the transition from asynchronous to synchronous states. Simulations lasted for 20s of biological time, and STDP and synaptic homeostasis were implemented after 1s of transient period. Simulation details are explained in **Methods**.

During STDP and synaptic homeostasis, synaptic efficacies and network dynamics interact with each other. To only investigate the influence of the network dynamics onto the efficacy variability without worrying about the change of dynamics caused by synaptic changes, we first recorded all the spikes of the excitatory population keeping the synaptic efficacies unchanged, then evolved the excitatory-to-excitatory (E-E) links according to the recorded spike patterns under the rules of STDP and synaptic homeostasis. We found that the variance of the efficacies of the E-E links experienced a sharp decrease when $\tau_{d,I}$ changed from 6ms to 7ms, where the network transited from asynchronous state to synchronous state, and got its smallest value at the weak synchronous state just after the asynchrony-to-synchrony transition.

To separately investigate the contribution of different aspects of pattern structure to the efficacy variability, we shuffled the recorded spike patterns using different methods to destroy specific aspects of pattern structure, and observed how the efficacy variance of the E-E links would change if they were evolved according to these shuffled spike patterns under the same STDP and synaptic homeostasis. The spike shuffling methods and their order to be implemented were carefully designed so that when one aspect of pattern structure was destroyed the other aspects remained largely intact (**Fig. 6**). For the two spike patterns before and after implementing a shuffling method, we compared their statistics which is closely relevant to the destroyed aspect of pattern structure, and also compared the variance of the efficacies when the E-E links were driven by each of them. In this way, we were able to obtain understanding on how different aspects of the pattern structure influence the efficacy variability, and compare this understanding with our results obtained from the motif studies.

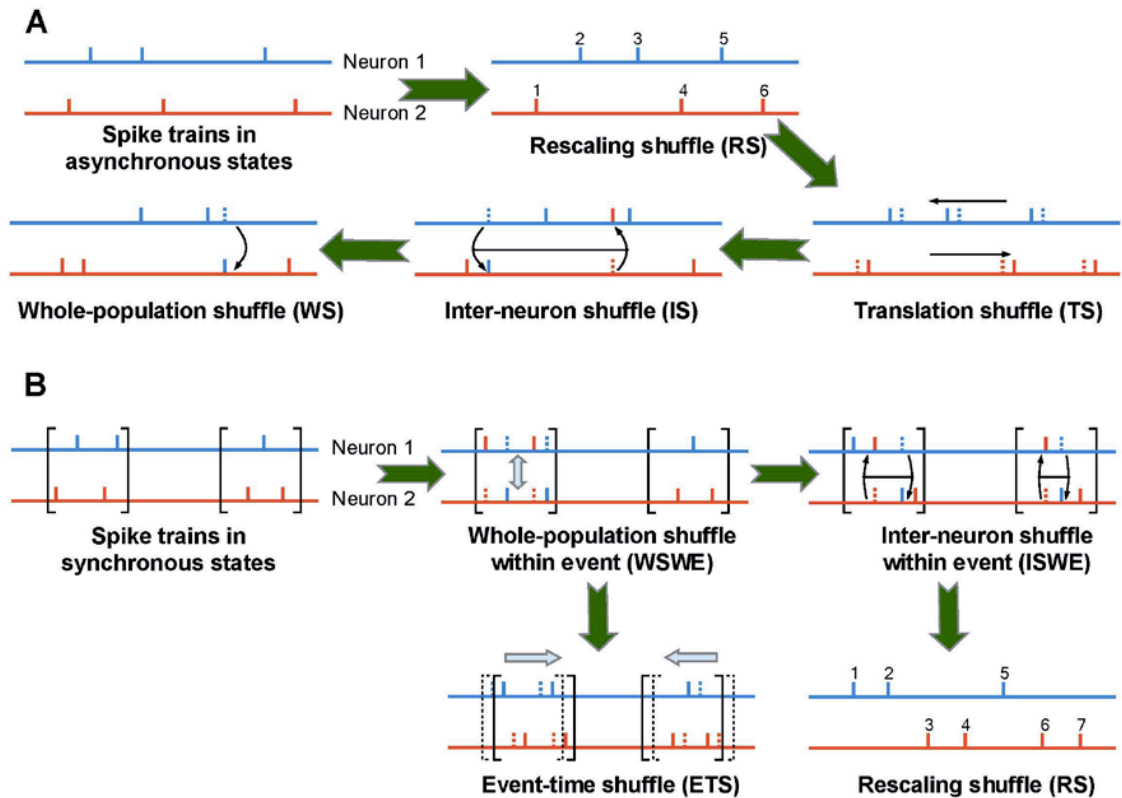

**Fig. 6. The spike shuffling methods implemented onto the recorded spike patterns of the excitatory population in the LIF network.** (**A**) For asynchronous states ($\tau_{d,I} \leq 6\text{ms}$). Rescaling shuffle (RS): all the spikes are ordered and uniformly assigned according to their indexes within the duration of the spike pattern, which destroys synchronous firing. Translation shuffle (TS): each spike train is translationally moved by a random displacement, and periodic boundary condition is used to deal with the spikes which are moved out of the boundaries of time, which destroys heterogeneity of cross-correlations. Inter-neuron shuffle (IS): the spike times of different neurons are randomly swapped, which destroys auto-temporal structure. Whole-population shuffle (WS): each spike is assigned to a randomly selected neuron, which destroys heterogeneity of rates. (**B**) For synchronous states ($\tau_{d,I} \geq 7\text{ms}$). Whole-population shuffle within event (WSWE): the spike sequences of pairs of randomly selected neurons within the same firing event are swapped, which destroys both heterogeneity of cross-correlations and heterogeneity of rates. For technical reasons, we used this method to destroy the two aspects of pattern structure simultaneously. Their individual contributions to the efficacy variability is estimated in **S1 Text Section S3** and **S12I Fig**. Inter-neuron shuffle within event (ISWE): the spike times of different neurons within the same firing event are randomly swapped, which destroys the auto-temporal structure within synchronization periods (i.e. $CV_{rescale}$ in **Fig. 3**). Event-time shuffle (ETS): all the spikes within a firing event are simultaneously moved by a random displacement while

keeping the order of firing events unchanged, which destroys the temporal structure of the occurrence of firing events (i.e. $CV_{events}$ in **Fig. 3**). Rescaling shuffle (RS) is the same as **A**. In **A-B**, Green arrows indicate the order to implement these shuffling methods. We compared the changes of efficacy variability and the changes of spike pattern statistics caused by these shuffling methods (**S1 Table**), and found that the comparison can be understood using the results obtained from the motifs studies (see **S1 Text Section S3** and **S11-S12 Figs** for details).

We used different spike shuffling methods for asynchronous states ($\tau_{d,I} \leq 6\text{ms}$) and synchronous states ($\tau_{d,I} \geq 7\text{ms}$) due to their sharp pattern difference (**Fig. 6**). We found that the influences of different aspects of patterns structure onto the efficacy variability are consistent with our results from the motifs research (**S1 Table**), and that the coupling of dendritic and axonal homeostasis is the main reason of the small efficacy variability in synchronous states. See **S1 Text Section S3** and **S11-S12 Figs** for modeling details.

## Biological Implications

In this section we will demonstrate the important biological implications of efficacy variability on the encoding and maintenance of connection patterns and on the early development of primary visual systems. We conducted simulations in which synaptic plasticity was implemented during self-organized neuronal activity. In this way, we can show that our previous results, which were obtained by studying how neuronal activity influenced synaptic plasticity without considering the feedback of synaptic changes onto network dynamics, can provide important insights into the biological meanings of the efficacy variability in the dynamics-synapse co-evolution situation.

**Encoding and Maintenance of Connection Patterns.** Efficacy variability reflects the variance of efficacies caused by uncontrollable noises during plasticity. Therefore, under spike patterns that cause large efficacy variability, connection patterns cannot be faithfully encoded into the network, and can be easily destroyed by the ongoing remnant plasticity during subsequent functioning (**Fig. 1A**). We used a similar LIF network as the previous section to examine the influence of spike pattern structure onto the encoding and maintenance of connection patterns in neuronal networks.

To do this, we created an artificial connection pattern by randomly assigning each E-E link either into the low efficacy group (low group, or LG) or into the high efficacy group (high group, or HG), then simulated the network with STDP as well as dendritic and axonal homeostasis being imposed on E-E links. For connection-pattern maintenance, the links in HG were assigned to a stronger weight than those in LG in the beginning, and the links within the same group had the same weight. For connection-pattern encoding, all the links had the same weight at the beginning, but LG and HG links were subject to different artificial drift velocities during plasticity, mimicking encoding processes. In reality, STDP can be both the power of connection-pattern encoding and the source of efficacy variability; but here, we separated the two processes, and controlled the encoding process using these two velocities, so that our simulation became more controllable. Despite the artificiality of the encoding process, we believe our simulation is able to provide sufficient insights onto the function of efficacy variability. After the simulation began, the efficacy distributions of both HG and LG got wider due to the efficacy variability (**Fig. 5CD, inset**), and we used signal-to-noise ratio of these two distributions to quantify the quality of the connection pattern. See **Methods** for modeling details.

The connection patterns we used kept the mean input excitatory efficacy to each neuron during on-going plasticity almost the same as that of the LIF network with uniform unchanged E-E links studied in the previous section (see **Methods** for model parameters). After implementing intrinsic homeostasis [33] by dynamically adjusting the threshold of the excitatory neurons to keep the firing rate of the excitatory population around 20Hz (see **Methods**), we found that the spike patterns of this plastic LIF network remained qualitatively the same as those of the LIF network with uniform unchanged E-E weights, so that we could compare the change of the signal-to-noise ratio with $\tau_{d,I}$ in this plastic network with that of the efficacy variability in the network with uniform unchanged E-E weights (**Fig. 5B**). Consistent with our analysis above (**Fig. 1A**), we found that the capability of this plastic LIF network for faithfully encoding and long-termly maintaining the connection patterns was inversely correlated with the efficacy variability in recurrent connections (**Fig. 5CD**).

Experimentally, it was observed that weak synchronous state is advantageous for memory. The absence of the weak gamma-band synchronization during memory encoding in hippocampus is detrimental to the performance in the subsequent recognition tasks [34], while epileptiform events can induce transient epileptic amnesia and accelerate long-term forgetting [35,36]. Our work suggests that working in weak synchronous state may be important for hippocampus to reduce its efficacy variability in recurrent connections, which requires experimental tests.

**Development of Primary Visual Systems Driven by Retinal Waves.** Next, we discuss the function of efficacy variability during the development of primary visual systems driven by retinal waves. Retinal waves are spontaneous bursts of action potentials that propagate in a wave-like fashion across the developing retina during prenatal and early postnatal period, and the retinal waves of the two eyes are not synchronized [37,38]. They induce strong synchrony within a patch of retinal ganglion cells (RGC) of the same eye that sharing similar receptive field (local RGCs), and induces weak synchrony between patches with different receptive fields or in different eyes (**Fig. 7A**). Retinal waves were found to be crucial to the formation of retinotopic map and eye-specific segregation in superior colliculus (SC) and dorsal lateral geniculate nucleus (dLGN) [39]. The developmental function of retinal waves has already been studied using Hebbian synaptic competition [40,41], but in this work we provide new understanding on this competition process using the concept of efficacy variability.

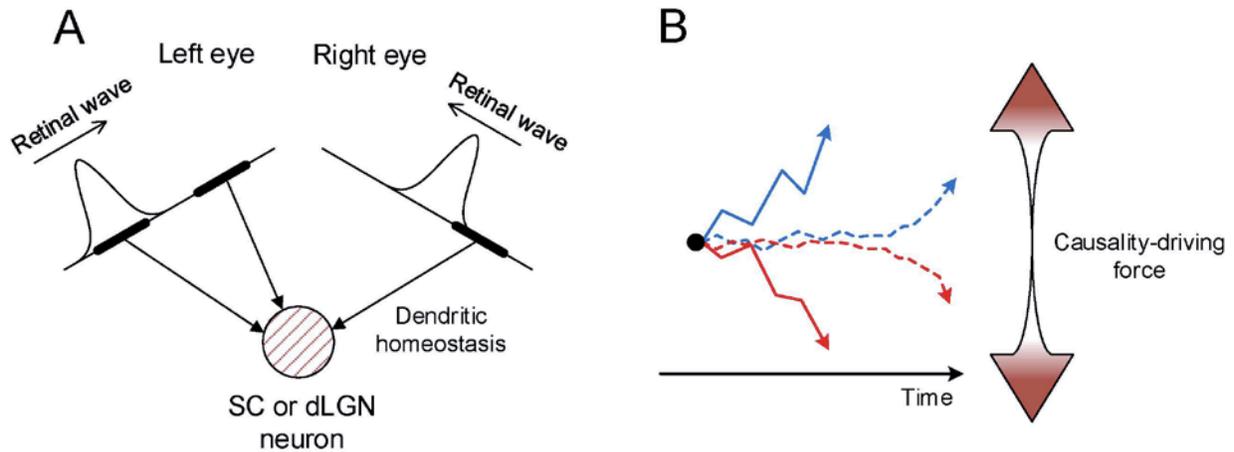

**Fig. 7**. **Schematic to understand the function of efficacy variability during the competition among RGCs induced by retinal waves.** (**A**) Retinal waves induce strong synchrony within a local RGC patch (represented by thick bars), but weak synchrony between patches with different receptive fields in the same eye or in different eyes. Connections from these patches to downstream neurons compete with each other under dendritic homeostasis. (**B**) Initially, synapses from different patches are homogeneous (indicated by the black dot), but noise-induced diffusion separates the synaptic efficacies from the two patches (zigzag arrows of two colors). As these two patches are separated farther, they will be pushed apart stronger by causality (big arrows). If the initial diffusion is strong (solid arrows), they will soon diffuse into the strong-causality range, and get separated quickly; if the initial diffusion is weak (dashed arrows), they will stay in the weak-causality range for a longer time, so that the separation process is hindered. A computational model to validate this mechanism is developed in **S1 Text Section S4.2** and **S13 Fig**.

    Initially, a neuron in SC or dLGN may homogeneously receive input from many patches of local RGCs. (The definition of local RGC patch is casual, but if two RGCs are close enough then their synchrony in retinal waves will become strong and their connections to downstream neurons will become similar, in which case they can be regarded to belong to the same patch.) The essential idea of the Hebbian-type competition [40,41] among these patches relies on a positive feedback, i.e. when the synapses from one patch become a little stronger than those from the other patches, the downstream neuron becomes more responsive to this patch than the others, and this *causality* then potentiates the synapses from this patch stronger, helping this patch compete over the others under synaptic homeostasis. However, before this causality is reliably established, noise-induced diffusion is the main source of the inter-patch separation. When this initial diffusion is strong, the difference between the synaptic efficacies from these patches can be quickly enlarged, so that the causality takes its effect early; if the initial diffusion is weak, the efficacies from different patches may wriggle around their common starting point for a long time before the causality reliably participates, which hinders the separation process (**Fig. 7B**).

    Large initial inter-patch diffusion requires a large efficacy variability between neurons in different patches over the efficacy variability between neurons in the same patch (**S1 Text eq. S24**). During a retinal wave, most local RGCs are activated synchronously, with similar spike

numbers and burst durations, which homogenizes the changes of the synapses coming from the same RGC patch during plasticity, thereby reducing the intra-patch efficacy variability. The inter-patch efficacy variability is always larger than that of intra-patch due to the weak inter-patch synchrony, especially between two patches with far-apart receptive fields or in different eyes. Therefore, retinal waves are able to introduce strong initial inter-patch diffusion, which helps to establish the causality as early as possible. When the inter-patch synchrony is strong, as is the case for two patches in the same eye with nearby receptive fields, the initial inter-patch diffusion is weak, and the causality is also weak (because when the downstream neuron responds to one patch, it also has a high probability to respond to the other one due to the strong inter-patch synchrony), so that the separation may not complete at the end of the critical period of development. In this case, the interaction of the inter-patch diffusion and the causality is able to produce a strong trial-to-trial variability of the difference between the mean efficacies coming from the two patches (**S13H Fig**). This suggests that a downstream neuron in SC or dLGN with a receptive field centered at $O$ can also receive inputs from RGCs with receptive fields centered near $O$, but the efficacies of these inputs should be widely distributed. How this may influence computation is an interest of future researches. We developed a computational model to show the contribution of efficacy variability to the inter-patch separation, see **S1 Text Section S4.2** and **S13 Fig** for modeling details.

## Discussions

In this paper, we provide clear evidences that efficacy variability is an important dimension of synaptic plasticity, and spike pattern structure has strong influences on efficacy variability. We systematically study the influences of four aspects of pattern structures, i.e. synchronous firing, auto-temporal structure, heterogeneity of rates and heterogeneity of cross-correlations, and their interactions (see **S1 Text Section S2**), using spike generating models in simple motifs and spike shuffling methods in LIF networks, and then show the functional importance of efficacy variability on the encoding and maintenance of connection patterns and on the early development of primary visual systems driven by retinal waves.

  The reason why we focus on these four aspects of pattern structures is because that under STDP they are the only four that mainly influence the lowest order of DiV and DrV, i.e. for DiV we suppose that all synapses have the same diffusion strength, and for DrV we do not consider correlations among drift velocities of synapses. Strictly speaking, heterogeneity of rates and heterogeneity of cross-correlations may not only make different synapses drift in different velocities, but also make them have different diffusion strengths. However, the contribution of the latter to the efficacy variability of the whole network is far less than that of the former, especially in the long run. For simplicity, we do not consider the heterogeneity of diffusion strengths in this study, but in principle it can be understood using the mechanisms introduced in this work, especially in **S6B Fig** (for heterogeneity of rates) and **S4A Fig** (for heterogeneity of cross-correlations).

  In this paper, we focus on an additive STDP model with linear accumulation of all possible pre- and post-synaptic spike pairs. Multiplicative STDP with linear weight dependence does not qualitatively change our results as long as the synaptic homeostasis is also multiplicative, as we can take logarithm before making discussions. However, physiologically, STDP may have varieties of complex realizations, depending on synaptic types, spike patterns and even locations of synapses on dendrites [42]. Therefore, the influence of pattern structures on the efficacy variability may be various, and our results need careful revisits before being implemented to understand real biological systems. For example, in the GABAergic synapses onto CA1 pyramidal neurons, pairing single pre- and post-synaptic spikes at short intervals leads to LTP regardless spike orders, and pairing spikes at long intervals leads to LTD [43]. This STDP rule can remove the mechanism of synapse splitting during synchronous firings with short durations and thus reduce the efficacy variability. As another example, in L2/3 synapses of visual cortical slices, the later spikes in each burst are found to be less effective in synaptic modification [44]. This discounts the synaptic changes contributed by per spike during bursts and thus may reduce the efficacy variability. Despite of its limitations, our work provides a comprehensive framework to understand the mechanisms how spike patterns influence the efficacy variability, and all these mechanisms should be carefully considered when dealing with more complicated situations.

Although the influence of spike patterns is various across systems, the concept of *efficacy variability* should be of general importance. The stochasticity of synapses and neuronal responses as well as the emergent heterogeneity of rates and cross-correlations in network dynamics together make efficacy variability an unavoidable nature of plasticity. Therefore, it is of great meaning to understand how animals make use of efficacy variability and get around of it in future researches. We believe that the concept of efficacy variability not only provides a new perspective to understand the function of plasticity, but is also a new angle to review our current knowledge on learning.

# Methods

Here we describe the plasticity rules and the LIF networks used in our simulations.

## STDP and Synaptic Homeostasis

The STDP updating caused by a pair of pre- and post-synaptic spike at $t_{pre}$ and $t_{post}$ is

$$\Delta w(t_{pre}, t_{post}) = \begin{cases} A_p \exp\left(-\dfrac{t_{post} - (t_{pre} + \tau_{delay})}{\tau_{STDP}}\right), & t_{post} > t_{pre} + \tau_{delay} \\ -A_d \exp\left(-\dfrac{(t_{pre} + \tau_{delay}) - t_{post}}{\tau_{STDP}}\right), & t_{post} < t_{pre} + \tau_{delay} \end{cases} \quad (4)$$

with $\tau_{delay}$ being the axonal delay. The contribution of all pairs of pre- and post-synaptic spikes are added together. $\tau_{STDP} = 20\text{ms}$, $\tau_{delay} = 1\text{ms}$ throughout the paper.

We used a dynamic bound to model synaptic homeostasis which dynamically maintained the mean efficacy of the incoming or out-going synapses of a neuron. In this model, the synaptic efficacies are updated every $\Delta T$ time according to

$$w_{ab}(t + \Delta T) = w_{ab}(t) + \varepsilon\left(w_{bound} - \frac{1}{N_a}\sum_{c=1}^{N_a} w_{ac}(t)\right), \quad \text{for dendritic homeostasis} \quad (5)$$

$$w_{ab}(t+\Delta T) = w_{ab}(t) + \varepsilon \left( w_{bound} - \frac{1}{N_b} \sum_{c=1}^{N_b} w_{cb}(t) \right), \quad \text{for axonal homeostasis} \tag{6}$$

with $N_a$ being the in-degree of the *a*th neuron, $N_b$ being the out-degree of the *b*th neuron, $w_{bound}$ being the ground line of synaptic homeostasis, and $\varepsilon$ being the plasticity rate. $\Delta T = 1$ms throughout the paper, the other parameters are indicated at relevant locations. In the motif studies, synaptic homeostasis is only imposed on the apical neuron of a motif.

## The LIF Neuronal Network

The network consists of 2000 excitatory and 500 inhibitory conductance-based LIF neurons, the links are randomly connected with probability 0.2. The dynamics of membrane voltage is

$$C_k \frac{dV_k}{dt} = g_{L,k}(V_{leak} - V_k) + \left( \sum_j g_{ext,k} s_{ext}(t-t_j) + \sum_j g_{E \to k} s_E(t-t_j) \right)(E_E - V_k)$$

$$+ \sum_j g_{I \to k} s_I(t-t_j)(E_I - V_k) \quad k = E, I \tag{7}$$

and

$$V_k \to V_r, \quad \text{when } V_k = \theta. \tag{8}$$

And the dynamics of synaptic conductance is

$$s_k(t) = \frac{\tau_k}{\tau_{d,k} - \tau_{r,k}} \left( \exp(-\frac{t-t_j}{\tau_{d,k}}) - \exp(-\frac{t-t_j}{\tau_{r,k}}) \right) \tag{9}$$

In the equations above, membrane time constants $\tau_E = C_E / g_{L,E} = 20$ms, $\tau_I = C_I / g_{L,I} = 10$ms; leakage conductances $g_{L,E} = g_{L,I} = 10$nS; inverse voltages $E_E = 0$mV, $E_I = -70$mV; link conductances $g_{E \to E} = 0.4$nS, $g_{I \to E} = 5.8$nS, $g_{E \to I} = 0.74$nS, $g_{I \to I} = 9.6$nS; firing threshold $\theta = -50$mV; reset voltage $V_r = -60$mV; refractory time $\tau_{ref,E} = 2$ms, $\tau_{ref,I} = 1$ms; rising time of synaptic conductance $\tau_{r,E} = \tau_{r,I} = 0.5$ms; decay time of excitatory conductance $\tau_{d,E} = 4$ms. The decay time of the inhibitory conductance $\tau_{d,I}$ takes 12 integer values from 3ms to 14ms. Each neuron also receives 1000Hz of external Poisson input, with external conductance $g_{ext,E} = c \times 0.53$nS, $g_{ext,I} = c \times 0.75$nS, where *c* is a coefficient adjusted to conserve the firing rate of the excitatory population at 20Hz, with values 3.13332, 3.28868, 3.38022, 3.44494, 1.63315, 1.50098, 1.30697, 1.06414, 0.845752, 0.636046, 0.421928, 0.327283 for $\tau_{d,I}$ as integer values from 3ms to 14ms. Axons have delay $\tau_{delay} = 1$ms. Parameters for STDP and synaptic homeostasis (see eq. 4-6): $A_p = A_d = 0.0012$nS, $w_{bound} = 0.4$nS, $\varepsilon = 0.001$. Simulations were performed using a second order Runge–Kutta scheme with fixed time step, $\delta t = 0.05$ms; and an interpolation scheme was also used for the determination of the firing times of the neurons [45]. We first recorded 20s of spike trains, and then evolved the E-E links under STDP and synaptic homeostasis when the activity of the excitatory population was according to the spike trains original or shuffled by different methods. STDP and synaptic homeostasis started after 1s of transient period. The efficacy variance shown in **Fig. 5B** was calculated at 20s of biological time.

The purpose of this work is to understand how dynamic patterns influence efficacy variability, instead of how dynamic properties change with model parameters, so averaging configurations of the random LIF networks does not help to gain more insight, only increasing complexity. Therefore, our study focused on a single typical configuration, thereby fixing dynamics at different $\tau_{d,I}$s, except that we chose different initial states and seeds of random generators for different trials, which resulted in trial-to-trial variability. However, we did check our results using other network configurations, and found qualitatively the same results.

## Connection-Pattern Maintenance in the LIF Neuronal Network

We created an artificial connection pattern by randomly assigning each E-E link either into the low efficacy group (low group, or LG) or the high efficacy group (high group, or HG). The links in LG or HG were assigned at 0.35nS or 0.45nS at the beginning, so that the mean efficacy of the E-E links (0.4nS) was the same as that of the LIF network mentioned in the previous subsection. We then simulated the network with STDP as well as dendritic and axonal homeostasis being imposed on the E-E links. We also chose the ground line of synaptic homeostasis $w_{bound} = 0.4\text{nS}$ (see eq. 5 and 6), so that the mean excitatory efficacy to each excitatory neuron was kept around 0.4nS during on-going plasticity. To conserve the firing rate of the excitatory population around 20Hz during plasticity, intrinsic homeostasis [33] was also implemented so that the threshold of all the excitatory neurons $\theta_E$ was adjusted every 10ms:

$$\theta_E(t) = \theta_E(t-10\text{ms}) + c(r(t) - r_0) \tag{10}$$

with $r(t)$ being the firing rate of the excitatory population in the past 1000ms, $r_0 = 20\text{Hz}$, $c = 0.001\text{mV} \cdot \text{s}$. In spite of the connection pattern and the on-going plasticity, we found that the dynamic pattern of this plastic LIF network remained qualitatively the same as that of the LIF network with uniform unchanged E-E links introduced in the previous subsection, so that we could use our understanding on the efficacy variability of the LIF network with uniform unchanged E-E links (**Fig. 5B**) to understand the performance of the connection-pattern maintenance in this plastic network.

Suppose after time *t*, the mean and variance of the efficacy distribution of LG are $\mu_{low}(t)$ and $\sigma^2_{low}(t)$, and those of HG are $\mu_{high}(t)$ and $\sigma^2_{high}(t)$. We quantified the quality of the connection pattern using

$$SNR(t) = \frac{\mu_{high}(t) - \mu_{low}(t)}{\sqrt{\sigma_{high}(t)\sigma_{low}(t)}} \tag{11}$$

and observed how *SNR(t)* changed with $\tau_{d,I}$ (**Fig. 5C**). All the E-E links were bounded within $[0.25\text{nS}, 0.55\text{nS}]$ using hard bounds, and we controlled the simulation time so that most efficacies were far from the boundaries. The influence of boundaries on connection-pattern maintenance is beyond the scope of the research.

The simulation lasted for 20s biological time, and plasticity (including STDP, synaptic homeostasis and intrinsic homeostasis) started after 1s of transient period. Parameters for STDP and synaptic homeostasis were the same as described in the previous subsection.

### Connection-Pattern Encoding in the LIF Neuronal Network

For connection-pattern encoding, we did not consider a detailed learning process, but instead modeled the encoding process generically by artificial drifts of synaptic efficacies. Specifically, each E-E link was also randomly assigned into LG or HG. Initially, both LG and HG links were 0.4nS, but they were subject to different drift velocities $v_{LG}$ and $v_{HG}$. So at time $t$, the efficacy of a LG link should be $w_{LG}(t) = w_{STDP}(t) + w_{hom}(t) + v_{LG}t$, with $w_{STDP}(t)$ being the contribution of STDP, $w_{hom}(t)$ being the contribution of dendritic and axonal homeostasis, and $v_{LG}$ being the drift velocity imposed on LG links; and the value of a HG link should be $w_{HG}(t) = w_{STDP}(t) + w_{hom}(t) + v_{HG}t$. The same as the study on connection-pattern maintenance, intrinsic homeostasis was also imposed on the excitatory neurons to keep their mean rate around 20Hz. During the simulation, the mean values of HG and LG were separated apart, while the distributions of HG and LG were continuously broadened (**Fig. 5D, inset**). We also used $SNR(t)$ to quantify the quality of the connection pattern at a given time (**Fig. 5D**).

$v_{HG} = (0.45 - 0.4)/(20 - 1)\text{nS} \cdot \text{s}^{-1}$ and $v_{LG} = (0.35 - 0.4)/(20 - 1)\text{nS} \cdot \text{s}^{-1}$ (note that $v_{HG} = -v_{LG}$, so that the mean efficacy of the E-E links was kept at 0.4nS during simulations). Simulations lasted for 20s biological time, and plasticity (including STDP, synaptic homeostasis and intrinsic homeostasis) started after 1s transient period. Parameters for STDP and synaptic homeostasis were the same as the previous subsection. All the E-E links were also bounded within [0.25nS, 0.55nS] using hard bounds, and we also controlled the simulation time so that most efficacies were far from the boundaries.

## Acknowledgments

CZ is partially supported by Hong Kong Baptist University (HKBU) Strategic Development Fund, NSFC-RGC Joint Research Scheme HKUST/NSFC/12-13/01, NSFC (Grant No. 11275027). HJZ is supported by the National Basic Research Program of China (grant number 2013CB932804) and the National Natural Science Foundations of China (grant numbers 11121403 and 11225526). ZB thanks Dongping Yang for kind helps on coding LIF networks at the beginning of the research. ZB also thanks David Hansel and Carl van Vreeswijk for helpful discussions during his visit in CNRS UMR 8119 in Université Paris Descartes in 2013.

## Author contribution

ZB and CZ conceived the idea, ZB designed and performed the research, HJZ supervised ZB on this work, ZB and CZ wrote the paper.

# Spike Pattern Structure Influences Efficacy Variability under STDP and Synaptic Homeostasis：
# SUPPLEMENTARY INFORMATION

**Authors:**  Zedong Bi[1,2]*, Changsong Zhou[2,3,4,5], Hai-Jun Zhou[1]

**Affiliations:**

[1]State Key Laboratory of Theoretical Physics, Institute of Theoretical Physics, Chinese Academy of Sciences, Beijing 100190, China

[2]Department of Physics, Hong Kong Baptist University, Kowloon Tong, Hong Kong, China

[3]Centre for Nonlinear Studies, and Beijing-Hong Kong-Singapore Joint Centre for Nonlinear and Complex Systems (Hong Kong), Institute of Computational and Theoretical Studies, Hong Kong Baptist University, Kowloon Tong, Hong Kong, China

[4]Beijing Computational Science Research Center, Beijing, China

[5]Research Centre, HKBU Institute of Research and Continuing Education, Virtual University Park Building, South Area Hi-tech Industrial Park, Shenzhen, China

*Correspondence to: zedong.bi@gmail.com.

## Section S1: The Decomposition of the Total Variance of the Synaptic Efficacies

*Key points of this section:*
1) *The efficacy variability can be decomposed into the variability caused by heterogeneity of drift velocities (DrV), and the variability caused by diffusion (DiV).*

The law of total variance says that if the probability space of *Y* is decomposed into several subspaces labeled by *X*, then the variance of *Y* in the whole space is equal to the summation of the variance of the expectations in these subspaces and the expectation of the variances in these subspaces, i.e.

$$\mathrm{Var}(Y) = \mathrm{Var}\big(\mathrm{E}(Y \mid X)\big) + \mathrm{E}\big(\mathrm{Var}(Y \mid X)\big) \tag{S1}$$

Now suppose that there is a matrix $\Delta \mathbf{W}$, each column of which represents the changes of the synaptic efficacies in a network after the plasticity in one trial, and different columns represent different trials. Then eq.S1 can be written as

$$\mathrm{Var}_{S,T}(\Delta \mathbf{W}) = \mathrm{Var}_S\big(\mathrm{E}_T(\Delta \mathbf{W})\big) + \mathrm{E}_S\big(\mathrm{Var}_T(\Delta \mathbf{W})\big) \tag{S2}$$

where the subscript *S* represents integrating over row index, i.e. structural disorder, and *T* represents integrating over column index, i.e. trial disorder.

Here, $\mathrm{E}_T(\Delta \mathbf{W})$ represents the trial expectations of the changes of all the synapses in the network; and $\mathrm{Var}_S\big(\mathrm{E}_T(\Delta \mathbf{W})\big)$ is the variance of these trial expectations, representing DrV. $\mathrm{Var}_T(\Delta \mathbf{W})$ represents the trial-to-trial variances caused by diffusion, and $\mathrm{E}_S\big(\mathrm{Var}_T(\Delta \mathbf{W})\big)$ is the average of these variance over all the synapses, representing DiV. Eq.S2 is the formal writing of eq.1 in the main text.

The law of total variance can decompose $\mathrm{Var}_{S,T}(\Delta \mathbf{W})$ in another way:

$$\mathrm{Var}_{S,T}(\Delta \mathbf{W}) = \mathrm{Var}_T\big(\mathrm{E}_S(\Delta \mathbf{W})\big) + \mathrm{E}_T\big(\mathrm{Var}_S(\Delta \mathbf{W})\big) \tag{S3}$$

Here $\mathrm{Var}_T\big(\mathrm{E}_S(\Delta \mathbf{W})\big)$ is the trial-to-trial variability of the mean synaptic change of the network, but a real biological process only allows a single trial, so this trial-to-trial variability cannot contribute to biological functions except for individual differences. Fortunately, $\mathrm{Var}_T\big(\mathrm{E}_S(\Delta \mathbf{W})\big) \sim \mathcal{O}(1/N)$, with *N* being the size of the network. So when *N* is large enough, eq.S3 becomes $\mathrm{Var}_{S,T}(\Delta \mathbf{W}) \approx \mathrm{E}_T\big(\mathrm{Var}_S(\Delta \mathbf{W})\big)$. Therefore, we can use the trial expectation of the efficacy variance in a network $\mathrm{E}_T\big(\mathrm{Var}_S(\Delta \mathbf{W})\big)$ to represent the total efficacy variance $\mathrm{Var}_{S,T}(\Delta \mathbf{W})$, which quantifies the efficacy variability. This is what we do in our simulations. Under this insight, eq.S2 becomes

$$\mathrm{E}_T\big(\mathrm{Var}_S(\Delta \mathbf{W})\big) \approx \mathrm{Var}_S\big(\mathrm{E}_T(\Delta \mathbf{W})\big) + \mathrm{E}_S\big(\mathrm{Var}_T(\Delta \mathbf{W})\big) \tag{S4}$$

In dendritic or axonal motifs (**Fig. 2D** in the main text), a synapse can be indexed by the non-apical neuron it links. If we use *a* to represent this index, eq.S4 can be written as

$$\mathrm{E}_T\left(\mathrm{Var}_a(\Delta w_a)\right) \approx \mathrm{Var}_a\left(\mathrm{E}_T(\Delta w_a)\right) + \mathrm{E}_a\left(\mathrm{Var}_T(\Delta w_a)\right) \tag{S5}$$

## Section S2: Motif Studies

### Section S2.1: Introduction to Motif Studies

*Key points of this subsection:*
1) *We controlled the activities of the neurons in motifs using spike generating models. This does not discount the generality of our results, and is also necessary for investigating the influence of different aspects of pattern structure onto the efficacy variability without worrying about the feedback of the synaptic changes onto spike patterns, which will happen in more biological models.*
2) *The results on efficacy variability obtained from studying dendritic motifs will be exactly the same as those obtained from studying axonal motifs if the spike patterns are statistically time-reversal invariant.*

We generated the spike trains of the neurons in dendritic and axonal motifs (**Fig. 2D** in the main text) using statistical models, and evolved the synapses according to the generated spike trains and additive STDP when dynamically conserving the mean synaptic efficacy into or out of the apical neuron of a motif using subtractive normalization (synaptic homeostasis), see **Methods** in the main text. Synaptic homeostasis does not influence the efficacy variance in a free motif, but may take its effect when dendritic and axonal motifs are coupled together.
    In reality, post-synaptic neurons generate spikes according to their inputs, but here spike trains are model-generated and imposed onto the neurons in motifs. Here we argue that the break of this causality does not discount the generality of our results. Firstly, a neuron in biological systems receives from and targets to various types of neurons with different functions, but the synapses in the motifs here only represent those synapses in biological systems with similar plasticity rule and strong homeostatic interactions. Thus the activities of biological pre-synaptic neurons which are represented by the pre-synaptic neurons in motifs may influence, but cannot fully determine, the activities of their post-synaptic neurons: noises from other sources can be very strong. Secondly, plasticity and spike evoking are two independent processes: synapses undergo the same plasticity under the same spike pattern, regardless whether the spike pattern is self-organized or model-generated. Therefore we can still gain insight onto the plastic process in self-organized systems even if our spike trains are model-generated. We also argue here that the break of this causality is very necessary for our research. In real systems, neural dynamics and plasticity are two interacting processes. Using spike-generating models, we can explicitly control different aspects of spike pattern structure while keeping population rate constant, thereby investigating the influences of pattern structure onto synaptic changes without worrying about the feedback of synaptic changes onto spike patterns. When discussing biological implications, we will use systems with self-organized dynamics, and show that our results from the motif studies can gain rich insights into the behavior of the system with self-organized dynamics to understand the important biological meanings of efficacy variability.

So what is the relationship between the efficacy variability in a dendritic motif and an axonal motif? The synapses in our motifs have axonal delay $\tau_{delay}$, and the STDP updatings in our model depend on the difference of the time of the post-synaptic spike and the time when the pre-synaptic spike *arrives* at the post-synaptic neuron. Suppose that in a dendritic motif the apical neuron fires at $t_{0,i}$ and the $a$th non-apical neuron fires at $t_{a,j}$, then the pre-synaptic spike arrives at the post-synaptic neuron at $t_{a,j} + \tau_{delay}$, so the time difference which the STDP updating depends on is $\Delta t_1 = t_{0,i} - (t_{a,j} + \tau_{delay})$. Now suppose that we play the whole spike pattern in a time-reversal way, just like showing the dendritic motif a backward movie, then the spike at the apical neuron will be at $-t_{0,i}$, and the spike at the $a$th non-apical neuron will be at $-t_{a,j}$, so after considering the axonal delay, the time difference will be $\Delta t_2 = -t_{0,i} - (-t_{a,j} + \tau_{delay})$. Now we reverse the direction of the links in the dendritic motif, so that it becomes an axonal motif, and show it the original *forward* spike trains, then the apical neuron becomes the pre-synaptic neuron, while the $a$th non-apical neuron becomes the post-synaptic one, so the time difference will be $\Delta t_3 = t_{a,j} - (t_{0,i} + \tau_{delay})$. We see that $\Delta t_2 = \Delta t_3$. This means that the backward spike train changes the synapses in the dendritic motif by the same values with the corresponding forward spike train in the axonal motif under STDP. Therefore, researches on efficacy variability using dendritic motifs and axonal motifs will get exactly the same results as long as the spike trains are statistically time-reversal invariant. Because of the time-reversal invariance of our statistical spike-generating models, we will only focus on dendritic motifs in the following. Results on axonal motifs will be exactly the same.

### Section S2.2: The Influence of Synchronous Firing

*Key points of this subsection:*
1) *Synchronous firing influences the efficacy variability through three mechanisms: spike gathering, synapse splitting and synaptic correlating.*

We evolved synapses in a dendritic motif according to spike trains generated by Model Sync 1 (**Section S5.1**), and recorded their variance per spike (variance divided by spike number per neuron) at the end to quantify the efficacy variability (**S1A Fig**). In this model, if a firing event happens during $[t_1, t_2]$ ($\tau_{cross} = t_2 - t_1$), then each neuron, including the apical neuron, in the dendritic motif will have a probability $p \in (0,1]$ to fire a spike within this interval. The rate of the occurrence of firing events is $r_0 / p$, so that the firing rate of each neuron is kept at $r_0$ when $p$ changes. Suppose that the spike time of the apical neuron during the firing event is $t_0$ ($t_1 \leq t_0 \leq t_2 = t_1 + \tau_{cross}$). If $\tau_{cross} < \tau_{delay}$, then we always have $t_0 < t_1 + \tau_{cross} < t_1 + \tau_{delay}$, which means that the apical neuron will not receive its afferents before its own firing. In this case, when $p$ is large, nearly all the synapses are depressed almost simultaneously under STDP, so the efficacy variability is reduced through *synapse correlating* (see the main text). If $\tau_{cross} > \tau_{delay}$, there are chances when $t_1 + \tau_{delay} < t_0 < t_2 + \tau_{delay}$, so that some synapses become potentiated and some other synapses become depressed after a firing event, causing *synapse splitting* (see the main text), so the efficacy variability gets increased. *Spike gathering* (see the main text) always increases with $p$, but its contribution on the increase of the efficacy variability is more obvious when $p$ is small, where both synapse correlating and synapse splitting are weak.

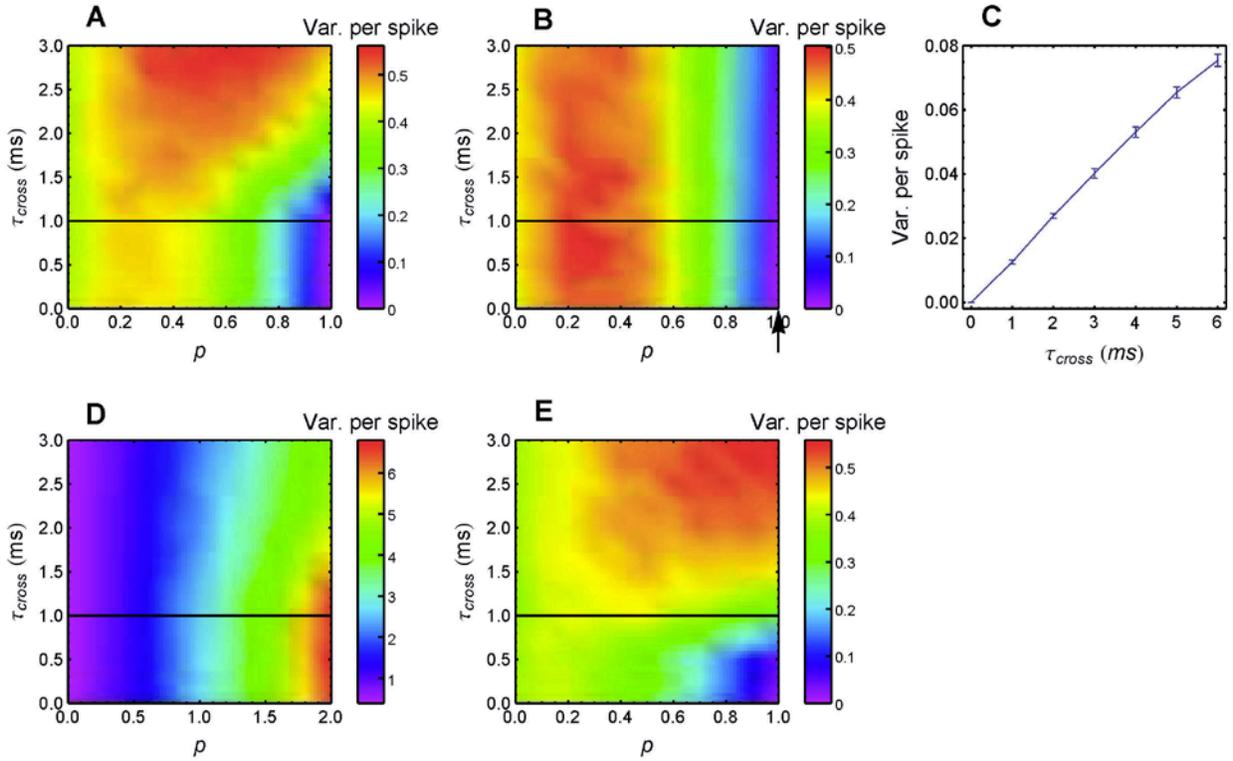

**S1 Fig. The influence of synchronous firing on the efficacy variability in a dendritic motif.** (**A**) Variance per spike (variance divided by spike number per neuron) as a function of $p$ and $\tau_{cross}$ when spike trains contained synchronous firing and a neuron fired no more than one spike in a firing event (Model Sync 1 in **S1 Text Section S5.1**). The horizontal black line represents the axonal delay $\tau_{delay} = 1\text{ms}$. (**B**) The same as **A**, but using spike patterns in which the apical neuron always fired its own spike after receiving all its afferents during a firing event (Model Sync 2 in **S1 Text Section S5.1**), thereby removing synapse splitting. (**C**) Although not apparent in **B**, variance per spike increases with $\tau_{cross}$ when $p = 1$ (indicated by the arrow in **B**), which is caused by the heterogeneity of synaptic updatings induced by the increasing time window $\tau_{cross}$. Error bars represent the standard error of the mean (s.e.m.). (**D**) The same as **A**, but spike trains were inhomogeneous Poisson processes (Model Sync 3 in **S1 Text Section S5.1**), so that synapse correlating was removed by the variety of spike numbers of different non-apical neurons during a firing event. Note the different horizontal scale for $p$ from the previous panels. (**E**) The same as **A**, but spike trains were generated based on a dichotomized Gaussian approach, so that they had near-maximal entropy (Model Sync 4 in **S1 Text Section S5.1**). In **A-E**, the dendritic motif has 200 non-apical neurons. Parameters for STDP: $A_p = A_d = 1$; parameters for synaptic homeostasis: $w_{bound} = 0\text{nS}$, $\varepsilon = 0.001$ (see **Methods** in the main text eq.4-6 for the meanings of these parameters).

All synaptic efficacies were 0 at the beginning, and simulations were run for 100s biological time, with 24 trials.

To further check the three mechanisms through which synchronous firing influences the efficacy variability, we next generated spike trains using two other statistical models. In the first model (Model Sync 2 in **Section S5.1**), the apical neuron can fire only at $t_2 + \tau_{delay}$ in a firing event, so that it fires only after receiving all its afferents, which removes synapse splitting. In this case, when *p* is small, the efficacy variability still increases with *p* because of spike gathering; when *p* is large, it decreases with *p* because of synapse correlating, and it does not strongly increase with $\tau_{cross}$ as it does for Model Sync 1 because of the removal of synapse splitting in this model (**S1BC Fig**). In the second model (Model Sync 3 in **Section S5.1**), spike trains are inhomogeneous Poisson processes, so that different non-apical neurons may fire very different numbers of spikes during a firing event, which results in very different efficacy changes on the corresponding synapses, thereby removing synapse correlating. Therefore, spike gathering makes the efficacy variability continuously increase with *p* even when $\tau_{cross} < \tau_{delay}$ (**S1D Fig**). Actually, for this model the efficacy variability for $\tau_{cross} < \tau_{delay}$ is even usually larger than that for $\tau_{cross} > \tau_{delay}$ (**S1D Fig**), which will be explained in the Miscellaneous (**Section S6.1**).

Experimentally, it was found that synchrony patterns in neuronal networks exhibit near-maximal entropy [1,2]. To check the universality of our results, we then generated spikes according to a model (Model Sync 4 in **Section S5.1**) based on the one introduced in [3], which was shown to possess near-maximal entropy [4]. We found that the efficacy variability changed in a similar way with *p* and $\tau_{cross}$ as it did in Model Sync 1 (**S1E Fig**).

### Section S2.3: The Interaction of Synchronous Firing and Heterogeneity of Rates

*Key points of this subsection:*
1) *Synchronous firing is able to change P-D imbalance.*
2) *Heterogeneity of rates makes use of P-D imbalance to change the efficacy variability in DrV manner.*

Next we add the ingredient of rate heterogeneity into the picture, and investigate its effect on the efficacy variability when it is interacted with synchronous firing. Suppose in a dendritic motif the rate of the *a*th non-apical neuron is $r_a$, the change of its axonal efficacy is $\Delta w_a$, then eq.S5 becomes

$$\mathrm{E}_T\left(\mathrm{Var}_a(\Delta w_a)\right) \approx \mathrm{E}_a\left(\mathrm{Var}_T(\Delta w_a \mid r_a)\right) + \mathrm{Var}_a\left(\mathrm{E}_T(\Delta w_a \mid r_a)\right) \quad (S6)$$

where we add the condition to $r_a$ to emphasize that different non-apical neurons have different firing rates. In this equation, $\mathrm{E}_a(\mathrm{Var}_T(\Delta w_a \mid r_a))$ represents DiV, and $\mathrm{Var}_a(\mathrm{E}_T(\Delta w_a \mid r_a))$ represents DrV (see the discussion in **Section S1**).

During STDP, both the strengths of potentiation and depression processes are proportional to the firing rates of the pre- and post-synaptic neurons. Therefore, the expectation of the efficacy change of the $a$th synapse

$$E_T(\Delta w_a) \propto (S_p - S_d) r_a r_0, \tag{S7}$$

with $r_0$ and $r_a$ being the rate of the apical and $a$th non-apical neuron, and $S_p - S_d$ quantifying the imbalance of potentiation and depression (*P-D imbalance*). Under P-D imbalance, if $r_a$ is heterogeneous, then $\text{Var}_a(E_T(\Delta w_a | r_a)) \propto (S_p - S_d)^2$ with non-zero coefficient, representing DrV.

For homogeneous Poisson processes, $S_p = A_p$ and $S_d = A_d$. After adding synchronous firing, $S_p$ and $S_d$ can also be influenced by the relative timing of the spike of the apical neuron within a firing event, thereby influencing DrV. To check this effect, we generated spike trains using Model Long Tail & Model Sync 3 (**Section S5.1**). In this model, spike trains are inhomogeneous Poisson processes, with the temporal fluctuation of the population rates determined by firing events. Time-averaged rates of the non-apical neurons are lognormal distributed with mean $r_0 = 20\text{Hz}$ and shape parameter $s$, and the rate of the apical neuron is kept at $r_0$. When $s = 1$, the distribution is of long tail; when $s = 0$, the distribution is $\delta$ function so that the firing rates of all the neurons are equal to $r_0$. We found that in this model, synchronous firing contributed to P-D imbalance by strengthening depression process (**S2A Fig**). This is because that the axonal delay in the dendritic motif tends to make the apical neuron receive its afferents from the non-apical neurons *after* its own spikes during a firing event (similar phenomenon has been observed in [5]).

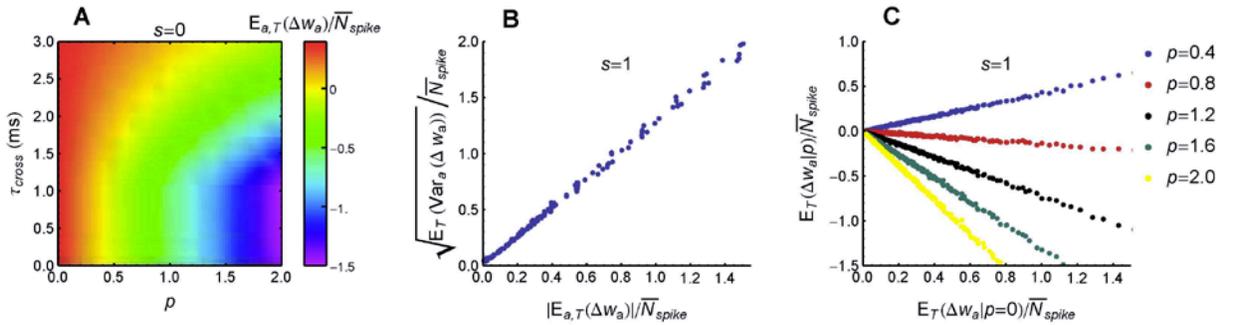

**S2 Fig. Synchronous firing influences P-D imbalance, and heterogeneity of rates makes use of P-D imbalance to change the efficacy variability in DrV manner. (A)** Mean efficacy change per spike as a function of $p$ and $\tau_{cross}$, when spike trains had synchronous firing and all neurons had the same firing rate (Model Long Tail & Model Sync 3 in **S1 Text Section S5.1**; $s = 0$). We can see that synchronous firing strengthens depression in this model. $\overline{N}_{spike}$ represents spike number per neuron. **(B)** Square root of the trial expectation of efficacy variance versus absolute mean efficacy change (both normalized by $\overline{N}_{spike}$) after 100s biological time, with heterogeneity of rates ($s = 1$). Dots represent different $(p, \tau_{cross})$ pairs uniformly sampled within the range in **A**. As efficacy variance is dominated by DrV in the long run, and absolute mean efficacy change

quantifies P-D imbalance, the shown linear relationship suggests that DrV caused by heterogeneity of rates is indeed due to P-D imbalance (see **S1 Text Section S2.3** for details). (**C**) Trial expectations of efficacy changes at $p \neq 0$ versus those at $p = 0$ under rate heterogeneity ($s = 1$). Dots sharing the same horizontal value represent the same synapse in the dendritic motif. This panel shows that under rate heterogeneity, synchronous firing changes $\mathrm{E}_T(\Delta w_a)$ proportionally. In **A-C**, $A_p = 2, A_d = 1$. The other parameters are the same as **S1 Fig**.

Now we check the influence of P-D imbalance on DrV. From eq. S7, we know that $\mathrm{DrV} = \mathrm{Var}_a(\mathrm{E}_T(\Delta w_a | r_a)) \propto (S_p - S_d)^2$. As $\mathrm{DrV} \propto t^2$ and $\mathrm{DiV} \propto t$, DrV will dominate in the total variance in the long run, so that $\mathrm{E}_T(\mathrm{Var}_a(\Delta w_a))$ (the left hand side of eq.S6) should become proportional to $(S_p - S_d)^2$ in the long run. From eq.S7, we also have $|\mathrm{E}_{a,T}(\Delta w_a)| \propto |S_p - S_d|$. These two facts make us expect that $\sqrt{\mathrm{E}_T(\mathrm{Var}_a(\Delta w_a))} \propto |\mathrm{E}_{a,T}(\Delta w_a)|$. As $|\mathrm{E}_{a,T}(\Delta w_a)|$ reflects the P-D imbalance, this proportional relationship reflects that the DrV caused by heterogeneity of rates is due to the P-D imbalance, which is indeed the case we found in our simulation (**S2B Fig**).

As $\mathrm{E}_T(\Delta w_a) \propto r_a r_0$, and synchronous firing influences the coefficient through P-D imbalance, we should expect that if the properties of synchronous firing changes while keeping $r_0$ and $r_a$ unchanged, then the change of $\mathrm{E}_T(\Delta w_a)$ will be proportional to $r_a$. Indeed, we found that $\mathrm{E}_T(\Delta w_a | p) - \mathrm{E}_T(\Delta w_a | p = 0) \propto r_a$, with $p$ controlling the strength of firing events (**S2C Fig**). This means that under rate heterogeneity, synchronous firing changes $\mathrm{E}_T(\Delta w_a)$ proportionally.

### Section S2.4: The Influence of Auto-temporal Structure

*Key points of this subsection:*
1) *Burstiness increases the efficacy variability mainly through the correlation of the efficacy changes caused by adjacent non-apical spikes and that caused by adjacent apical spikes.*
2) *Strong regularity increases the efficacy variability through transient cross-correlation.*

We used Gamma processes to model the auto-temporal structure of spike trains (Model Auto in **Section S5.1**). We changed the shape parameter $\alpha$ of the Gamma process while conserving the firing rates. The coefficient of variance (*CV*) of a Gamma process is $CV = 1/\sqrt{\alpha}$. When *CV* gets larger, spike trains are burstier, when *CV* gets smaller, spikes are more regular. We found that both burstiness and strong regularity in auto-temporal structure increase efficacy variability, and efficacy variability gets its minimal value when *CV* is around 0.3~0.7 (**S3A Fig**).

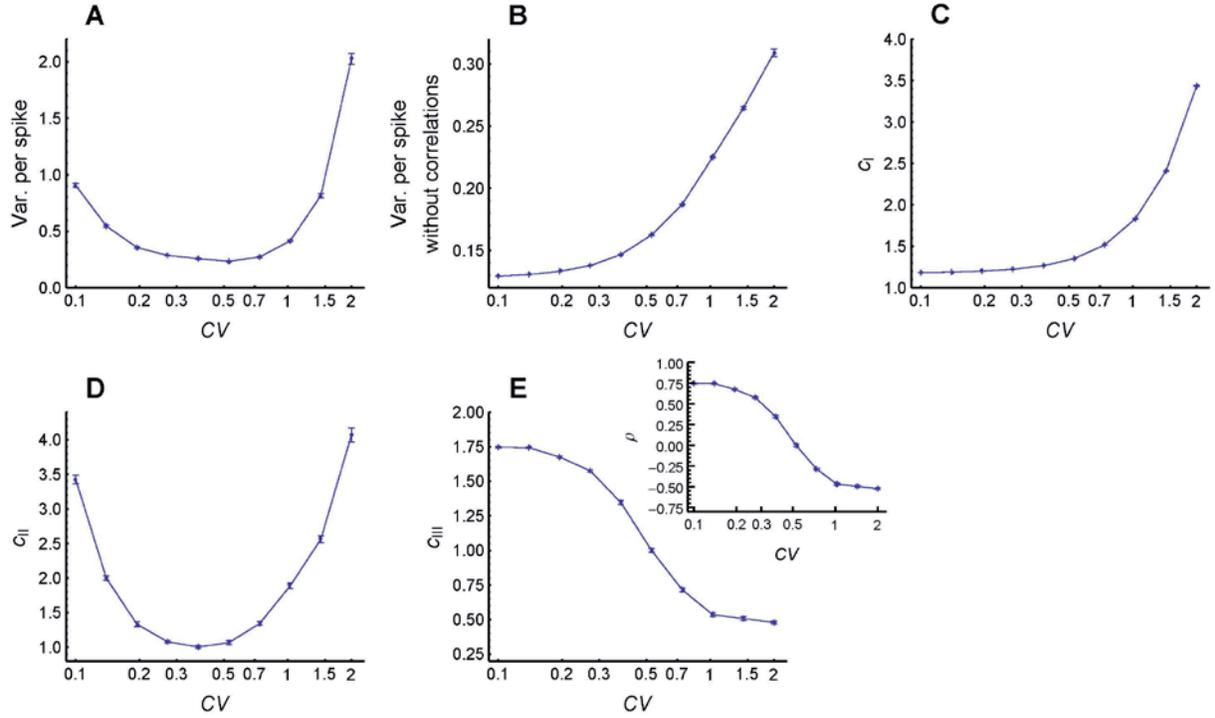

**S3 Fig. Both burstiness and strong regularity increase the efficacy variability.** (**A**) Variance per spike as a function of *CV*. Spike trains were Gamma processes with conserved rate when *CV* changed (Model Auto in **S1 Text Section S5.1**). (**B**) Vertical coordinate is $\mathrm{E}_i\left(\sum_j\sum_k\left(\mathrm{Var}_a\left(\Delta w_{a,k}(t_i,t_{a,j})\right)\right)\right)$, representing variance per spike when all the three types of correlations are absent (**S1 Text eq. S9**). (**C**) $c_{\mathrm{I}}$ quantifies the contribution of Type I correlation to the efficacy variability (**S1 Text eq. S12**). (**D**) $c_{\mathrm{II}}$ quantifies the contribution of Type II correlation to the efficacy variability (**S1 Text eq. S11**). (**E**) $c_{\mathrm{III}}$ quantifies the contribution of Type III correlation to the efficacy variability (**S1 Text eq. S10**). Inset: $\rho$ is the correlation coefficient of the total potentiation and depression values imposed on the same synapse during STDP (**S1 Text eq. S13, S14**). In **A-E**, error bars represent s.e.m., the other parameters are the same as **S1 Fig**.

Next, we want to understand the reason why efficacy variability changes with *CV*. To do this, we write the total change of the *a*th synapse in a dendritic motif as

$$\Delta w_a = \sum_{k=p,d}\Delta w_{a,k} = \sum_{k=p,d}\sum_i\sum_j \Delta w_{a,k}(t_i,t_{a,j}) \qquad (\text{S8})$$

with $\Delta w_{a,p}$ being the total potentiation value, $\Delta w_{a,p}(t_i,t_{a,j})$ being the potentiation value pairing the *i*th spike of the apical neuron and the *j*th spike of the *a*th non-apical neuron, and $\Delta w_{a,d}$ and $\Delta w_{a,d}(t_i,t_{a,j})$ being the corresponding depression values. For the consistency with previous simulations, we also add axonal delay $\tau_{delay}$, and the sign of STDP depends on the timing difference between $t_i$ and $t_{a,j} + \tau_{delay}$ (**Fig. 2B** in main text). The index of $i$ starts from the beginning of the spike train of the apical neuron. The indexing of $j$, however, depends on $a$, $i$ and $k$. For the potentiation process ($k = p$), the index of $j$ starts from the spike immediately before $t_i - \tau_{delay}$ in the spike train of the $a$th non-apical neuron, and goes backward along the spike train; for the depression process ($k = d$), its index starts from the spike immediately after $t_i - \tau_{delay}$, and goes forward along the spike train.

Using eq. S8, we can rewrite the variance of efficacy changes like this:

$$\text{Var}_a(\Delta w_a) = c_{\text{III}} c_{\text{II}} c_{\text{I}} \cdot \sum_i \sum_j \sum_k \left( \text{Var}_a \left( \Delta w_{a,k}(t_i, t_{a,j}) \right) \right) \tag{S9}$$

where

$$c_{\text{III}} = \frac{\text{Var}_a \left( \sum_i \sum_j \sum_k \Delta w_{a,k}(t_i, t_{a,j}) \right)}{\sum_k \left( \text{Var}_a \left( \sum_i \sum_j \Delta w_{a,k}(t_i, t_{a,j}) \right) \right)}, \tag{S10}$$

$$c_{\text{II}} = \frac{\sum_k \left( \text{Var}_a \left( \sum_i \sum_j \Delta w_{a,k}(t_i, t_{a,j}) \right) \right)}{\sum_i \sum_k \left( \text{Var}_a \left( \sum_j \Delta w_{a,k}(t_i, t_{a,j}) \right) \right)}, \tag{S11}$$

$$c_{\text{I}} = \frac{\sum_i \sum_k \left( \text{Var}_a \left( \sum_j \Delta w_{a,k}(t_i, t_{a,j}) \right) \right)}{\sum_i \sum_j \sum_k \left( \text{Var}_a \left( \Delta w_{a,k}(t_i, t_{a,j}) \right) \right)}. \tag{S12}$$

To understand the three coefficients, first note that

$$\text{Var}_a \left( \sum_i \sum_j \sum_k \Delta w_{a,k}(t_i, t_{a,j}) \right) = \sum_{k=p,d} \text{Var}_a \left( \sum_i \sum_j \Delta w_{a,k}(t_i, t_{a,j}) \right)$$

$$+ 2\rho \sqrt{\text{Var}_a \left( \sum_i \sum_j \Delta w_{a,p}(t_i, t_{a,j}) \right) \text{Var}_a \left( \sum_i \sum_j \Delta w_{a,d}(t_i, t_{a,j}) \right)}, \tag{S13}$$

with $\rho$ being the correlation coefficient between the total potentiation and depression value imposed on the same synapse. Thus,

$$c_{\text{III}} = 1 + \frac{2\rho \sqrt{\text{Var}_a \left( \sum_i \sum_j \Delta w_{a,p}(t_i, t_{a,j}) \right) \text{Var}_a \left( \sum_i \sum_j \Delta w_{a,d}(t_i, t_{a,j}) \right)}}{\sum_k \left( \text{Var}_a \left( \sum_i \sum_j \Delta w_{a,k}(t_i, t_{a,j}) \right) \right)}, \tag{S14}$$

which quantifies the contribution to the efficacy variability by the correlation between the total potentiation and depression value imposed on the same synapse (*Type III correlation*). Similarly, $c_{\text{II}}$ quantifies the contribution by the correlation between the STDP updatings caused by adjacent spikes of the apical neuron (*Type II correlation*), $c_\text{I}$ quantifies the contribution by the correlation between the STDP updatings caused by adjacent spikes of the non-apical neurons (*Type I correlation*); and $\sum_i \sum_j \sum_k \left( \text{Var}_a \left( \Delta w_{a,k}(t_i, t_{a,j}) \right) \right)$ represents the efficacy variability when all of the three correlations are absent.

Now we analyze the reason why $\text{Var}_a(\Delta w_a)$ is smallest when *CV* is around 0.3~0.7, and gets larger when spike trains are burstier or more regular. We find that $\sum_i \sum_j \sum_k \left( \text{Var}_a \left( \Delta w_{a,k}(t_i, t_{a,j}) \right) \right)$ increases with *CV* (**S3B Fig**). However, comparing to $\text{Var}_a(\Delta w_a)$, both its value and increase are small, which means that it alone is far from sufficient to understand the value and change of $\text{Var}_a(\Delta w_a)$. $c_\text{III} > 1$ when spike trains are regular, and $c_\text{III} < 1$ when spike trains are busty (**S3E Fig**), which means that Type III correlation contributes positively to $\text{Var}_a(\Delta w_a)$ when spike trains are regular, and contributes negatively when spike trains are bursty. However, $c_\text{III}$ changes steepest when *CV* is around 0.3~0.7 with $\text{Var}_a(\Delta w_a)$ being flat, and is flat when *CV* gets outside this range while $\text{Var}_a(\Delta w_a)$ changing steeply (**S3AE Fig**), which means that Type III correlation is not the main contribution to the change of $\text{Var}_a(\Delta w_a)$ with *CV*. $c_\text{II}$ gets large when spike trains get bursty or regular (**S3D Fig**), which means that Type II correlation contributes to $\text{Var}_a(\Delta w_a)$ when spike trains are both bursty or regular. $c_\text{I}$ is large when spike trains are bursty, and monotonically decreases when *CV* decreases (**S3C Fig**), which means that Type I correlation significantly contributes to $\text{Var}_a(\Delta w_a)$ when spike trains are bursty, but does not contribute to the increase of $\text{Var}_a(\Delta w_a)$ when spike trains get regular. From our analysis above, we know that both Type II and Type I correlation contributes to the increase of $\text{Var}_a(\Delta w_a)$ when spike trains are bursty, and only Type II correlation contributes to its increase when spike trains are regular.

The physical pictures of how Type II and Type I correlation contribute to $\text{Var}_a(\Delta w_a)$ have already been explained in the main text. The physical pictures of how $\sum_i \sum_j \sum_k \left( \text{Var}_a \left( \Delta w_{a,k}(t_i, t_{a,j}) \right) \right)$ and Type III correlation change with *CV* will be explained in the Miscellaneous (**Section S6.2, S6.3**).

### Section S2.5: The Influence of Heterogeneity of Cross-correlations

*Key points of this subsection:*
1) *Heterogeneity of cross-correlations can induce heterogeneity of diffusion strengths among the synapses in a network.*
2) *The same degree of heterogeneity of cross-correlations may induce different DrV, depending on the positions of the cross-correlation time windows relative to the STDP time window. DrV is large when the cross-correlations aggregate near the sharp change point of the STDP time window, and is small when the cross-correlations distribute far away from this point.*

We define the cross-correlation between the $a$th non-apical neuron and the apical neuron in a dendritic motif as

$$C_{cross,a}(\tau) = \frac{\langle r_a(t)r_0(t+\tau)\rangle - \langle r_a(t)\rangle\langle r_0(t)\rangle}{\sqrt{\langle r_a(t)\rangle\langle r_0(t)\rangle}} \quad (S15)$$

with $r_0(t)$ and $r_a(t)$ being the firing rates of the apical neuron and the $a$th non-apical neuron, and $\langle \cdot \rangle$ representing time average. So when the two neurons have the same stationary firing rate $r_0$, the expectation of the increment of the $a$th synapse caused by STDP per unit time is

$$\frac{d\langle \Delta w_a \rangle}{dt} = r_0 \int_{-\infty}^{\infty} C_{cross,a}(\tau) H(\tau) d\tau \quad (S16)$$

with $H(\tau)$ being the STDP time window whose sharp change point is at the axonal delay $\tau_{delay}$ (**Fig. 2B** in the main text). The heterogeneity of $d\langle \Delta w_a \rangle/dt$ for different $a$ is what to cause DrV.

The diffusion strength of the $a$th synapse has a strong dependence on the width of the time window of $C_{cross,a}(\tau)$. For example, suppose both the time windows of $C_{cross,1}(\tau)$ and $C_{cross,2}(\tau)$ are symmetric around $\tau_{delay}$, but the time window of $C_{cross,1}(\tau)$ is narrower than that of $C_{cross,2}(\tau)$ (**S4A Fig**). Then if $H(\tau)$ is strictly asymmetric around $\tau_{delay}$, we have $d\langle \Delta w_1 \rangle/dt = d\langle \Delta w_2 \rangle/dt = 0$, thereby DrV being zero. However, as the single-step change of the 1st synapse is larger than that of the 2nd synapse during STDP, the 1st synapse tends to diffuse farther away from its initial value than the 2nd synapse.

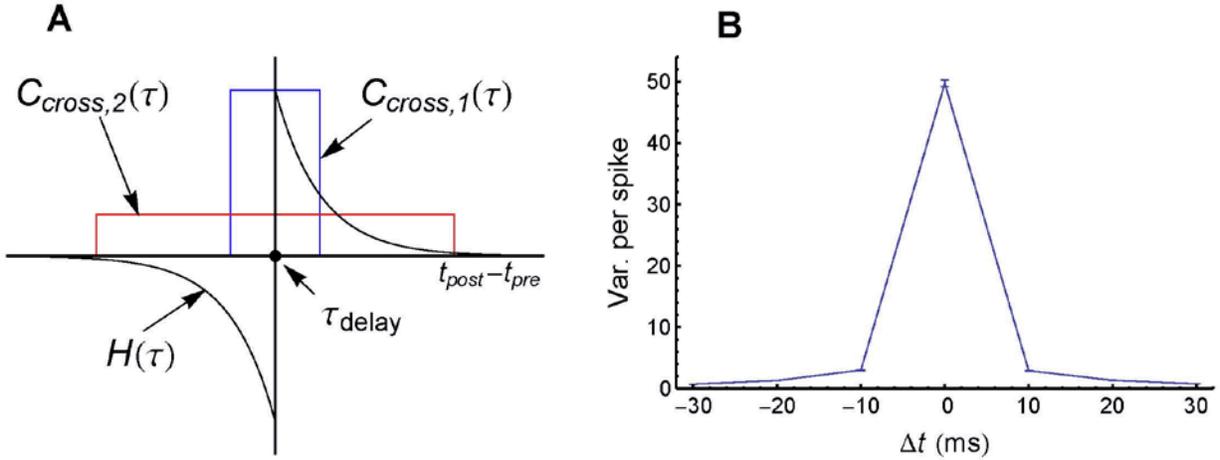

**S4 Fig. The influence of heterogeneity of cross-correlations on the efficacy variability.** (**A**) Schematic on how heterogeneity of cross-correlations causes heterogeneity of diffusion strengths. The STDP window $H(\tau)$ is represented by the black curve. Two cross-correlations $C_{cross,1}(\tau)$ and $C_{cross,2}(\tau)$, indicated by the blue and red curve respectively, are symmetric around $\tau_{delay}$, but have different widths. Both of them cause zero drift velocity of synaptic efficacies, but the diffusion strength of the 1st synapse is stronger than that of the 2nd one. (**B**) The same degree of heterogeneity of cross-correlations may induce different DrV, depending on the positions of cross-correlation time windows relative to the STDP time window. Spikes were generated according to Model Cross-correlation (**S1 Text Section S5.1**), with $q = 0.2$ representing the cross-correlation strength, $\varepsilon_0 = 10\text{ms}$ representing the degree of heterogeneity of cross-correlations (see **S1 Text Section S5.1** for modeling details), and $\Delta t$ as the horizontal coordinate here representing the average position of the cross-correlation time windows relative to the STDP time window. When $\Delta t = 0$, the potentiation and depression processes of STDP most split synapses to different directions, so that DrV gets its maximal value. Error bars represent s.e.m. Other parameters are the same as **S1 Fig**.

However, as we explained in the Discussion in the main text, we do not seriously consider the heterogeneity of diffusion strengths among synapses in this work, but only consider the heterogeneity of drift velocities in details. Therefore, we constructed spike patterns in which $C_{cross,a}(\tau)$ were $\delta$ functions (Model Cross-correlation in **Section S5.1**):

$$C_{cross,a}(t - \tau_{delay}) = q\delta(t - \tau_a) \tag{S17}$$

with $q$ being the strength of the cross-correlation, and $\tau_a$ indicating its location. We let $\tau_a$ uniformly distributed within $[\Delta t - \varepsilon_0, \Delta t + \varepsilon_0]$ for different $a$, with $\varepsilon_0$ quantifying the heterogeneity of cross-correlations and $\Delta t$ being the average position of the cross-correlations. When $q$ and $\varepsilon_0$ were fixed, DrV got its maximal value when $\Delta t = 0$ (**S4B Fig**), where the potentiation and depression processes of STDP most split synapses to different directions; DrV got weaker as the cross-correlations gradually moving farther away from the sharp change point $\tau = \tau_{delay}$ of the STDP time window $H(\tau)$. Therefore, the same degree of heterogeneity of cross-correlations may cause different DrV, depending on their positions relative to the STDP time window.

### Section S2.6: The Interaction of Synchronous Firing and Auto-temporal Structure

*Key points of this subsection:*
1) *Synchronous firing and auto-temporal structure are coupled together based on time-rescaling transform (**Fig. 3** in the main text). Auto-temporal structure then has two aspects: the auto-temporal structure of the spikes in the rescaled time (represented by $CV_{rescale}$), and the temporal structure of the occurrence of firing events (represented by $CV_{events}$).*
2) *Burstiness in $CV_{events}$ increases the efficacy variability; strong regularity in it does not have significant effect on the efficacy variability.*
3) *Burstiness in $CV_{rescale}$ increases the efficacy variability; strong regularity in it usually decreases the efficacy variability, except when the amplitudes of firing events are so weak that the spike pattern approaches to asynchrony.*

4) *If the firing order of neurons in adjacent firing events are correlated, the efficacy variability may be enlarged by transient cross-correlation.*

The spike pattern of real neural populations possesses both synchronous firing and auto-temporal structure, so it is desirable to know how these two pattern structures interact to influence the efficacy variability. To investigate this issue, we couple them together based on time-rescaling transform (**Fig. 3** in the main text): the population rate $r(t)$ is determined by the occurrence of firing events, then a rescaled time is defined as the cumulative function of $r(t)$ [6]

$$\Lambda(t) = \int_0^t r(s) \mathrm{d}s, \tag{S18}$$

which stretches the inter-spike intervals in proportion to the firing rate. Auto-temporal structure comes into the picture in two ways: the auto-temporal structure of the spikes in the rescaled time, represented by $CV_{rescale}$, and the temporal structure of the occurrence of firing events, represented by $CV_{events}$ (**Fig. 3** in the main text).

To understand the interaction of synchronous firing and auto-temporal structure, here we consider a model in a dendritic motif in which both the occurrence of firing events and the spike trains in the rescaled time are Gamma processes (Model Sync-Auto in **Section S5.1**).

The efficacy variability increases with $CV_{events}$ almost monotonically (**S5AB Fig**). It increases when $CV_{events}$ represents burstiness, but does not change much when $CV_{events}$ is in the range of strong regularity. To understand this, note that the burstiness of the occurrence of firing events is able to gather spikes closer, thereby enhancing the mechanism of spike gathering caused by synchronous firing (see **Fig. 4** in the main text); when the occurrence of firing events is regular, firing events are far apart, in which case the efficacy variability becomes less sensitive to the mechanism of spike gathering caused by $CV_{events}$, because of the exponential decay of the STDP time window.

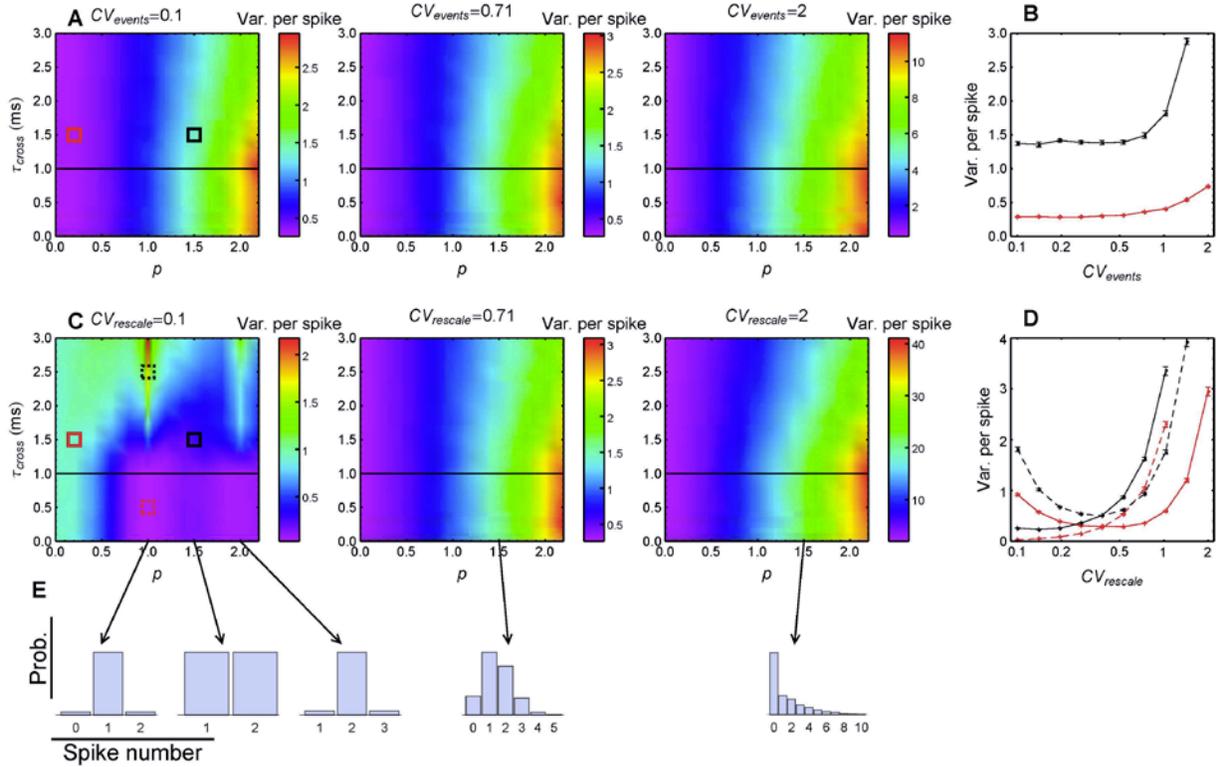

**S5 Fig. How the interaction of synchronous firing and auto-temporal structure influences the efficacy variability.** (A) Variance per spike as a function of $p$ and $\tau_{cross}$ at different $CV_{events}$ values. Spike trains were generated in the scenario of **Fig. 3**, in which both the occurrence of firing events and the spike trains in the rescaled time were Gamma processes (Model Sync-Auto in **S1 Text Section S5.1**). $CV_{rescale} = 0.71$. The horizontal black line represents the axonal delay $\tau_{delay} = 1$ms. (B) Variance per spike as a function of $CV_{events}$ for $(p, \tau_{cross})$ pairs marked in **A** (with the corresponding colors). Error bars represent s.e.m. (C) Variance per spike as a function of $p$ and $\tau_{cross}$ at different $CV_{rescale}$ values. $CV_{events} = 0.71$. (D) Variance per spike as a function of $CV_{rescale}$ for $(p, \tau_{cross})$ pairs marked in **C**. Error bars represent s.e.m. (E) Probability distribution of the spike numbers of the non-apical neurons in a firing event at different $p$ and $CV_{rescale}$ values indicated by the starting points of arrows. In **A-E**, the other parameters are the same as **S1 Fig**.

    Comparing to $CV_{events}$, $CV_{rescale}$ influences the efficacy variability in a more complicated way. When the amplitudes of firing events are weak, the efficacy variability increases both when $CV_{rescale}$ is too large or too small (**S5CD Fig**), which is similar to how auto-temporal structure influences the efficacy variability when synchronous firing is absent (**S3A Fig**). When firing events are not weak, the key concept to understand its influence is the distribution of the spike numbers of the non-apical neurons in a firing event. When $CV_{rescale}$ represents burstiness, this

distribution is wide (**S5E Fig**), so that different non-apical neurons in a dendritic motif may fire different number of spikes in a firing event, which makes the STDP updatings of the corresponding synapses heterogeneous, thereby increasing the efficacy variability. When $CV_{rescale}$ reduces into the range of strong regularity, this distribution becomes narrow and also sensitive to $p = r_0 / r_{events}$ (**S5E Fig**), with $r_0$ being the firing rate of neurons, and $r_{events}$ the rate of the occurrence of firing events. On the one hand, this narrow distribution is able to induce synapse correlating (see **Fig. 4** in the main text), which significantly reduces the efficacy variability when $\tau_{cross} < \tau_{delay}$ (**S5C Fig**). On the other hand, when $p = r_0 / r_{events}$ becomes integer, this distribution becomes further narrowed and strongly peaked at a single number (**S5E Fig**). This single-number peaked distribution has two effects. Firstly, it further enhances the mechanism of synapse correlating, which may further reduce the efficacy variability when $\tau_{cross} < \tau_{delay}$. Secondly, it makes the firing order of neurons almost unchanged in adjacent firing events: for example, if the firing order in a firing event is $a \to b \to c$ (here the letters represent neuronal indexes), then the order is very likely to be also $a \to b \to c$ in the next firing event. This effect induces transient cross-correlation (see **Fig. 4** in the main text), which can increase the efficacy variability especially if it couples with the mechanism of synapse splitting (see **Fig. 4** in the main text) when $\tau_{cross} > \tau_{delay}$, when the non-apical neurons whose spikes arrive at the apical neuron before the firing of the apical neuron itself in several sequential firing events potentiate their synapses in these firing events, while the non-apical neurons whose spikes arrive at the apical neuron after the firing of the apical neuron itself depress their synapses in these sequential firing events (**S5CD Fig**).

In biological systems, the amplitudes of firing events may exhibit strong variability [7,8], which further increases the complexity of the problem. Our simulation suggested that the variability of amplitudes may increase the efficacy variability, as the firing events with large amplitudes can gather more spikes closer; and if these amplitudes are also temporal-correlated so that strong firing events tend to appear sequentially, the efficacy variability may be further increased (data not shown). However, we argue that this varying-amplitude situation may be included in the constant-amplitude scenario using $CV_{events}$, after noting that strong firing events can be regarded as the burstiness of many small firing events. For simplicity, we will not consider this varying-amplitude situation in our following discussion.

### Section S2.7: The Interaction of Auto-temporal Structure and Heterogeneity of Rates

*Key points of this subsection:*
1) *Heterogeneity of rates does not influence the DiV caused by auto-temporal structure when spike trains are bursty.*
2) *When spike trains are regular, heterogeneity of rates destroys transient cross-correlation, thereby decreasing the efficacy variability.*

As a reminder for the readers, we rewrite eq. S6 here

$$\mathrm{E}_T \left( \mathrm{Var}_a (\Delta w_a) \right) \approx \mathrm{E}_a \left( \mathrm{Var}_T (\Delta w_a \mid r_a) \right) + \mathrm{Var}_a \left( \mathrm{E}_T (\Delta w_a \mid r_a) \right) \tag{S19}$$

where *a* is the index of non-apical neuron, *T* represents integrating over different trials, and the conditioning to firing rate $r_a$ emphasizes that different non-apical neurons have different firing rates. $E_a\left(\text{Var}_T(\Delta w_a \mid r_a)\right)$ represents DiV, and $\text{Var}_a\left(E_T(\Delta w_a \mid r_a)\right)$ represents DrV.

We modeled spike patterns in a dendritic motif with the interaction of auto-temporal structure and heterogeneity of rates as Gamma processes of different rates (Model Auto & Model Long Tail in **Section S5.1**). Differently from synchronous firing (**S2 Fig**), auto-temporal structure does not induce P-D imbalance to change DrV (to understand this, note that as we suppose that the spike trains of the apical neuron and a non-apical neurons are independent, the distribution of the intervals between the spikes of the apical neuron and the spikes of the non-apical neuron is *independent* on the auto-temporal structure of their spike trains in the long run, so the strengths of both the potentiation process and the depression process of STDP remain the same when the auto-temporal structure changes). So $\text{Var}_a\left(E_T(\Delta w_a \mid r_a)\right) = 0$ when the potentiation and depression strength of the STDP time window is equal (i.e. $A_p = A_d$). So we investigated how the heterogeneity of rates influences DiV, i.e. $E_a\left(\text{Var}_T(\Delta w_a \mid r_a)\right)$.

In a random walk, the diffusion variance is proportional to the step number, so if we regard the STDP process as a random walk, we should expect $\text{Var}_T(\Delta w_a \mid r_a) \propto r_a$, which is indeed the case when spike trains are bursty (**S6B Fig, upper panels**). This makes the heterogeneity of $r_a$ does not change $E_a\left(\text{Var}_T(\Delta w_a \mid r_a)\right)$ much as long as the mean of $r_a$ conserves (**S6A Fig**, note that $E_T\left(\text{Var}_a(\Delta w_a)\right) \approx E_a\left(\text{Var}_T(\Delta w_a \mid r_a)\right)$ here). However, when spike trains are regular, $\text{Var}_T(\Delta w_a \mid r_a)$ peaks at $r_a = r_0$ due to transient cross-correlation (see **Fig. 4** in the main text) with $r_0$ being the firing rate of the apical neuron (**S6B Fig, lower panels**). Therefore, compared to the case when $r_a = r_0$ homogeneously, $E_a\left(\text{Var}_T(\Delta w_a \mid r_a)\right)$ is small when $r_a$ becomes heterogeneous (**S6A Fig**).

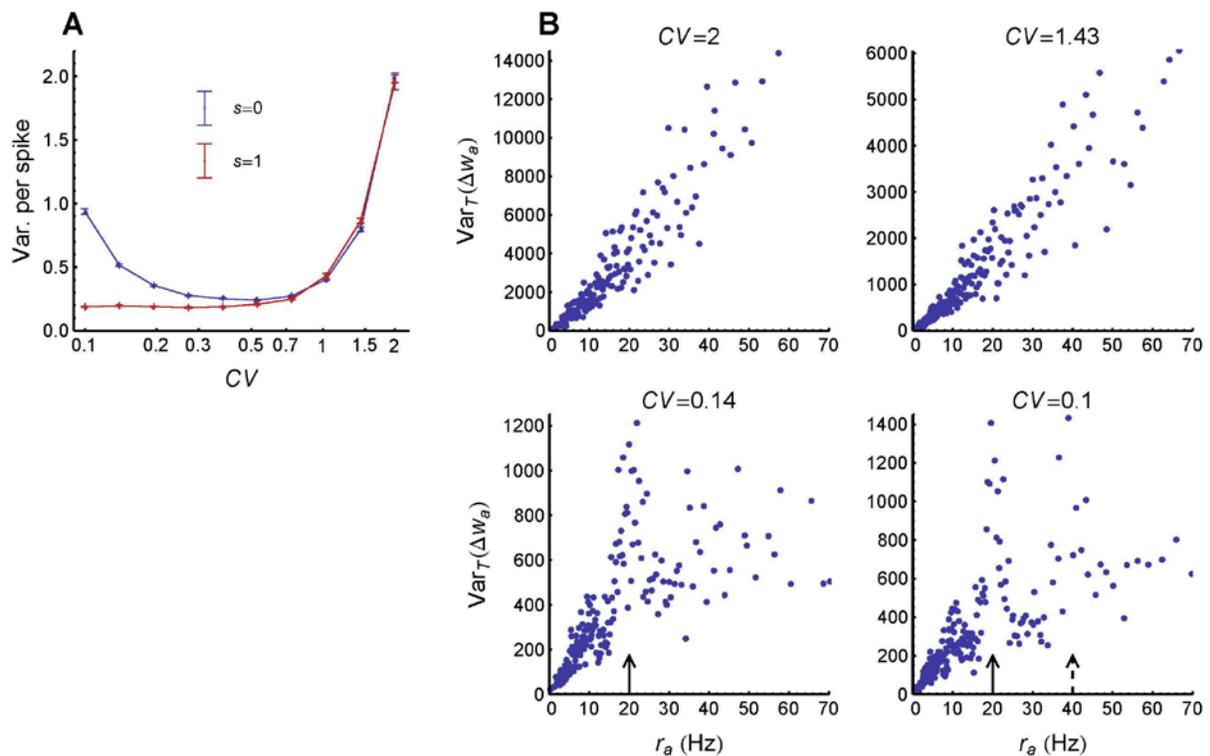

**S6 Fig. How the interaction of heterogeneity of rates and auto-temporal structure influences the efficacy variability.** (**A**) Heterogeneity of rates does not significantly influence the efficacy variability when the spikes are bursty, but removes the increase of the efficacy variability caused by strong regularity. Spike trains were homogeneous Gamma processes with lognormal rate distributions (Model Long Tail & Model Auto in **S1 Text Section S5.1**). Error bars represent s.e.m. (**B**) The diffusion of a synapse depends on the firing rate $r_a$ of the non-apical neuron it links. The firing rate of the apical neuron was kept at $r_0 = 20\text{Hz}$. When the spike trains are bursty (*CV*=2 or 1.43), the diffusion of a synapse linearly correlates with the firing rate $r_a$ of the corresponding non-apical neuron. When the spike trains are regular (*CV*=0.14 or 0.1), the diffusion peaks when $r_a$ is equal to the firing rate of the apical neuron $r_0 = 20\text{Hz}$ (indicated by solid arrows), because of transient cross-correlation. Note that when the spikes are very regular (*CV*=0.1), the diffusion can even peak at $2r_0$ (dashed arrow). The other parameters are the same as **S1 Fig**.

### Section S2.8: The Interaction of Auto-temporal Structure, Synchronous Firing and Rate Heterogeneity

*Key points of this subsection:*

1) $CV_{rescale}$ does not influence P-D imbalance. The efficacy variability increases with $CV_{rescale}$ in DiV manner, so this increase is significant when potentiation and depression are balanced so that DrV is zero, but gets negligible when they are imbalanced so that DrV is nonzero.
2) $CV_{events}$ not only changes DiV but also influences P-D imbalance, thereby changing DrV with the existence of heterogeneity of rates. The dependence of the efficacy variability on $CV_{events}$ is complicated.

As we discussed (**S2 Fig**), synchronous firing may change P-D imbalance, and rate heterogeneity can make use of this imbalance to change DrV. Now we discuss the case when auto-temporal structure is added into the picture.

As shown in **Fig. 3** in the main text, auto-temporal structure is represented by $CV_{rescale}$ and $CV_{events}$. $CV_{rescale}$ increases DiV by inducing variety of spike numbers of non-apical neurons in a firing events (**S5E Fig**), but it hardly influences P-D imbalance (**S7A Fig, lower panel**). So when the strength of firing events is adjusted so that the potentiation and depression almost balance each other (so that $DrV = 0$), the efficacy variability increases with $CV_{rescale}$ because of the increase of DiV (**S7A Fig, upper panel**). However, when the potentiation and depression are imbalanced, the efficacy variability becomes almost independent of $CV_{rescale}$ (**S7A Fig, upper panel**). This is because $CV_{rescale}$ hardly influences P-D imbalance so that hardly changes DrV: as $DrV \propto t^2$ while $DiV \propto t$, the efficacy variability becomes almost the same in the long run when DrV is the same even though DiV is different.

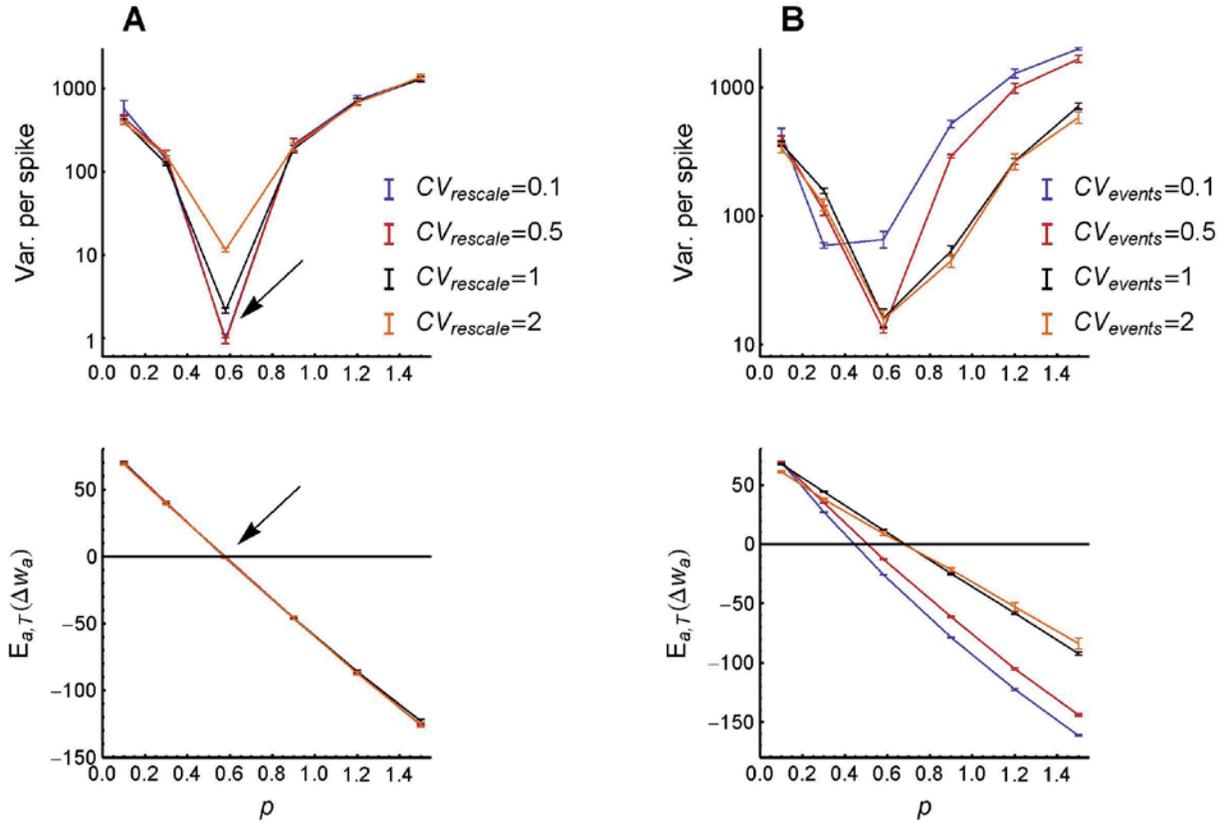

**S7 Fig. The influence of auto-temporal structure onto the efficacy variability when both synchronous firing and rate heterogeneity exist.** (**A**) Upper: Variance per spike as a function of $p$ at different $CV_{rescale}$, keeping $CV_{events} = 0.71$ and $s = 1$ (Model Long Tail & Model Sync-Auto in **S1 Text Section S5.1**). Lower: the corresponding mean efficacy changes, which represents P-D imbalance. Note that $CV_{rescale}$ hardly changes P-D imbalance, and thus hardly changes DrV with the existence of heterogeneity of rates. The arrows indicate the $p$ value at which the mean efficacy change is almost zero (indicated by the horizontal black line), so that DrV=0 due to P-D balance. At this point, the efficacy variability significantly increases with $CV_{rescale}$ (upper panel) due to DiV effect. Error bars represent s.e.m. in normal scale and relative errors corresponding to s.e.m. in log scale. (**B**) The same as **A** except that different lines represent different $CV_{events}$ values, keeping $CV_{rescale} = 0.71$. Note that $CV_{events}$ is able to change P-D imbalance (lower panel), and thus change DrV with the existence of heterogeneity of rates, so that the dependence of the efficacy variability on $CV_{events}$ is complicated (upper panel). In **A-B**, $\tau_{cross} = 2\text{ms}$, $A_p = 2$, $A_d = 1$. The other parameters are the same as **S1 Fig**.

However, the influence of $CV_{events}$ is complicated, because $CV_{events}$ not only changes DiV, but also changes P-D imbalance (**S7B Fig, lower panel**), thereby changing DrV. When the temporal structure of the occurrence of firing events is not considered, we can just say that the synapses are potentiated or depressed according to the relative timings of the spikes of the apical neuron and the non-apical neurons in a firing event. But to be exact, the values of synaptic depression or potentiation also depend on the occurrence of nearby firing events, which is here controlled by $CV_{events}$. Because of this reason, we find that the efficacy variability depends on $CV_{events}$ in a complex non-monotonic way (**S7B Fig, upper panel**). We leave detailed discussions of this issue to future researches.

### Section S2.9: The Interaction of Heterogeneity of Cross-correlations With the Other Pattern Structures

*Key points of this subsection:*
1) *Heterogeneity of cross-correlations is able to control the drift velocities of synaptic efficacies in a much finer way than heterogeneity of rates. They together determine the drift velocities when they coexist.*
2) *We believe that when adding heterogeneity of cross-correlations, especially when the heterogeneity of cross-correlations is weak, the mechanisms how the other pattern structures influence the efficacy variability remain valid, so their influences should remain qualitatively unchanged.*

Without heterogeneity of cross-correlations, $E_T(\Delta w_a | r_a) \propto r_a r_0$ with almost the same proportional coefficient for different *a*, and synchronous firing can only change this coefficient simultaneously for all *a* (**S2C Fig**). However, heterogeneity of cross-correlations is able to change $E_T(\Delta w_a)$ in a much finer way, depending on the details of the cross-correlation structure in spike patterns. When heterogeneity of cross-correlations coexists with synchronous firing and rate heterogeneity, they together determine $E_T(\Delta w_a)$.

Due to its complexity, we do not explicitly examine the case when heterogeneity of cross-correlations interacting with the other pattern structures. We have already explained the physical mechanism how these pattern structures influence the efficacy variability in the previous sections, and we believe that their influences on the efficacy variability remain qualitatively the same as long as these mechanisms remain valid, which is the case especially when the heterogeneity of cross-correlations is weak.

### Section S2.10: The Coupling of Dendritic and Axonal Homeostasis

*Key points of this subsection:*
1) *Comparing to dendritic homeostasis alone, the coupling of dendritic and axonal homeostasis changes the efficacy variability through the correlation of the STDP change of a link with the mean STDP change in the axonal motif that the link belongs to (represented by $\text{Corr}(\Delta w_{ab}, \Delta \overline{w}_b)$ ).*
2) *Both synchronous firing and heterogeneity of rates individually decrease the efficacy variability through the coupling of dendritic and axonal homeostasis, heterogeneity of cross-correlations changes (not necessarily decreases) the efficacy variability, auto-temporal structure hardly has effect when the other aspects of pattern structure are absent.*

3) *Synchronous firing increases* $\text{Corr}(\Delta w_{ab}, \Delta \bar{w}_b)$ *by increasing the variance of the correlated component as* $\mathcal{O}(t)$ *order with time, which is of the same order as diffusion noises. Heterogeneity of rates and heterogeneity of cross-correlations increase the variance of the correlated component as* $\mathcal{O}(t^2)$ *order, thus will dominate over the effect of synchronous firing in the long run.*
4) *When synchronous firing, auto-temporal structure and heterogeneity of rates coexist,* $\text{Corr}(\Delta w_{ab}, \Delta \bar{w}_b)$ *decreases with* $CV_{rescale}$ *especially at P-D balance; the influence of* $CV_{events}$ *is complicated.*

Synaptic homeostasis may be simultaneously imposed onto the synapses afferent to or efferent from each neuron. In our model, both STDP and synaptic homeostasis are additive, so the efficacy of the link $b \to a$ relative to the mean efficacy of all the links in a network is

$$\Delta w_{ab,total} = \Delta w_{ab} - \Delta \tilde{w}_a - \Delta \tilde{w}_b \tag{S20}$$

with $\Delta w_{ab}$ being the relative efficacy contributed by STDP, $\Delta \tilde{w}_a$ and $\Delta \tilde{w}_b$ being the compensating value due to the synaptic homeostasis imposed on the dendritic and axonal motif that the link belongs to. Now let us investigate what $\Delta \tilde{w}_a$ and $\Delta \tilde{w}_b$ are. Suppose $\Delta \bar{w}_a$ being the mean relative efficacy of the dendritic motif $\mathcal{D}_{b \to a}$ that the link $b \to a$ belongs to, while $\Delta \bar{w}_b$ being that of the axonal motif $\mathcal{A}_{b \to a}$. Then the synaptic homeostasis imposed on all the axonal motifs coupling with $\mathcal{D}_{b \to a}$ changes $\Delta \bar{w}_a$ by order $\mathcal{O}(1/\sqrt{N_{a,in}})$, with $N_{a,in}$ being the in-degree of neuron $a$; similarly, all the dendritic motifs coupling with $\mathcal{A}_{b \to a}$ contribute $\Delta \bar{w}_b$ by $\mathcal{O}(1/\sqrt{N_{b,out}})$, with $N_{b,out}$ being the out-degree of neuron $b$. So when $N_{a,in}$ and $N_{b,out}$ are sufficiently large, all the motifs coupling with $\mathcal{D}_{b \to a}$ or $\mathcal{A}_{b \to a}$ cannot change $\Delta \bar{w}_a$ or $\Delta \bar{w}_b$ too much, which makes $\Delta \tilde{w}_a \approx \Delta \bar{w}_a$ and $\Delta \tilde{w}_b \approx \Delta \bar{w}_b$ in the network. In this case, the variance of the efficacies that belong to $\mathcal{D}_{b \to a}$ is

$$\text{Var}_b(\Delta w_{ab,total}) \approx \text{Var}_b(\Delta w_{ab} - \Delta \bar{w}_b)$$
$$= \text{Var}_b(\Delta w_{ab}) + \text{Var}_b(\Delta \bar{w}_b) - 2\text{Corr}(\Delta w_{ab}, \Delta \bar{w}_b)\sqrt{\text{Var}_b(\Delta w_{ab}) \cdot \text{Var}_b(\Delta \bar{w}_b)} \tag{S21}$$

So under the requirement that the efficacy variance under the coupling of dendritic and axonal homeostasis is smaller than that under dendritic homeostasis alone, i.e. $\text{Var}_b(\Delta w_{ab,total}) < \text{Var}_b(\Delta w_{ab})$, we have

$$\text{Corr}(\Delta w_{ab}, \Delta \bar{w}_b) > \frac{1}{2}\sqrt{\frac{\text{Var}_b(\Delta \bar{w}_b)}{\text{Var}_b(\Delta w_{ab})}} \tag{S22}$$

Because $-1 \leq \text{Corr}(\Delta w_{ab}, \Delta \bar{w}_b) \leq 1$, this condition requires

$$\frac{\text{Var}_b(\Delta \bar{w}_b)}{\text{Var}_b(\Delta w_{ab})} < 4 \tag{S23}$$

to realize.

Because $\text{Corr}(\Delta w_{ab}, \Delta \bar{w}_b)$ is the key reason why the efficacy variability further changes due to the coupling of dendritic and axonal homeostasis, we use it to quantify the ability of different aspects of pattern structure to change the efficacy variability through dendritic-axonal coupling.

To investigate the influence of dendritic-axonal coupling onto efficacy variability, we studied a tree-structural motif in which dendritic motif and axonal motifs were coupled together (**S8A Fig**). We compared the efficacy variability in the coupled dendritic motif with that in a free one, and also calculated $\text{Corr}(\Delta w_{0a}, \Delta \bar{w}_a)$, with $\Delta w_{0a}$ being the efficacy change only contributed by STDP (without counting synaptic homeostasis) at the $a$th synapse in the coupled dendritic motif, and $\Delta \bar{w}_a$ being the mean efficacy change contributed by STDP in the $a$th axonal motif. Our simulation suggested that synchronous firing and rate heterogeneity can individually increase $\text{Corr}(\Delta w_{0a}, \Delta \bar{w}_a)$ and accordingly reduce the efficacy variability in the coupled dendritic motif, heterogeneity of cross-correlations can also influence (not necessarily increase) $\text{Corr}(\Delta w_{0a}, \Delta \bar{w}_a)$, but auto-temporal structure hardly has effect when the other aspects of pattern structure are absent (**S8B-E Fig**).

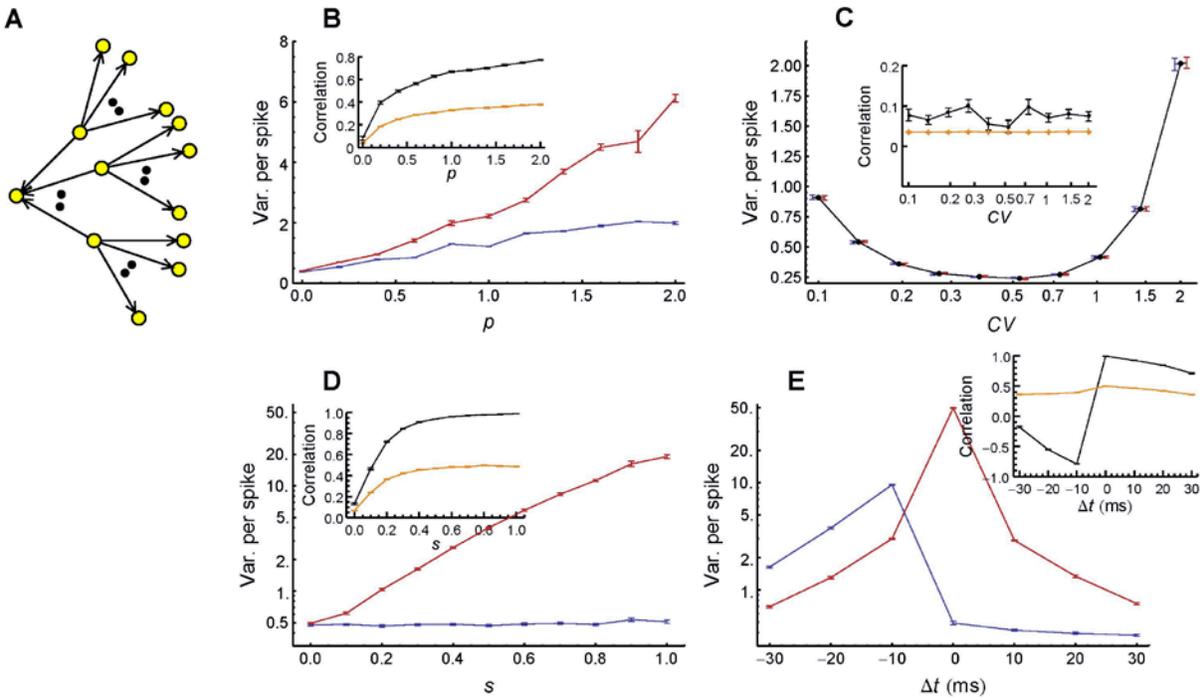

**S8 Fig. How different aspects of pattern structures individually influence the efficacy variability through the coupling of dendritic and axonal homeostasis.** (**A**) The tree-structural motif used in our simulations, in which a dendritic motif is coupled with many axonal motifs (the same as **Fig. 2E** in the main text). (**B**) The influence of synchronous firing. Variance per spike as a function of $p$ in free (red) or coupled (blue) dendritic motif, when spike trains were generated by Model Sync 3 (**S1 Text Section S5.1**). $\tau_{cross} = 2\text{ms}$. Inset: $\text{Corr}(\Delta w_{0a}, \Delta \bar{w}_a)$ (black) and $\frac{1}{2}\sqrt{\text{Var}_a(\Delta \bar{w}_a)/\text{Var}_a(\Delta w_{0a})}$ (orange)

as functions of $p$ (see **S1 Text eq. S22**). Different colors in the following panels have the same meanings. $A_p = A_d = 1$. (**C**) The influence of auto-temporal structure. Spike trains were generated by Model Auto (**S1 Text Section S5.1**). Theoretically, the small $\text{Corr}(\Delta w_{0a}, \Delta \bar{w}_a)$ causes only about 1% change of the variance, which can be easily overwhelmed by trial-to-trial variability. $A_p = A_d = 1$. (**D**) The influence of heterogeneity of rates. Spikes were generated by Model Long Tail (**S1 Text Section S5.1**). $A_p = 1.2$, $A_d = 1$. (**E**) The influence of heterogeneity of cross-correlations. Spike trains were generated by Model Cross-correlation (**S1 Text Section S5.1**). $q = 0.2$, $\varepsilon_0 = 10\text{ms}$. $A_p = A_d = 1$. Note that heterogeneity of cross-correlations can make $\text{Corr}(\Delta w_{0a}, \Delta \bar{w}_a)$ positive or negative, thus reduces or increases the efficacy variability in the coupled dendritic motif, depending on the details of the cross-correlation structure of spike patterns. In **B-E**, both the apical neuron in the dendritic motif and the apical neurons in the axonal motifs coupling with the dendritic motif connect with 200 neurons. Error bars represent s.e.m. in normal scale, and relative errors corresponding to s.e.m. in log scale. Simulation were run for 100s biological time, with 24 trials. Parameters for synaptic homeostasis: $w_{bound} = 0\text{nS}$, $\varepsilon = 0.001$ (see **Methods** in the main text eq.5-6 for the meanings of these parameters).

The mechanisms of how synchronous firing, heterogeneity of rates and heterogeneity of cross-correlations individually influence $\text{Corr}(\Delta w_{0a}, \Delta \bar{w}_a)$ have been already explained in the main text. Here we emphasize that the underlying mechanism of the first one is fundamentally different from those of the latter two. Heterogeneity of rates and heterogeneity of cross-correlations introduce correlation between the drift velocities of $\Delta w_{0a}$ and $\Delta \bar{w}_a$. As $\text{DrV} \propto t^2$, this correlation will dominate in the long term, but in the short term, it may be buried inside the diffusion noises of DiV, and this makes $\text{Corr}(\Delta w_{0a}, \Delta \bar{w}_a)$ gradually increase before it saturates (**S9 Fig**). The correlation caused by synchronous firing, however, is due to the correlation of the diffusion noises between $\Delta w_{0a}$ and $\Delta \bar{w}_a$, so $\text{Corr}(\Delta w_{0a}, \Delta \bar{w}_a)$ saturates almost at the beginning (**S9 Fig**). When several aspects of pattern structure coexist, $\Delta w_{0a}$ and $\Delta \bar{w}_a$ may be correlated separately by these two mechanisms shortly or long after the beginning, so the correlated component of $\Delta w_{0a}$ and $\Delta \bar{w}_a$ may be changed during a long run.

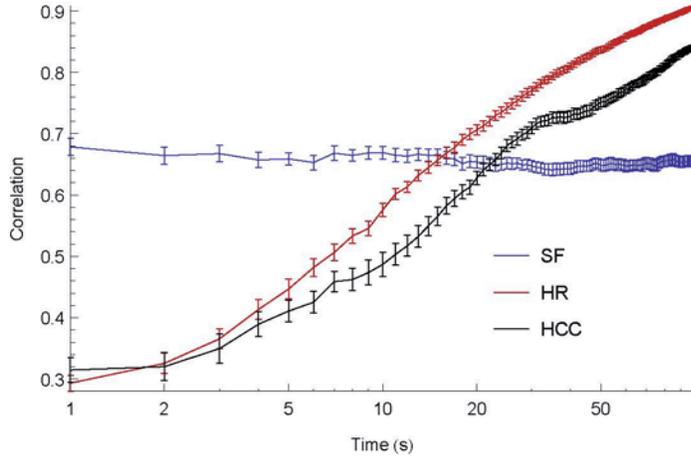

**S9 Fig. The evolution of $\mathrm{Corr}(\Delta w_{0a}, \Delta \bar{w}_a)$ when it is caused by synchronous firing (SF), heterogeneity of rates (HR) or heterogeneity of cross-correlations (HCC).** Note that both HR and HCC gradually increase $\mathrm{Corr}(\Delta w_{0a}, \Delta \bar{w}_a)$, but SF saturates $\mathrm{Corr}(\Delta w_{0a}, \Delta \bar{w}_a)$ almost instantaneously. For SF, spike trains were generated using Model Sync 3 (**S1 Text Section S5.1**), $A_p = A_d = 1$, $p = 1$, $\tau_{cross} = 2\mathrm{ms}$. For HR, spike trains were generated using Model Long Tail (**S1 Text Section S5.1**), $A_p = 1.2$, $A_d = 1$, $s = 0.4$. For HCC, spike trains were generated using Model Cross-correlation (**S1 Text Section S5.1**), $A_p = A_d = 1$, $q = 0.2$, $\Delta t = 20\mathrm{ms}$, $\varepsilon_0 = 10\mathrm{ms}$. Error bars represent s.e.m. The other parameters are the same as **S8 Fig**.

Now let's consider the case when synchronous firing, heterogeneity of rate and auto-temporal structure coexist and after a long run. When potentiation and depression are imbalanced, $\mathrm{Corr}(\Delta w_{0a}, \Delta \bar{w}_a)$ is mainly contributed by heterogeneity of rates after a long run; when potentiation and depression are balanced, $\mathrm{Corr}(\Delta w_{0a}, \Delta \bar{w}_a)$ is contributed by synchronous firing. Auto-temporal structure can influence DiV noises which are of $\mathcal{O}(t)$ order. In P-D balanced cases, $\mathrm{Corr}(\Delta w_{0a}, \Delta \bar{w}_a)$ decreases with $CV_{rescale}$, because $CV_{rescale}$ hardly influences P-D imbalance to change DrV, and its burstiness induces variety of spike numbers of non-apical neurons in a firing event, thereby inducing stronger diffusion noises to destroy the correlation (**S10A Fig**). The effect of $CV_{events}$, however, is more complicated because $CV_{events}$ itself can influence P-D imbalance, which contributes to $\mathrm{Corr}(\Delta w_{0a}, \Delta \bar{w}_a)$ through the heterogeneity of rates (**S10B Fig**).

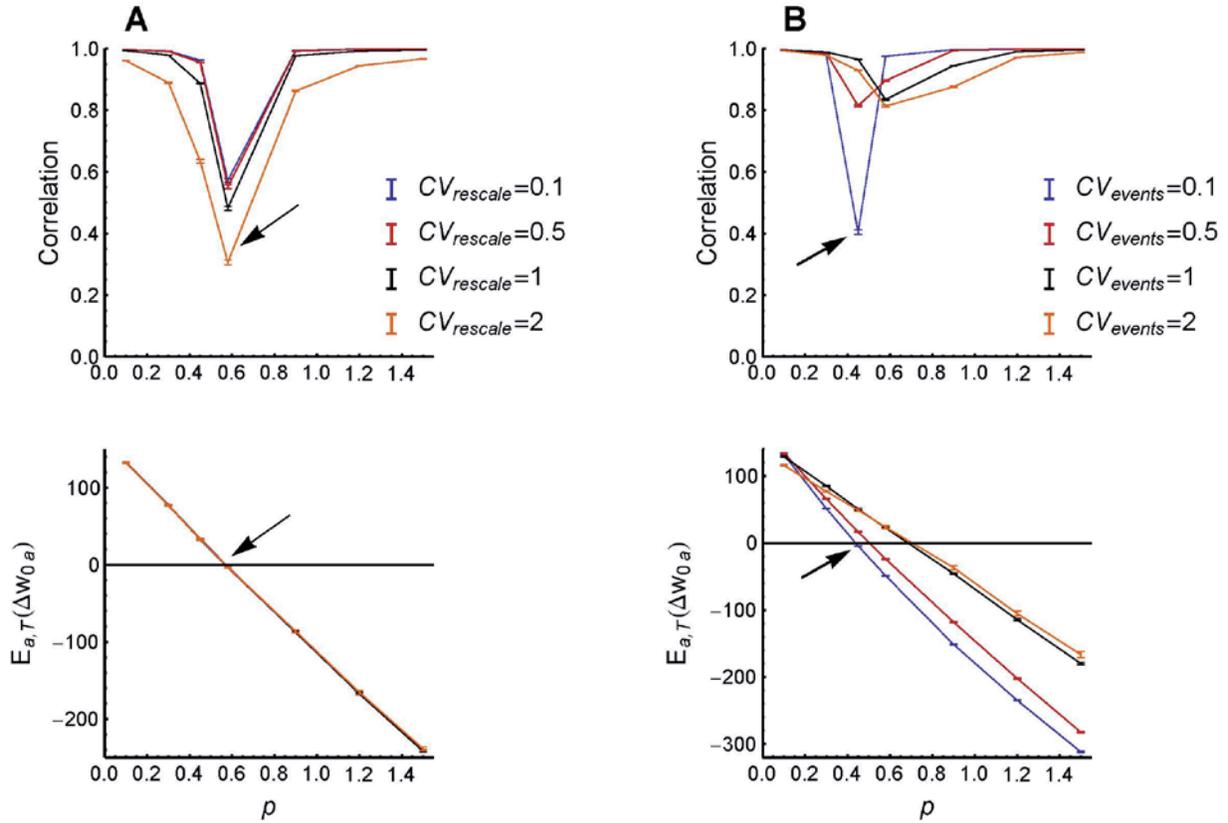

**S10 Fig. The influence of auto-temporal structure onto $\mathrm{Corr}(\Delta w_{0a}, \Delta \bar{w}_a)$ in the dendritic motif coupling with axonal motifs (S8A Fig) when both synchronous firing and heterogeneity of rates exist.** (**A**) Upper: $\mathrm{Corr}(\Delta w_{0a}, \Delta \bar{w}_a)$ as a function of $p$ for different $CV_{rescale}$, keeping $CV_{events} = 0.71$ and $s = 1$ (Model Long Tail & Model Sync-Auto in **S1 Text Section S5.1**). Lower: the corresponding mean efficacy changes within the coupled dendritic motif, representing P-D imbalance. Note that $CV_{rescale}$ hardly changes P-D imbalance, and thus hardly changes DrV with the existence of heterogeneity of rates. The two arrows indicate the $p$ value at which the mean efficacy change is almost zero (indicated by the horizontal black line), so that DrV=0 due to P-D balance. At this point, the correlation significantly decreases with $CV_{rescale}$ (upper panel) due to the DiV noises $CV_{rescale}$ induces. (**B**) The same as **A**, but different lines represent different $CV_{events}$ values, keeping $CV_{rescale} = 0.71$. The two arrows indicate the $p$ value at which the mean efficacy change is almost zero when $CV_{events} = 0.1$, and increases with $CV_{events}$ for $CV_{events}=0.1, 0.5, 1$. As $\mathrm{Corr}(\Delta w_{0a}, \Delta \bar{w}_a)$ increases with P-D imbalance in the long run, $\mathrm{Corr}(\Delta w_{0a}, \Delta \bar{w}_a)$ increases with $CV_{events}$ (for $CV_{events}=0.1, 0.5, 1$) at this $p$ value, which is reversed comparing to other $p$ values. In **A-B**, $A_p = 1.2$, $A_d = 1$, $\tau_{cross} = 2$ms. To increase precision, means and 95% confidence intervals (error bars) of correlations were calculated from 240 trials using Fisher z-transform. Error bars for mean efficacy changes represent s.e.m. calculated from 240 trials. Simulations were run for 20s biological times. The other parameters are the same as **S8B Fig**.

## Section S3: Efficacy Variability in LIF Network

*Key points of this section:*
1) *We used different spike shuffling methods to destroy different aspects of pattern structure in the recorded spike patterns generated by a LIF network, and observed how the efficacy variability changed when E-E links were evolved according to the original or shuffled spike patterns under STDP when dendritic and axonal homeostasis are imposed alone or both. We found that this change of the efficacy variability can be understood using our results from the motifs studies.*

The LIF network we used consisted of 2000 excitatory and 500 inhibitory conductance-based LIF neurons, and the links were randomly connected with probability 0.2. As we mentioned in the main text, during the simulation of the LIF network, we first recorded all the spikes of the excitatory population while keeping the synaptic efficacies unchanged; and then evolved the excitatory-to-excitatory (E-E) links according to the spike pattern originally recorded or shuffled by different methods, under the rules of STDP and synaptic homeostasis. We did this to investigate how different aspects of pattern structure influence the efficacy variability without worrying about the feedback of synaptic changes onto network dynamics. Details of the LIF model are presented in **Methods** in the main text.

We used different spike shuffling methods for asynchronous and synchronous states due to their sharp pattern difference. The spike shuffling methods are in the following:

*Rescaling Shuffle (RS)*:

This shuffling method aims to destroy the pattern structure of synchronous firing.
Spike times are first projected to the rescaled time through the accumulative function of firing rate (eq. S18)

$$\Lambda(t) = \int_0^t r(s) \mathrm{d}s,$$

then are projected back to the normal time using $\Lambda_0^{-1}(s)$, where $\Lambda_0(t)$ is the linear function connecting $(0,0)$ with $(T, \Lambda(T))$, with *T* being the duration of the spike train. Given a spike pattern, $\Lambda(t)$ is calculated by accumulating spike number at the times of spikes. So technically, this shuffling method is to first order all the *M* spikes in the pattern, then set the time of the *i*th spike at $iT/M$ (**Fig. 6** in the main text).

*Inter-neuron Shuffle (IS)*:

This shuffling method is used in asynchronous states, and it aims to destroy the auto-temporal structure in the spike train from each neuron.

The idea of this method is to randomly swap the spike times of different neurons (**Fig. 6** in the main text). Technically, this is realized by first randomly shuffling the firing order of the neurons during the spike pattern, and then assigning the spike times to the shuffled order.

*Translation Shuffle (TS)*:

This shuffling method is used in asynchronous states, and it aims to destroy heterogeneity of cross-correlations.

Each spike train is translationally moved by a random displacement. Periodic boundary condition is used to deal with the spikes which are moved out of the boundaries of time (**Fig. 6** in the main text).

*Whole-population Shuffle (WS)*:

This shuffling method is used in asynchronous states, and it aims to destroy heterogeneity of firing rates.

Each spike in the spike pattern is assigned to a randomly selected neuron (**Fig. 6** in the main text).

*Whole-population Shuffle within Event (WSWE)*:

This shuffling method is used in synchronous states, it destroys both heterogeneity of rates and heterogeneity of cross-correlations, while keeping the spike number in each firing event unchanged.

The idea of this method is to swap the spike sequences of pairs of randomly selected neurons in the same firing event (**Fig. 6** in the main text). Technically, this is realized by first randomly shuffling neuronal indexes for each firing event, and then assigning the spike sequences within each firing event to the shuffled neuronal indexes.

Numerically, a firing event was defined like this: we first calculated temporal rates of the excitatory population in bins of 0.1ms, then filtered these binned rates using Gaussian window of $\sigma_{window} = 2\text{ms}$; a firing event was defined as sequential bins in which the filtered rates were above a small threshold 0.0001.

*Inter-neuron Shuffle Within Event (ISWE)*:

This shuffling method is used in synchronous states, and it aims to destroy the auto-temporal structure in the rescaled time, while keeping the spike number in each firing event unchanged.

The idea of this method is to randomly swap spike times of different neurons within the same firing event (**Fig. 6** in the main text). Technically, this is realized by first randomly shuffling the firing order of neurons during each firing event, and then assigning the spike times during each firing event to the shuffled order.

*Event Time Shuffle (ETS)*:

This shuffling method is used in synchronous states, and it aims to destroy the auto-temporal structure of the occurrence of firing events.

The idea of this method is that all the spikes within the same firing events are translationally moved by a random displacement, while keeping the order of firing events unchanged (**Fig. 6** in the main text). Technically, this is realized by first randomly selecting $N_{event}$ points in the duration $[0,T]$, then set the mean spike time of the *i*th firing event at the *i*th point. Here $N_{event}$ is the number of the firing events, and $T$ is the duration of the spike train.

*Notes on the order to implement the shuffling methods*:

As some shuffling methods may destroy more than one aspects of pattern structures, the order of these methods to be implemented must be carefully designed so that when one aspects of pattern structure is destroyed the others remain largely intact. The pattern structures destroyed by each shuffling method are listed in **S1 Table**, and their order to be implemented is shown in **Fig. 6** in the main text.

| | Names of Spike Shuffling Methods | Aspects of pattern structure to destroy | | | | Effects on efficacy variability (**S11A-C Fig**) | Reason |
|---|---|---|---|---|---|---|---|
| | | SF | HCC | AT | HR | | |
| Asynchronous states ($\tau_{d,I} \leq 6\text{ms}$) | RS | × | | | | Slightly decrease, compared with original patterns | Before RS, the weak fluctuation of the population rates in asynchronous states increases the efficacy variability through spike gathering (**S12A Fig**). |
| | RS+TS | × | × | | | Decrease, compared with RS shuffled patterns | The heterogeneity of cross-correlations is reduced by TS. (**S12E Fig**). |
| | RS+TS+IS | × | × | × | | Decrease, compared with RS+TS shuffled patterns | $CV_{rescale}$ is decreased by IS (**S12C Fig**). |
| | RS+TS+WS | × | × | × | × | No obvious change, compared with RS+TS+IS shuffled patterns | The P-D *balance* in asynchronous states caused by $A_p = A_d$ in our model (**S12H Fig**) disables heterogeneity of rates to change the efficacy variability through DrV. |
| Synchronous states ($\tau_{d,I} \geq 7\text{ms}$) | WSWE | | × | | × | Strongly decrease, compared with original patterns | 1) Before WSWE, synchronous firing imbalances potentiation and depression (**S12H Fig**), which increases the efficacy variability through heterogeneity of rates (**S12G Fig**). 2) WSWE reduces the heterogeneity of cross-correlations (**S12F Fig**). |

| | | | | | | |
|---|---|---|---|---|---|---|
| | WSWE+ISWE | | × | × ($CV_{rescale}$) | × | Increase, compared with WSWE shuffled patterns | $CV_{rescale}$ is increased by ISWE (**S12C Fig**). |
| | WSWE+ISWE+RS | × | × | × ($CV_{rescale}$) | × | Decrease, compared with WSWE+ISWE shuffled patterns | Synchronous firing is strong before RS, and synapse correlating is impossible to reduce the efficacy variability before RS because of $\tau_{cross} > \tau_{delay} = 1\text{ms}$ (**S12B Fig**), therefore the efficacy variability before RS is large because of spike gathering and synapse splitting. |
| | WSWE+ETS | | × | × ($CV_{events}$) | × | Increase, compared with WSWE shuffled patterns | $CV_{events}$ is increased by ETS (**S12D Fig**). |

**S1 Table. Spike shuffling methods, the aspects of pattern structure that they destroy, and their influences on efficacy variability.** Abbreviations for pattern structures: SF, synchronous firing; HCC, heterogeneity of cross-correlations; AT, auto-temporal structure; HR, heterogeneity of rates. Spike shuffling methods are explained in **Fig. 6** in the main text and **S1 Text Section S3**.

The change of efficacy variance caused by different spike shuffling methods are qualitatively the same with the existence of dendritic homeostasis, axonal homeostasis alone or both (**S11A-C Fig**). The change of pattern statistics caused by different shuffling methods are shown in **S12 Fig** with respect to $\tau_{d,I}$. We compare the change of the efficacy variance with the change of these statistics caused by different shuffling methods in **S1 Table**, and find that they are consistent with our results from the motif studies. Because we kept the firing rate of the excitatory population to be constant for different $\tau_{d,I}$ (see **Methods** in the main text), we can also compare how the efficacy variance and these statistics change with $\tau_{d,I}$ without worrying about the influence caused by the change of firing rates.

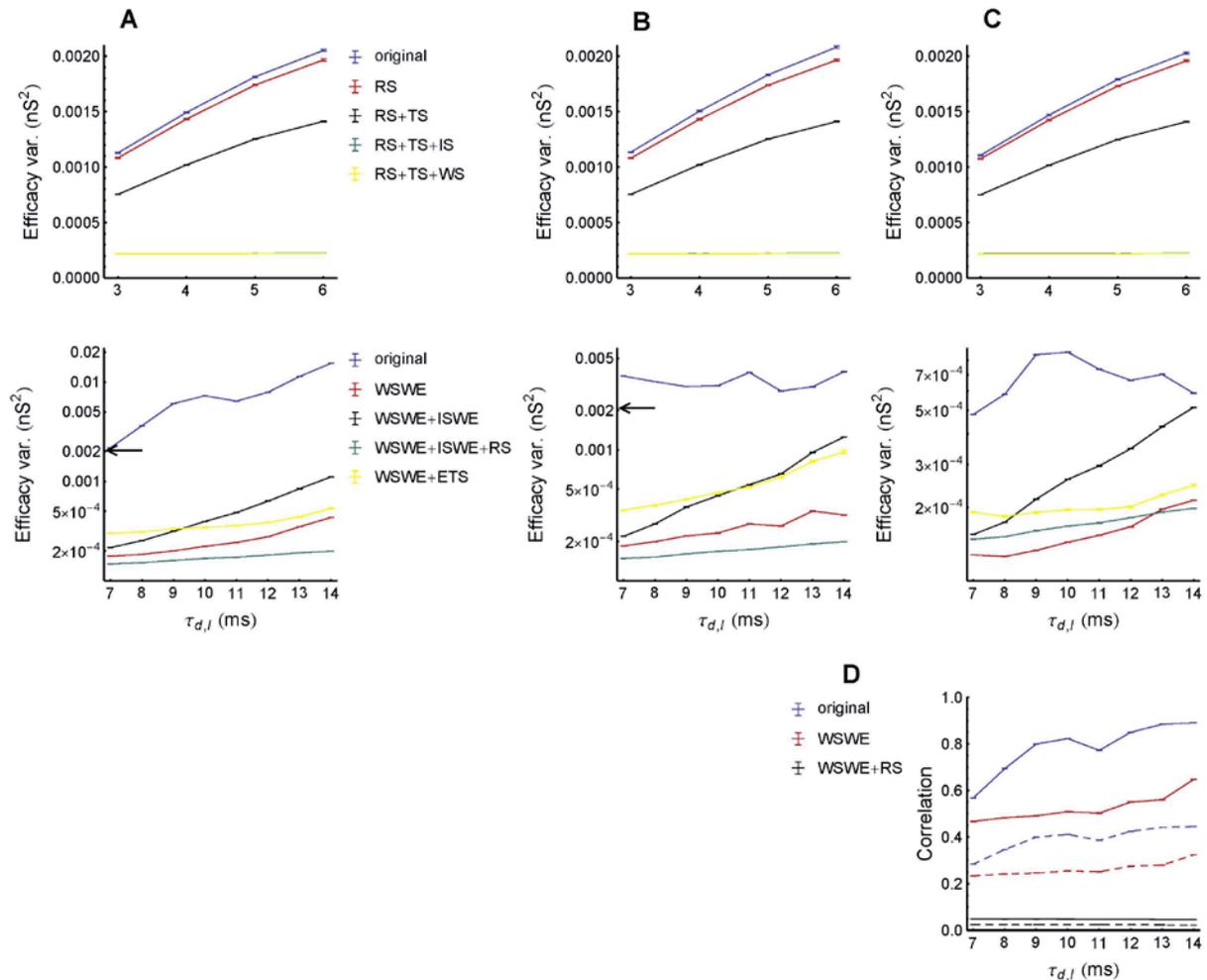

**S11 Fig. The efficacy variance of all the E-E links in the LIF network when the activity of the excitatory population is according to the spike patterns original or shuffled by different methods under STDP and synaptic homeostasis.** (**A**) Efficacy variance as a function of $\tau_{d,I}$ when the spike patterns were original or shuffled by different methods (see **S1 Text Section S3** and **Fig. 6** in the main text for details) when only dendritic homeostasis was imposed. Upper: asynchronous states ($3\text{ms} \leq \tau_{d,I} \leq 6\text{ms}$); lower: synchronous states ($7\text{ms} \leq \tau_{d,I} \leq 14\text{ms}$). Note that in asynchronous states (upper panel), the efficacy variances caused by RS+TS+IS and RS+TS+WS almost overlap. In synchronous states (lower panel), the efficacy variances span a great range, so we use log scale to better show their changes. To help readers compare the efficacy variances just before and after the asynchrony-to-synchrony transition, we use an arrow to indicate the efficacy variance caused by the original spike pattern at $\tau_{d,I} = 6\text{ms}$ in the lower panel. As we used different shuffling methods for asynchronous and synchronous states, the changes of the efficacy variances caused by the shuffled spike patterns are not comparable before and after the transition. (**B**) The same as **A**, except that only axonal homeostasis was imposed. (**C**) The same as **A**, except that both dendritic and axonal homeostasis were imposed. Note the sharp decrease of the efficacy variance

when the spike patterns transit from asynchronous to synchronous states in this case. As the efficacy variances in synchronous states are much smaller than those in asynchronous states, we do not mark an arrow in the lower panel. (**D**) $\text{Corr}(\Delta w_{ab}, \Delta \overline{w}_b)$ when the LIF network operates in synchronous states. Dashed lines represent $\frac{1}{2}\sqrt{\text{Var}_b(\Delta \overline{w}_b)/\text{Var}_b(\Delta w_{ab})}$ (**S1 Text** eq. S22). In asynchronous states, $\text{Corr}(\Delta w_{ab}, \Delta \overline{w}_b)$ is close to zero because of the P-D balance caused by $A_p = A_d$ in our model (not shown). In **A-D**, simulations were run for 20s biological time with 24 trials, and STDP and synaptic homeostasis were imposed after the first 1s of transient period. Error bars represent s.e.m.

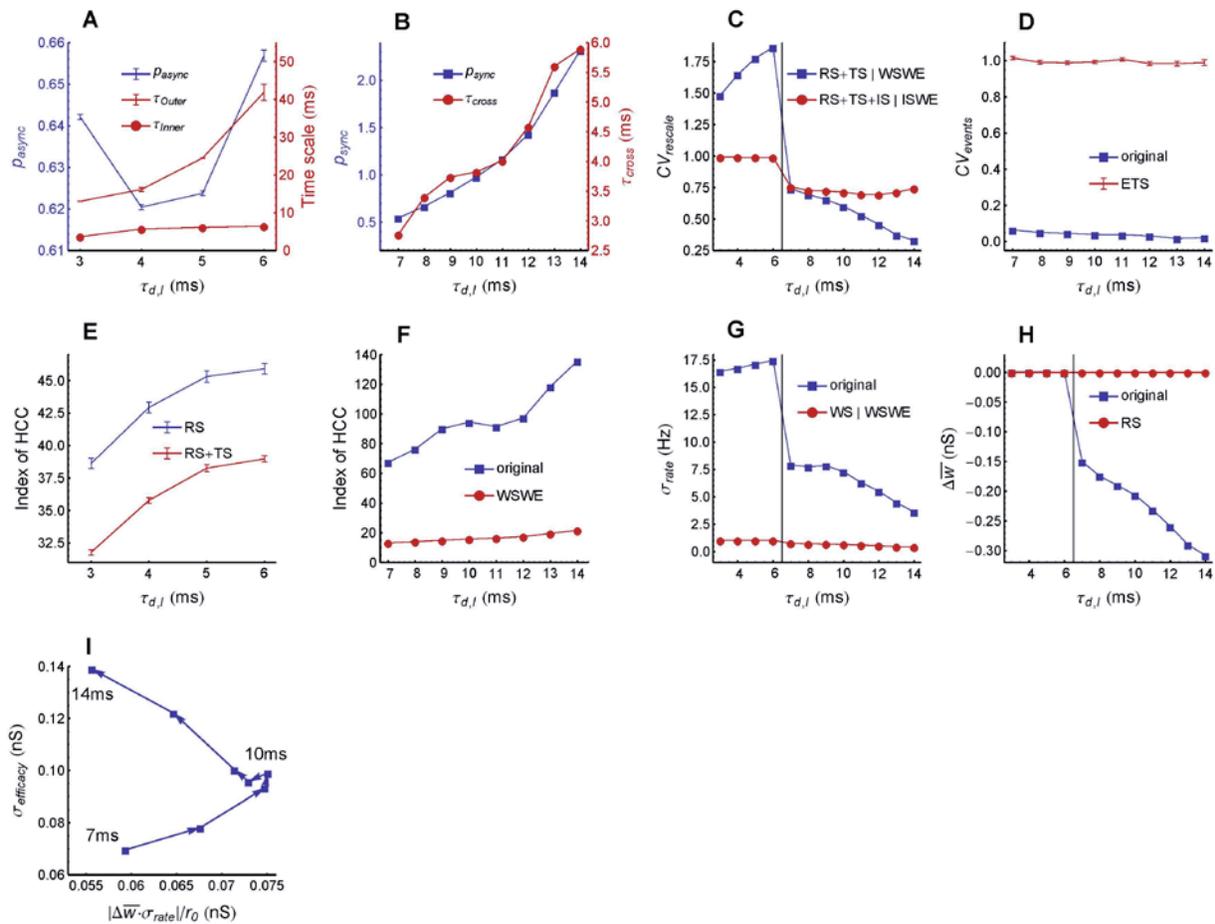

**S12 Fig. Statistical analysis of the spike patterns of the excitatory population of the LIF network, original or shuffled by different methods.** (**A**) The amplitudes $p_{async}$ of the fluctuation of the population rates and the two time scales ($\tau_{outer}$, $\tau_{inner}$) of the oscillating decaying auto-correlation of the population rates in the original patterns of asynchronous states (see methods in **S1 Text Section S5.2**). The fluctuation of population rates in asynchronous states can be regarded as weak firing events, which can increase the efficacy variability by spike gathering (**Fig. 4** in the main text). (**B**) The

strengths $p_{sync}$ and durations $\tau_{cross}$ of the firing events in the original patterns of synchronous states (see methods in **S1 Text Section S5.2**). Note that $\tau_{cross} > \tau_{delay} = 1\text{ms}$, therefore synapse correlating (**Fig. 4** in the main text) can hardly take its effect to reduce the efficacy variance. (**C**) $CV_{rescale}$ in the patterns shuffled by different methods, indicating the auto-temporal structure in the rescaled time (see methods in **S1 Text Section S5.2**). The black vertical line indicates the transition from asynchronous states to synchronous states. Note that we used different shuffling methods for these two states. (**D**) $CV_{events}$ in synchronous states in the original and ETS shuffled patterns, indicating the temporal structure of the occurrence of firing events. (**E**) The index of heterogeneity of cross-correlations (HCC) as a function of $\tau_{d,I}$ (see methods in **S1 Text Section S5.2**) in asynchronous states. The difference between the indexes for RS shuffled and RS+TS shuffled patterns represents the above-chance heterogeneity of cross-correlations in the original spike patterns. (**F**) The index of HCC in synchronous states in the original and WSWE shuffled patterns. (**G**) The standard deviation of rate $\sigma_{rate}$ in the original and shuffled patterns, indicating the heterogeneity of rates. (**H**) The mean efficacy change caused by STDP $\Delta\bar{w}$ in the original and RS shuffled patterns, indicating P-D imbalance. Note that $\Delta\bar{w}$ is almost zero when $\tau_{d,I} \leq 6\text{ms}$ and after RS because of the P-D *balance* in asynchronous states caused by $A_p = A_d$ in our model. (**I**) The vertical coordinate $\sigma_{efficacy}$ means the s.d. of the efficacy changes caused by STDP, and the horizontal coordinates $|\Delta\bar{w} \cdot \sigma_{rate}|/r_0$ quantifies the s.d. of the efficacy changes contributed by the interaction of P-D imbalance and heterogeneity of rates; both of the two coordinates are for the original patterns in synchronous states. Arrows indicate the increasing of $\tau_{d,I}$ from 7ms to 14ms, taking integer values. When $\tau_{d,I} \leq 10\text{ms}$, $|\Delta\bar{w} \cdot \sigma_{rate}|/r_0$ contributes a significant part of $\sigma_{efficacy}$, but this contribution gets weaker when $\tau_{d,I} \geq 11\text{ms}$. As heterogeneity of rates increases the efficacy variability through DrV, the factor can overwhelm its contribution in the long run should also be of DrV nature, which can only be heterogeneity of cross-correlations (see **S1 Text Section S3** for more explanations). In **A-I**, error bars represent s.e.m., which may not be seen when the error bars are smaller than the symbol sizes. Simulations were run for 20s biological time with 24 trials, and the first 1s of spike trains were regarded as transient period, and excluded from analysis.

**S1 Table** is straightforward to understand, except that WSWE destroys both heterogeneity of rates and heterogeneity of cross-correlations in synchronous states, so it is unclear how each of them contributes to the reduction of the efficacy variability caused by WSWE (**S11A-C Fig, lower panels**). To test their individual contributions, we plotted $\sigma_{efficacy}$ versus $|\Delta\bar{w} \cdot \sigma_{rate}|/r_0$ at different $\tau_{d,I}$s for the original patterns (**S12I Fig**), with $\sigma_{efficacy}$ being the standard deviation of the efficacy changes only caused by STDP (without counting synaptic homeostasis), $\Delta\bar{w}$ being the mean efficacy changes caused by STDP over all the links of the network, and $r_0 = 20\text{Hz}$ and

$\sigma_{rate}$ respectively being the mean and standard deviation of the firing rates of the excitatory neurons. Because the efficacy change of a link $a \to b$ is proportional to the firing rate of neuron $a$ and neuron $b$, and the proportion coefficient is determined by P-D imbalance, we used $|\Delta \overline{w} \cdot \sigma_{rate}|/r_0$ to estimate the efficacy variability caused by heterogeneity of rates and P-D imbalance only through STDP (without counting synaptic homeostasis). We found that when $\tau_{d,I} \leq 10\text{ms}$, $|\Delta \overline{w} \cdot \sigma_{rate}|/r_0$ contributed a significant part of $\sigma_{efficacy}$, but this contribution got weaker when $\tau_{d,I} \geq 11\text{ms}$ (**S12I Fig**). As heterogeneity of rates increases the efficacy variability through DrV, the factor can overwhelm its contribution in the long run should also be of DrV nature, which can only be heterogeneity of cross-correlations. This deduction above requires that DrV should dominate in $\sigma_{efficacy}$ at the end of the simulations. As WSWE destroys both heterogeneity of rates and heterogeneity of cross-correlations, which are the two sources of DrV, we can estimate the contribution of DiV by calculating $\sigma_{efficacy}$ for the WSWE-shuffled patterns. We found that WSWE shrank $\sigma_{efficacy}$ by 80%~85% when $\tau_{d,I}$ went from 7ms to 14ms (data not shown), which suggests that DrV indeed dominates in the $\sigma_{efficacy}$ for the original patterns.

When the dendritic or axonal homeostasis was imposed alone, the transition from asynchronous state to synchronous state did not induce sharp change of efficacy variance, and the efficacy variance did not decrease with $\tau_{d,I}$ in synchronous states (**S11AB Fig**). However, when these two synaptic homeostasis coexisted, the efficacy variance was strongly reduced at the asynchrony-to-synchrony transition point, and also decreased when $\tau_{d,I}$ was large enough and the network went into synchronously bursting states (**S11C Fig**). This means that the coupling of dendritic and axonal homeostasis is the key reason for the reduction of the efficacy variability in synchronous states. To understand this reduction, we recorded $\text{Corr}(\Delta w_{ab}, \Delta \overline{w}_b)$ both in the original spike pattern and in the spike patterns treated by different shuffling methods (**S11D Fig**), with $\Delta w_{ab}$ being the efficacy change on link $b \to a$ only caused by STDP (without counting synaptic homeostasis), and $\Delta \overline{w}_b$ being the mean efficacy change (also only caused by STDP) of the axonal motif that the link $b \to a$ belonged to. We found that $\text{Corr}(\Delta w_{ab}, \Delta \overline{w}_b)$ was far from zero and continuously increased with $\tau_{d,I}$ in the synchronous states (**S11D Fig**), which explains both the sharp reduction of the efficacy variability at the asynchrony-to-synchrony transition point, and why the efficacy variability decreases with $\tau_{d,I}$ when the network goes into synchronously bursting states when dendritic and axonal homeostasis coexist (**S11C Fig, lower panel**).

We also found that WSWE could significantly reduce $\text{Corr}(\Delta w_{ab}, \Delta \overline{w}_b)$, and a further RS could significantly reduce it again (**S11D Fig**). The effect of WSWE represents the roles played by heterogeneity of rates and heterogeneity of cross-correlations, and effect of RS represents that of synchronous firing. The reduction of correlation after WSWE is because that heterogeneity of rates and heterogeneity of cross-correlations correlate $\Delta w_{ab}$ and $\Delta \overline{w}_b$ through the correlation of the drift velocities (DrV manner), so that the variance along the correlated component increases with time as $\mathcal{O}(t^2)$ order; but synchronous firing correlates them through the correlation of diffusion (DiV manner), so that the variance along the correlated component *only* increases as

$\mathcal{O}(t)$ order, which is of the same order as diffusion noises. After a further RS, $\text{Corr}(\Delta w_{ab}, \Delta \overline{w}_b)$ was reduced to almost zero (**S11D Fig**), which is because of the absence of synchronous firing in the spike pattern.

An important point to note is the P-D *balance* in asynchronous states (also including the spike patterns after RS) induced by $A_p = A_d$ in our model (**S12H Fig**). Due to this P-D balance, heterogeneity of rates cannot increase the efficacy variability nor increase $\text{Corr}(\Delta w_{ab}, \Delta \overline{w}_b)$ in asynchronous states. Therefore, WS (which destroys the heterogeneity of rates) in asynchronous states ($\tau_{d,I} \leq 6\text{ms}$) cannot reduce the efficacy variability (note that the line for RS+TS+IS almost overlaps with the line for RS+TS+WS in the upper panels of **S11A-C Fig**), and $\text{Corr}(\Delta w_{ab}, \Delta \overline{w}_b)$ in asynchronous states is so weak that the coupling of dendritic and axonal homeostasis hardly reduces the efficacy variability (compare the upper panels of **S11AB Fig** with the upper panel of **S11C Fig**). If $A_p \neq A_d$, there will be P-D imbalance in asynchronous states, so that on the one hand, WS in asynchronous states will be able to significantly reduce the efficacy variance by destroying the heterogeneity of rates, and on the other hand, $\Delta w_{ab}$ and $\Delta \overline{w}_b$ will also be correlated in asynchronous states through the heterogeneity of rates, which makes the coupling of dendritic-axonal homeostasis significantly reduce the efficacy variability in asynchronous states; these are indeed what we found in our simulations (data not shown).

To help readers better understand the dynamics of our model, we will explain some phenomena observed in **S11 Fig** and **S12 Fig** in the Miscellaneous (**Section S6.4**).

## Section S4: Biological Implications

### Section S4.1: Maintenance and Encoding of Connection Patterns

*Key points of this subsection:*
1) *A connection pattern was designed so that the dynamics of the LIF network studied in the previous section remained qualitatively unchanged after encoding the connection pattern.*
2) *The capability of the LIF network for faithfully encoding and long-termly maintaining the connection pattern is inversely correlated with the efficacy variability.*

We used a similar LIF network as the previous section to examine the influence of spike pattern structures on the performance of the maintenance and encoding of connection patterns of neuronal networks. Key results have been pointed out in the main text, and model details have been presented in **Methods** in the main text.

### Section S4.2: Developmental Functions of Retinal Waves

*Key points of this subsection:*
1) *A two-layered feedforward network model was built to understand the developmental function of retinal waves (**S13A Fig**). The first layer contained two groups, whose intra-*

*group and inter-group synchrony was controlled by spike-generating models; the second layer was a LIF neuron.*
2) *The neuron in the second layer initially received equally from the two groups of the first layer, but under the competition caused by STDP and synaptic homeostasis, it might eventually respond to a single group. The difference of the intra- and inter-group efficacy variability controls the initial separation of these two groups before the LIF neuron is reliably more responsive to one group than the other (causality). The larger the difference, the larger the initial separation, and the sooner this causality will be established.*

To understand the developmental functions of retinal waves, we built up a two-layered feedforward network model (**S13A Fig**). In this model, the first layer was divided into two groups, representing two local RGC patches. Their activities were determined by a spike generating model, which explicitly controlled the probability of a neuron to fire during a firing event $p_{intra}$, and the portion $p_{inter}$ of inter-group events within all the firing events happening in one group (**S13B Fig**). $p_{intra}$ represents the synchrony within the patch of RGCs in the same eye that sharing similar receptive field (local RGCs), and $p_{inter}$ represents the synchrony among patches with different receptive fields or in different eyes. Here, we used a single spike to represent the bursting activity of a RGC during a retinal wave. We also jittered the spikes in a firing event by $[-\tau_{cross}/2, \tau_{cross}/2]$ to model the slight difference of the bursting times of local RGCs caused by, say, propagation of retinal waves. The second layer was a single LIF neuron, modeling a downstream neuron in SC or dLGN.

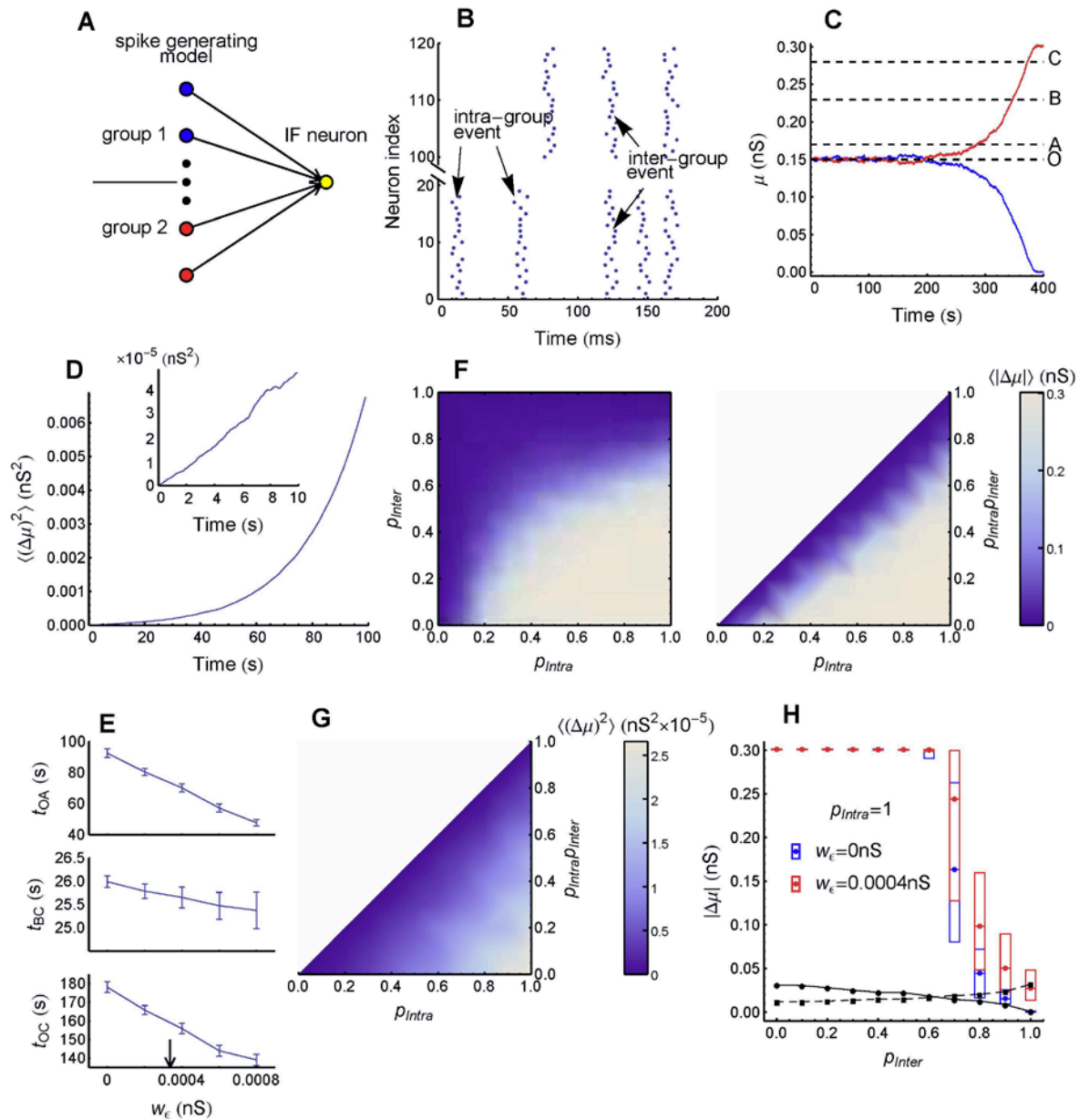

**S13 Fig. Initial inter-patch diffusion promotes inter-patch separation in the competition between local RGC patches caused by retinal waves.** (**A**) Network architecture. Blue and red dots in the first layer are two groups of neurons whose activities are controlled by a spike generating model (**S1 Text Section S4.2**, **S5.3**). The yellow dot in the second layer is a LIF neuron. (**B**) The model-generated spike pattern of the first layer. (**C**) One trial of evolution of the mean synaptic efficacies coming from the two groups, under STDP and dendritic homeostasis, when $p_{intra} = 0.8$, $p_{inter} = 0.2$, $\tau_{cross} = 2\text{ms}$. Note that at the beginning the mean efficacies from these two groups crossed over each other several times, suggesting the diffusion-driven nature of their separation. (**D**) Trial average of $(\Delta\mu)^2$ grows linearly with time at the beginning (inset),

but grows supra-linearly afterwards. This suggests that the inter-group separation is caused by diffusion at the beginning, but is gradually influenced by causality as the separation grows. (**E**) $t_{OA}$ represents the time that the mean efficacy of the eventually stronger group takes to grow from point $O$ to point $A$ (marked in **C**), $t_{BC}$ and $t_{OC}$ have similar meanings. Upper: during $O \to A$, the inter-group separation is mainly induced by diffusion, so that $t_{OA}$ decreases with the artificially added inter-group diffusion strength $w_\varepsilon$ (**S1 Text Section S4.2**). Middle: during $B \to C$, the separation is mainly induced by causality, so that $t_{BC}$ does not significantly change with $w_\varepsilon$. Lower: the total time $t_{OC}$ decreases with $w_\varepsilon$. $p_{intra} = 0.8$, $p_{inter} = 0.2$, $\tau_{cross} = 2\mathrm{ms}$. The arrow indicates the $w_\varepsilon$ value which increases $\langle(\Delta\mu)^2\rangle$ at early times (here, 4s) by the same amount as the $\langle(\Delta\mu)^2\rangle$ value itself when $w_\varepsilon = 0$, which gives readers the sense how strong noises we added through $w_\varepsilon$. Error bars represent s.e.m. (**F**) Left: The difference $|\Delta\mu|$ of the mean synaptic efficacies coming from the two groups after evolving for 400s biological time, with $p_{intra}$ and $p_{inter}$ taking different values and $\tau_{cross} = 2\mathrm{ms}$. Right: the same as left, but with $|\Delta\mu|$ being a function of the synchrony between two neurons in the same group $p_{intra}$ and the synchrony between two neurons in different groups $p_{intra}p_{inter}$. (**G**) Initial inter-group diffusion (represented by $\langle(\Delta\mu)^2\rangle$ at early time, here, 4s) as a function of $p_{intra}$ and $p_{intra}p_{inter}$. Comparing to **F**, we see that the large initial inter-group diffusion contributes to the large group separation when $p_{intra}$ is large and $p_{inter}$ is small. (**H**) The distribution of $|\Delta\mu|$ after 400s at different $p_{inter}$ values while keeping $p_{intra} = 1$. Dots represent median values, and error bars represent quartiles. Note the wide distribution of $|\Delta\mu|$ when $p_{inter}$ takes moderate values. Without causality-driving force, s.d. of $|\Delta\mu|$ only caused by diffusion are marked by the circles along the solid line, which are much smaller than the widths of the observed distributions of $|\Delta\mu|$. After adding $w_\varepsilon$, mean values of $|\Delta\mu|$ increase in the whole range of parameter. The mean increases of $|\Delta\mu|$ only caused by diffusion (if the causality-driving force is absent) are marked by the squares along the dashed line, which are also much smaller than the observed increases. This panel shows that the interaction of diffusion and causality-driving force increases trial-to-trial variability and promotes inter-group separations when the separations are not completed. In **D-H**, simulations were run for 240 trials.

Initially, all the synapses had equal strength, so the LIF neuron in the second layer responded to both groups equally. Then the synapses were evolved according to STDP and dendritic homeostasis when intrinsic homeostasis was also implemented on the LIF neuron to conserve its firing rates (see **Section S5.3** for simulation details). After the simulation began, the synaptic strengths coming from the two groups started to separate, and we tried to understand the competition induced by retinal waves by investigating the properties of this inter-group separation.

At the beginning of the simulation, inter-group separation is largely driven by diffusion (**S13C Fig**). For two synapses $x_1$ and $x_2$ coming from the same group, the expectation of the square of their difference at time $t$ should be $\langle (x_1 - x_2)^2 \rangle = \sigma_{intra}^2 t$, with $\sigma_{intra}^2$ quantifying the intra-group efficacy variability. Similarly, for two synapses $y_1$ and $y_2$ coming from different groups, $\langle (y_1 - y_2)^2 \rangle = \sigma_{inter}^2 t$, with $\sigma_{inter}^2$ quantifying the inter-group efficacy variability. Then the difference $\Delta\mu$ between the mean values of the synapses coming from the two groups should be

$$\langle (\Delta\mu)^2 \rangle = \langle (y_1 - y_2)^2 \rangle - \langle (x_1 - x_2)^2 \rangle = (\sigma_{inter}^2 - \sigma_{intra}^2) t \tag{S24}$$

which means that the initial inter-group separation depends on the difference between the inter- and intra-group efficacy variability.

In our model, the synchrony between two neurons in the same group is $p_{intra}$, and the synchrony between two neurons in different groups is $p_{intra} p_{inter}$. As the activity of the LIF neuron is driven by the firing events of the first layer, and the jitter time window $\tau_{cross}$ is also short, synchrony controls the correlation between synaptic updatings, which is inversely correlated with the efficacy variability (the mechanism of synapse correlating, see **Fig. 4** in the main text). Initially, the synapses from the two groups have the same strength, so that the LIF neuron responds to both groups equally. In this case, the inter-group separation can hardly be driven by the stronger causality of the LIF neuron to one group than to the other one, so the separation is caused by diffusion, so $\langle (\Delta\mu)^2 \rangle \propto t$; as the simulation goes on, the synaptic strengths from the two groups gradually separate apart, so that the LIF neuron becomes more responsive to one group than the other, and this causality makes $\langle (\Delta\mu)^2 \rangle$ grow with $t$ supra-linearly (**S13D Fig**).

If the initial diffusion, i.e. $\langle (\Delta\mu)^2 \rangle$, is large, the causality will take its effect soon; if it is weak, the causality will participate at later time, which hinders the separation process (**Fig. 7B** in the main text). To check this effect, we artificially added a small efficacy value $\delta w$ to all the synapses coming from the first group, and added $-\delta w$ to all the synapses coming from the second group every 50ms during our simulation, with $\delta w$ being drawn uniformly from the interval $[-w_\varepsilon, w_\varepsilon]$. In this way, we could increase the inter-patch diffusion, while the causality was intact after a long-term average. Consistently with our argument (**Fig. 7B** in the main text), we found that increasing $w_\varepsilon$ significantly promotes separation in the initial diffusion-dominating range, and hardly has effect in the later causality-dominating range; and the total time needed for this separation also reduces with $w_\varepsilon$ (**S13E Fig**).

After simulation for 400s biological time, the separation of the two groups depends on the values of $p_{intra}$ and $p_{inter}$. We found that good separation in our model was realized in the large $p_{intra}$ and small $p_{inter}$ range (**S13F Fig**). This is consistent with the dynamic pattern and developmental function of retinal waves: 1) retinal waves induce strong synchrony within a local RGC patch of the same eye, and weak synchrony between patches with different receptive fields or in different eyes; 2) RGCs with different receptive fields or in different eyes target to different parts of SC and dLGN (the formation of retinotopic map and eye-specific segregation). Notably, the initial separation $\langle(\Delta\mu)^2\rangle$ also got its largest value in the large $p_{intra}$ and small $p_{inter}$ range (**S13G Fig**), which suggests that large initial diffusion positively contributes to the inter-patch separation in the real physiological process.

When the inter-patch synchrony is strong, as is the case when two patches in the same eye have nearby receptive fields, the initial inter-patch diffusion is weak, and the causality is also weak (because when the downstream neuron responds to one patch, it also has a high probability to respond to the other one), so that the inter-patch separation may not complete at the end of the critical period of development. We found that in this case, the inter-group separation $|\Delta\mu|$ in the end of the simulation exhibited strong trial-to-trial variability (**S13H Fig**). This variability is due to the interaction between the inter-patch diffusion and the causality: the stochastic nature of diffusion can induce different initial $|\Delta\mu|$ values in different trials, and the two groups can be pushed apart stronger by the causality in those trials with larger initial $|\Delta\mu|$, thereby further increasing the difference of $|\Delta\mu|$ from the trials with smaller initial $|\Delta\mu|$. In the situations when the separation does not complete at the end of the simulation, we also added artificial diffusion and found that the inter-group separation got strongly promoted (**S13H Fig**), which validates our previous argument that the initial inter-group diffusion can significantly promote the inter-group separation under its interaction with the causality (**Fig. 7B** in the main text, **S13E Fig**).

## Section S5: Supplementary Methods

### Section S5.1: Spike Generating Models

*Model Sync 1*:

This model generates synchronous firing patterns with *spike uniqueness*, which means that a neuron can fire no more than one spike in a firing event.

Suppose the probability of a neuron to fire in a firing event is $p$, then the occurrence of firing events in this model is a Poisson process of rate $r_0/p$, with $r_0 = 20\text{Hz}$ being the rate of each neuron. Suppose the middle time of the *i*th firing event is at $t_i$, then the spike times of neurons within this firing event are randomly chosen within $[t_i - \tau_{cross}/2, \ t_i + \tau_{cross}/2]$, with $\tau_{cross}$ being the length of the time window of firing events. Each neuron can fire within a firing event no more than once, so $0 < p \leq 1$.

*Model Sync 2*:

This model generates synchronous firing patterns with spike uniqueness, but without *synapse splitting* (see **Fig. 4** in the main text).

In a dendritic motif, spikes of the non-apical neurons are generated in the same way as Model Sync 1; but if the apical neuron fires in a firing event whose middle time is at $t_i$, then its spike must be at $t_i + \tau_{cross}/2 + \tau_{delay}$, with $\tau_{delay}$ being the axonal time delay. In this way, all the spikes of the non-apical neurons always arrive at the apical neuron *before* the firing of the apical neuron itself, thereby removing synapse splitting.

*Model Sync 3*:

This model generates synchronous firing patterns without spike uniqueness. In these patterns, for the non-apical neurons which fire in a firing event, their spike numbers in the firing event can be different, so that *synapse correlating* is removed because of the dissimilarity of STDP updatings among the synapses (see **Fig. 4** in the main text).

Spikes trains are inhomogeneous Poisson processes with time average rate $\langle r_0(t) \rangle = 20\text{Hz}$. The time-dependent rate $r_0(t)$ is constructed using the occurrence of firing events. Specifically, $r_0(t)$ is the summation of the square-shaped functions of width $\tau_{cross}$ and area $p$ contributed by all the firing events. And similar to Model Sync 1, the occurrence of firing events is also a Poisson process of rate $\langle r_0(t) \rangle / p$ in this model. Note that in this model $p$ can be larger than 1.

*Model Sync 4*:

This model generates synchronous firing patterns with exponentially decaying cross-correlation, based on a model which can generate spike trains with near-maximal entropy.

The spike trains of rate $r_0 = 20\text{Hz}$ and synchrony strength $p$ are first generated by a dichotomized Gaussian approach [3], which was shown to have near-maximal entropy [4]. Then the spikes are jittered according to a distribution $f(t)$ which should satisfy

$$\int_{-\infty}^{\infty} f(t)f(t+\Delta t)\mathrm{d}t = \frac{1}{\tau_{cross}} e^{-\frac{|\Delta t|}{\tau_{cross}/2}}$$

so that the spike trains will have exponentially decaying cross-correlation of time scale $\tau_{cross}/2$. To calculate the function $f(t)$, we construct Toeplitz matrix of $f(t)$ and $\frac{1}{\tau_{cross}} e^{-\frac{|\Delta t|}{\tau_{cross}/2}}$, denoting as $F$ and $X$. So the equation above can be written as

$$FF^T = X$$

where both $F$ and $X$ are symmetric. If $X$ can be diagonalized as

$$X = P\Lambda P^{-1}$$

then the desired matrix $F$ will be

$$F = P\sqrt{\Lambda} P^{-1}$$

and then the middle row of $F$ is taken as the desired $f(t)$. To make this method work, diagonal elements of $\Lambda$ must be non-negative. The Fourier bases are the eigenvectors of the Toeplitz matrix $X$ when its size is sufficiently large, and the eigenvalues are just

$$\int_{-\infty}^{\infty} e^{-i\omega t} \frac{1}{\tau_{cross}} e^{-\frac{|t|}{\tau_{cross}/2}} dt = \frac{1}{1+\tau_{cross}^2 \omega^2/4} > 0$$

Thus $X$ here is always positive definite.

*Model Auto*:

Spikes trains are Gamma processes with inter-spike intervals following the distribution

$$p(x\mid\alpha,\beta) = \frac{1}{\Gamma(\alpha)\beta^\alpha} x^{\alpha-1} e^{-x/\beta}$$

The rate of the Gamma process is $\beta/\alpha$, and the coefficient of variance is $1/\sqrt{\alpha}$.

We use $\alpha$ to control the burstiness/regularity of the spike train, while adjusting $\beta$ to keep the firing rate at 20Hz. The spike train becomes more bursty when $\alpha$ is smaller, and more regular when $\alpha$ is large.

*Model Sync-Auto*:

This model generates synchronous firing patterns with controllable auto-temporal structure.
Spike trains are generated in a way similar to Model Sync 3, except that the spike trains are inhomogeneous Gamma processes with shape parameter $\alpha_{rescale}$ and time-averaged rate $\langle r_0(t)\rangle = 20\text{Hz}$, and the occurrence of firing events is a Gamma process with shape parameter $\alpha_{events}$ and rate $\langle r_0(t)\rangle/p$. $\beta$ values of the Gamma processes are adjusted to keep their rates unchanged at different $\alpha$. $CV_{events} = 1/\sqrt{\alpha_{events}}$, $CV_{rescale} = 1/\sqrt{\alpha_{rescale}}$.

Inhomogeneous Gamma processes with time-averaged rate $\langle r_0(t)\rangle$, shape parameter $\alpha_{rescale}$ and duration $T$ are generated as follows. Suppose $\Lambda(t) = \int_0^t r_0(\tau)d\tau$ is the accumulative function of the firing rate $r_0(t)$, we first generate homogeneous Gamma processes of rate $\Lambda(T)\langle r_0(t)\rangle/T$, shape parameter $\alpha_{rescale}$ and duration $\Lambda(T)$ in the rescaled time (see **Section S2.5** and eq.S18), then project these Gamma processes to the normal time using $\Lambda^{-1}(t)$.

*Model Long Tail*:

This model generates long-tailed distributed firing rates for the non-apical neurons in a dendritic or axonal motif, the firing rate of the apical neuron is always kept at $r_0 = 20\text{Hz}$.

The firing rates of the non-apical neurons are lognormal distributed as

$$p(x\mid m, s) = \frac{1}{sx\sqrt{2\pi}} \exp\left(-\frac{(\ln x - m)^2}{2s^2}\right)$$

The mean of this distribution is at $\exp(m+\frac{s^2}{2})$. Parameter $s$ is used to control the shape, while $m$ is accordingly adjusted to keep the mean at $r_0 = 20\text{Hz}$. This distribution is a $\delta$ function when $s = 0$, and gradually becomes long tailed when $s$ increases.

This model can be also combined with Model Sync 3, Model Auto or Model Sync-Auto to introduce heterogeneity of rates into the spike patterns with other aspects of pattern structure.

*Model Cross-correlation*:

This model generates spike trains in which the cross-correlations between the apical neuron and different non-apical neurons are heterogeneous.

The spike train $\mathcal{T}_0$ of the apical neuron in a dendritic motif is a Poisson process of rate $r_0 = 20\text{Hz}$. To generate the spike train $\mathcal{T}_a$ of the $a$th non-apical neuron, we do as follows: for each spike at time $t_i$ in $\mathcal{T}_0$, $\mathcal{T}_a$ has a probability $q$ to have a spike at $t_i - \tau_{delay} - \tau_a$, with $\tau_{delay}$ being the axonal delay, and $\tau_a \in [-\varepsilon_0 + \Delta t, \varepsilon_0 + \Delta t]$ being a fixed value for $a$th non-apical neuron, and then a Poisson train of rate $(1-q)r_0$ is superimposed onto $\mathcal{T}_a$. In this way, all the neurons in the dendritic motif have rate $r_0$ and the cross-correlation between the $a$th non-apical neuron and the apical neuron is $C_{cross}(t - \tau_{delay}) = q\delta(t - \tau_a)$.

In the case of a dendritic motif coupling with many axonal motifs (**S8E and S9 Figs**), we first generate the spike trains of the neurons in the coupled dendritic motif, i.e. $\mathcal{T}_0$ and $\mathcal{T}_a$ $(a = 1, 2, \cdots)$ (defined in the paragraph above), according to the method above. To generate the spike train $\mathcal{T}_{ba}$ of the $b$th non-apical neuron in the $a$th axonal motif, we do as follows: for each spike at time $t_i$ in $\mathcal{T}_a$, $\mathcal{T}_{ba}$ has a probability $q$ to have a spike at $t_i + \tau_{delay} + \tau_a$ or $t_i - \tau_{delay} - \tau_a$ for $\Delta t \geq 0$ or $\Delta t < 0$ in **S8E Fig**, and then a Poisson train of rate $(1-q)r_0$ is superimposed onto $\mathcal{T}_{ba}$. In this way, the cross-correlation between the apical neuron of the $a$th axonal motif and all its non-apical neurons is uniformly $C_{cross,a}(t - \tau_{delay}) = q\delta(t - \tau_a)$ or $C_{cross,a}(t + \tau_{delay}) = q\delta(t + \tau_a)$ for $\Delta t \geq 0$ or $\Delta t < 0$ respectively. The cross-correlation between the $a$th non-apical neuron and the apical neuron in the coupled dendritic motif is $C_{cross}(t - \tau_{delay}) = q\delta(t - \tau_a)$ (see the previous paragraph), so under STDP, the weight change of the $a$th link in the coupled dendritic motif $\Delta w_{0a}$ is positively or negatively correlated with the mean change in the $a$th axonal motif $\Delta \bar{w}_a$ for $\Delta t \geq 0$ or $\Delta t < 0$ (**S8E Fig**).

### Section S5.2: Spike Pattern Analysis

Here are the methods we used to analyze the pattern structure of the excitatory population in the LIF network (**S12 Fig**).

*Synchronous Firing*:

We used three parameters $p_{async}$, $\tau_{inner}$ and $\tau_{outer}$ to quantify the rate fluctuation of the excitatory population in asynchronous states (**S12A Fig**). For $p_{async}$, we first calculated the temporal firing rate of the excitatory population according to the spike numbers within bins of 0.1ms, then defined $p_{async}$ as the standard deviation of the binned firing rates versus their mean value. For $\tau_{inner}$ and $\tau_{outer}$, we first calculated the connected auto-correlation $C_{auto,+}(\tau) = \langle r(t)r(t+\tau)\rangle - \langle r(t)\rangle^2$ using these binned firing rates above. We found that $C_{auto,+}(\tau)$ oscillated, and the oscillation amplitude gradually decayed to zero as $|\tau|$ increased. Therefore, we used $\tau_{inner}$ to quantify the time scale of its oscillation, and used $\tau_{outer}$ to quantify the time scale of the decay of its amplitude. $\tau_{inner}$ was defined as the duration between the two times at which $C_{auto,+}(\tau)$ first dropped below 10% of $C_{auto,+}(0)$ toward positive and negative directions; and $\tau_{outer}$ was defined as the duration between the two times at which $C_{auto,+}(\tau)$ last dropped below 10% of $C_{auto,+}(0)$ toward these two directions.

For synchronous states, we used $p_{sync}$ and $\tau_{cross}$ to quantify the mean strength and duration of firing events (**S12B Fig**). We first calculated the temporal rate of the excitatory population in bins of 0.1ms, then filtered these data using Gaussian window of $\sigma_{window} = 2\text{ms}$. Numerically, *firing events* were defined as sequential bins in which the filtered rates were above a small threshold 0.0001. $p_{sync}$ was estimated as the average spike number per neuron within a single firing event, and $\tau_{cross}$ was defined as the average duration between the two bins at which the unfiltered binned firing rate dropped below 10% of its peak value within a firing event.

*Auto-temporal Structure*:

To calculate $CV_{rescale}$ (**S12C Fig**), we first ordered all the spikes in the population (essentially shuffled the spike trains using Rescaling Shuffle, see **Section S3**), then averaged the *CV* values of the ordered indexes over all the neurons which fired more than 5 spikes during the simulation.

$CV_{events}$ in synchronous states was defined as the *CV* value of the mean times of the firing events (**S12D Fig**).

*Heterogeneity of Cross-correlations*:

HCC (abbreviation for heterogeneity of cross-correlation) index (**S12EF Fig**) was used to quantify the heterogeneity of cross-correlations. It was defined and calculated as follows: for link $a \to b$ and each spike $t_i$ of neuron $a$, we denoted $\Delta n_{i,a\to b}$ as the spike number of neuron $b$ within the interval $[t_i + \tau_{delay}, t_i + \tau_{delay} + \tau_{STDP})$ minus the spike number of neuron $b$ within the interval $[t_i + \tau_{delay} - \tau_{STDP}, t_i + \tau_{delay})$. We then defined $\Delta n_{a\to b} = \sum_i n_{i,a\to b}$, which quantifies the tendency that neuron $b$ fire after neuron $a$ within the time scale of STDP. And HCC index was defined as the standard deviation of $\Delta n_{a\to b}$ over all the links in the network, which quantifies the heterogeneity of cross-correlations.

## Section S5.3: Developmental Functions of Retinal Waves

The network that we used is a two-layered feedforward network (**S13A Fig**). The first layer contains two groups, with 100 neurons in each group. Activities of these neurons are controlled by a spike generating model. In this model, the occurrence of firing events in each group is a Poisson process with rate $r_0 / p_{intra}$, with $p_{intra}$ being the probability that a neuron fire in a firing event, and the firing rate of neurons are kept at $r_0 = 20\text{Hz}$ when $p_{intra}$ changes. Within all the firing events of a group, $p_{inter}$ portion of them occurs simultaneously with a firing event in the other group (see **S13B Fig** for the spike pattern). Technically, the middle times of the firing events of the two groups are generated like this: two Poisson processes of rate $(1 - p_{inter})r_0 / p_{intra}$ are first generated, then a Poisson process of rate $p_{inter} r_0 / p_{intra}$ is generated and superimposed onto both of them. All the spikes in a firing event are then jittered around the middle time of the firing event by a randomly chosen value within $[-\tau_{cross}/2, \tau_{cross}/2]$. In this study, we fix $\tau_{cross}$ at 2ms. The second layer is a LIF neuron with the same parameter as the excitatory neurons in **Methods** in the main text except for the refractory period $\tau_{refractory} = 1\text{ms}$. The axons also have delay $\tau_{delay} = 1\text{ms}$, and intrinsic homeostasis [9] is also implemented to keep the firing rate of the LIF neuron around 20Hz by adjusting the firing threshold of the LIF neurons $\theta_E$ every 10ms:

$$\theta_E(t) = \theta_E(t - 10\text{ms}) + c(r(t) - r_0)$$

where $r(t)$ is the firing rate of the LIF neuron in the past 1000ms, $r_0 = 20\text{Hz}$, $c = 0.001\text{mV}\cdot\text{s}$. The initial conductance between the two layers is 0.15nS, the STDP parameters are $A_p = A_d = 3.75 \times 10^{-4}\text{nS}$, the parameters for dendritic homeostasis are $w_{bound} = 0.15\text{nS}$, $\varepsilon = 0.01$. Intrinsic homeostasis started immediately at the beginning of the simulation, while STDP and dendritic homeostasis started after 10s of transient period, waiting for the adjustment of $\theta_E$ by intrinsic homeostasis.

## Sections S6: Miscellaneous

To help readers better understand our simulation results, here we explain the physical pictures behind some phenomena observed in our simulations.

### Section S6.1: Why the efficacy variability for $\tau_{cross} < \tau_{delay}$ is usually larger than that for $\tau_{cross} > \tau_{delay}$, if spikes trains are generated using Model Sync 3 (S1D Fig)?

When spikes trains are generated using Model Sync 1, the efficacy variability for $\tau_{cross} > \tau_{delay}$ is large because of synapse splitting, and the efficacy variability for $\tau_{cross} < \tau_{delay}$ is small because of synapse correlating. Model Sync 3 destroys synapse correlating by introducing variety of the spike numbers of the non-apical neurons in each firing event. However, we found that the efficacy variability for $\tau_{cross} < \tau_{delay}$ usually surpassed that for $\tau_{cross} > \tau_{delay}$ after using Model Sync 3 (**S1D Fig**). To understand this, note that in Model Sync 3, the number of spikes fired by a non-apical neuron during a firing event follows Poisson distribution $Poi(\lambda_1)$, with $\lambda_1$ being the mean and variance of this distribution. When $\tau_{cross} < \tau_{delay}$, all these spikes depress the corresponding synapse under STDP. If for simplicity, we suppose that every spike depresses the synapse by the same value $-\Delta w_1$, then the variance of the total depression value after a firing event is $\text{Var}(\tau_{cross} < \tau_{delay}) = \lambda_1 (\Delta w_1)^2$. When $\tau_{cross} > \tau_{delay}$, some non-apical spikes potentiate the synapse by $\Delta w_2$ on average, while the others depress the synapse by $-\Delta w_2$ on average; if again for simplicity, we suppose that the apical neuron always fire at a fixed relative position within a firing event, say, the middle point, then the variance of the total potentiation (depression) value after a firing event is $\lambda_p (\Delta w_2)^2$ ($\lambda_p (\Delta w_2)^2$), with $\lambda_p$ ($\lambda_d$) being the mean number of spikes which potentiate (depress) the synapse during a firing event, and $\lambda_p + \lambda_d = \lambda_1$. Therefore, the total variance of the STDP updatings after a firing event can be estimated as the summation of the variances contributed by the potentiation and depression processes, which is $\text{Var}(\tau_{cross} > \tau_{delay}) \approx (\lambda_p + \lambda_d)(\Delta w_2)^2 = \lambda_1 (\Delta w_2)^2$. Therefore, synapse splitting does not help to increase the efficacy variability when $\tau_{cross} > \tau_{delay}$ in Model Sync 3, and the difference between $\text{Var}(\tau_{cross} < \tau_{delay})$ and $\text{Var}(\tau_{cross} > \tau_{delay})$ mainly depends on the difference between $\Delta w_1$ and $\Delta w_2$. When $\tau_{cross} > \tau_{delay}$ especially when $\tau_{cross}$ becomes large, $\Delta w_2$ gets small because of the exponentially decaying STDP time window. This is the reason why the efficacy variability for $\tau_{cross} < \tau_{delay}$ usually surpasses that for $\tau_{cross} > \tau_{delay}$ after using Model Sync 3 (**S1D Fig**).

### Section S6.2: Why the regularity of spike trains increases the correlation between the total potentiation and total depression values (S3E Fig)?

**S3E Fig** shows that $\rho$ decreases with *CV*, which means that the regularity of spike trains increases the correlation between the total potentiation and total depression values. To understand this, consider three adjacent spikes of the apical neuron in a dendritic motif $\{t_{0,1}, t_{0,2}, t_{0,3}\}$ and two adjacent spikes of the *a*th non-apical neuron $\{t_{a,1}, t_{a,2}\}$. Because the firing rates of these two neurons are the same and the spike trains are regular, if $t_{0,1} < t_{a,1} < t_{0,2}$ then it is very likely that $t_{0,1} < t_{a,1} < t_{0,2} < t_{a,2} < t_{0,3}$. If we move $t_{a,1}$ a little earlier, then the STDP depression caused by pairing $(t_{0,1}, t_{a,1})$ gets stronger, while the potentiation caused by pairing $(t_{a,1}, t_{0,2})$ gets weaker, which induces positive correlation between depression and potentiation values. As spike trains are regular, moving $t_{a,1}$ earlier also moves $t_{a,2}$ earlier at the same time, so that the depression caused by pairing $(t_{0,2}, t_{a,2})$ also gets stronger, and the potentiation caused by pairing

($t_{a,2}, t_{0,3}$) also gets weaker. Thus, we see that regularity of spike trains increases the correlation between total depression and potentiation values under homogeneous firing rate.

## Section S6.3: Why burstiness increases $\sum_i \sum_j \sum_k \left( \text{Var}_a \left( \Delta w_{a,k}(t_i, t_{a,j}) \right) \right)$, i.e. the variance of the synapses when all the three types of correlations induced by auto-temporal structure are absent (S3B Fig)?

Suppose a spike $t_i$ of the apical neuron, and the spikes $\{t_{a,1}\}_a$ of all the non-apical neurons which are immediately after $t_i - \tau_{delay}$ (see **Section S2.4** for the indexing of $j$, here $j=1$). Suppose the mean value of the inter-spike intervals is $\overline{\Delta t}$, then if the spike trains approach strictly regular, $t_{a,1} - t_i$ will be uniformly distributed within $[0, \overline{\Delta t}]$ across $a$. When the spike trains get burstier, the distribution of $t_{a,1} - t_i$ get wider, which increases the variance of the depression value $\Delta w_{a,d}(t_i, t_{a,1})$ across $a$. Similar reason also applies to the other indexes of $j$ as well as the potentiation process.

## Section S6.4: Explanations of the dynamic patterns of the LIF network

To help readers understand the dynamics of the LIF network, we briefly explain some phenomena shown in **S11** and **S12 Figs**.

The burstiness of spikes in asynchronous states (**S12C Fig**) may be due to the strong excitatory and inhibitory couplings in our network model [10]. The regularity represented by small $CV_{rescale}$ in synchronous states (**S12C Fig**) is because of the regular firing due to the fixed refractory period and the supra-threshold input in each synchronization period. The large rate heterogeneity in asynchronous states (**S12G Fig**) is due to the quenched Gaussian distribution input in random networks and the nonlinear conductance-rate relationship in balanced state [11]; and the reduction of rate heterogeneity in synchronous states is because that in each synchronization period fast excitatory currents and slow inhibitory currents cause transient supra-threshold inputs, which transiently push neurons into the regime of linear conductance-rate relation [12], and even saturate their rates at $1/\tau_{refractory}$ when the inputs are too strong. The reason why synchronous states tend to depress synaptic strength (**S12H Fig**) is already explained in **Section S2.3**, also see [5]. The heterogeneity of cross-correlations in asynchronous states (**S12E Fig**) is due to the cellular response properties and the network structure, such as unidirectional connection, common inputs etc.[13,14].

Another interesting phenomenon is the asymmetry of the rising and decaying phases of the synchronization periods in synchronous states. After carefully looking at the spike patterns in synchronous states, we found that neurons tend to start to fire one by one at the rising phases of the synchronous periods, while they tend to shut down simultaneously at the decaying phases. To understand this, note that at the early rising phase, inhibitory neurons do not fire, with the inhibitory currents into the excitatory neurons decaying with time. Therefore, the neurons which receive larger number of excitatory connections and smaller number of inhibitory connections tend to start to fire before those which receive smaller number of excitatory connections and larger number of inhibitory connections. As inhibitory neurons have smaller membrane time scale, their firing rates can quickly arise once most excitatory neurons start to fire; and then the suddenly increased inhibitory currents quickly shut down all excitatory neurons.

One consequence of this rising-decaying asymmetry is the difference of the efficacy variability between under only dendritic homeostasis and under only axonal homeostasis (**S11AB Fig, lower panels**). As we discussed in **Section S2.1**, the dendritic and axonal homeostasis have exactly the same effect as long as the spike pattern is statistically time-reversal invariant, but this rising-decaying asymmetry apparently destroy this time-reversal symmetry. In our model, we suppose that axons have delay $\tau_{delay}$, and STDP depends on the difference between the time when the post-synaptic neuron fires and the time when the pre-synaptic spike arrives at the post-synaptic neuron. Therefore, because of the axonal delay, in a dendritic motif, the spike of the apical neuron in a firing event is closer to the "zigzag" rising phase of the spikes of the non-apical neurons during the firing event, which enlarges the efficacy variability; however, in an axonal motif, the spike of the apical neuron is closer to the "clear-cut" decaying phase of the spikes of the non-apical neurons during the firing event, which reduces the efficacy variability. This is why the efficacy variability is larger under dendritic homeostasis than under axonal homeostasis in synchronous states (**S11AB Fig, lower panels**).

Another consequence of this rising-decaying asymmetry is the large heterogeneity of cross-correlations in synchronous states (**S12F Fig**). As we discussed above, the neurons which receive larger number of excitatory connections and smaller number of inhibitory connections tend to start to fire before those which receive smaller number of excitatory connections and larger number of inhibitory connections in a synchronization period. This results in different cross-correlations between neuron pairs, depending on the connection details. WSWE reduces the heterogeneity of cross-correlations (**S12F Fig**) by randomly shuffling the spike sequences of all neurons in each synchronization period.

Another interesting problem is how the heterogeneity of cross-correlations in synchronous states contributes to $\text{Corr}(\Delta w_{ab}, \Delta \bar{w}_b)$ (**S11D Fig**). From **S8 Fig**, we see that synchronous firing and heterogeneity of firing rates always positively contributes $\text{Corr}(\Delta w_{ab}, \Delta \bar{w}_b)$, thereby decreasing the efficacy variability through the coupling of dendritic and axonal homeostasis; but the effect of heterogeneity of cross-correlations depends on the cross-correlation details in the spike patterns. As discussed above, cross-correlations in synchronous states come from the "zigzag" rising phases of synchronous periods, which emerges from the underlying connection details, therefore each neuron starts to fire at an almost fixed relative time during the "zigzag" rising phases. Now we suppose an excitatory neuron $b$ in the network. If neuron $b$ fires early in each synchronous period, then its cross-correlations with most neurons it targets to tend to increase the strengths of the synapses between them under STDP; on the contrary, if neuron $b$ fires later in each synchronous period, then its cross-correlations with most neurons it targets to tend to decreases the strengths of these synapses. Therefore, the heterogeneity of cross-correlations here actually positively contributes to $\text{Corr}(\Delta w_{ab}, \Delta \bar{w}_b)$, thereby decreasing the efficacy variability through the coupling of dendritic and axonal homeostasis. Therefore, in the spike patterns of our LIF network, synchronous firing, heterogeneity of rates and heterogeneity of cross-correlations together decrease efficacy variability through the coupling of dendritic and axonal homeostasis, which is the reason why the efficacy variability is so small in synchronous states when dendritic and axonal homeostasis coexist (**S11C Fig, lower panel**).